\documentclass [12pt]{article}
\RequirePackage{epsfig}
\RequirePackage{wrapfig}
\RequirePackage{feynmf}
\usepackage{amssymb}

\begin{document}
 
\newenvironment{malist}
         {\begin{list}{\tiny{$\bullet$}}
         {\setlength{\parsep}{3pt}
         \addtolength{\leftmargin}{0mm}
         \setlength{\topsep}{0pt}
         \setlength{\itemsep}{0pt}}}{\end{list}}

\newcommand{\etal}  {\mbox{\it et al.}}
\def\nim#1#2#3  {{\em Nucl. Instr. Meth.} {\bf#1} (#2) #3.}
\def\prev#1#2#3 {{\em Phys. Rev.} {\bf#1} (#2) #3.}
\def\np#1#2#3   {{\em Nucl. Phys.} {\bf#1} (#2) #3.}
\def\pl#1#2#3   {{\em Phys. Lett.} {\bf#1} (#2) #3.}
\def\prl#1#2#3  {{\em Phys. Rev. Lett.} {\bf#1} (#2) #3.}
\def\zp#1#2#3   {{\em Zeit. Phys.} {\bf#1} (#2) #3.}
\def\ppc#1      {{\em Preprint CERN,} CERN-PPE/#1}
\def\ppcep#1      {{\em Preprint CERN,} CERN-EP/#1}
\def\NPB#1#2#3{{\rm Nucl.~Phys.} {\bf{B#1}} (19#2) #3}
\def\PLB#1#2#3{{\rm Phys.~Lett.} {\bf{B#1}} (19#2) #3}
\def\PRD#1#2#3{{\rm Phys.~Rev.} {\bf{D#1}} (19#2) #3}
\def\PRL#1#2#3{{\rm Phys.~Rev.~Lett.} {\bf{#1}} (19#2) #3}
\def\ZPC#1#2#3{{\rm Z.~Phys.} {\bf C#1} (19#2) #3}
\def\PTP#1#2#3{{\rm Prog.~Theor.~Phys.} {\bf#1}  (19#2) #3}
\def\MPL#1#2#3{{\rm Mod.~Phys.~Lett.} {\bf#1} (19#2) #3}
\def\PR#1#2#3{{\rm Phys.~Rep.} {\bf#1} (19#2) #3}
\def\RMP#1#2#3{{\rm Rev.~Mod.~Phys.} {\bf#1} (19#2) #3}
\def\HPA#1#2#3{{\rm Helv.~Phys.~Acta} {\bf#1} (19#2) #3}
\def\NIMA#1#2#3{{\rm Nucl.~Instr.~and~Meth.} {\bf#1} (19#2) #3}
\def\CPC#1#2#3{{\rm Comp.~Phys.~Comm.} {\bf#1} (19#2) #3}
\def \GUT { grand unified theories }
\def \wrt { with respect to }
\def \cft { conformal field theory }
\def \dof { degrees of freedom }
\def \bef { $\beta -$function }
\def \bet { beta function }
\def \mow { modular weights }
\def\loe { low energy }
\def \km { Kac-Moody }
\def \susy { supersymmetry }
\def \susyq { supersymmetric }
\def \sig { $\sigma -$model }
\def \ws { world sheet }
\def \tres { threshold corrections }
\def\L {\Lambda }
\def\l {\lambda } 
\def\ka {\kappa } 
\def \t {\theta }
\def \vt {\vartheta }
\def\a {\alpha }
\def\dh {\partial }
\def \d {\delta }
\def \D {\Delta }
\def \bq {\bar q }
\def \bQ {\bar Q}
\def \g {\gamma }
\def \G {\Gamma }
\def \O {\Omega }
\def \o {\omega }
\def \b {\beta }
\def \S {\Sigma }
\def \s {\sigma }
\def \e {\epsilon }
\def \cc { coupling constant }
\def \ccs {coupling constants }
\def \ph { phenomenology } 
\def \pheny { phenomenologically } 
\def \rhs { right hand side } 
\def \lhs {left  hand side } 
\def \ren { renormalization } 
\def \rge { renormalization group equations}
\def \ud { {1 \over 2} } 
\def \ut { {1 \over 3} } 
\def \td {{3 \over  2 }} 
\def \SM { Standard Model }
\def \q {\eqno  }
\def \mssm { minimal supersymmetric standard model } 
\def \he  { high energy }
\def \st { string theory }
\def \ft { field theory }
\def \bea {\begin{equation}}
\def \eea {\end{equation}}
\def \tchi  {{\tilde \chi }}
\def \Eslash {E \kern-.5em\slash}
\def \pslash {p \kern-.5em\slash}
\def \kslash {k \kern-.5em\slash}
\def \jmp { Jour. Math. Phys. }
\def \ptp { Prog. Theor. Phys. }
\def \ap {Ann. Phys. (NY) }
\def \z {Z. Phys. } 
\def \pr  { Phys. Rev. }
\def \np { Nucl. Phys. }  
\def \arr { Ann. Rev. Nucl. Sci. } 
\def \ar { Ann. Rev. Nucl. Part. Sci. } 
\def \pp { Phys. Reports } 
\def \je { Sov. Phys. JETP } 
\def \prl { Phys. Rev. Lett. } 
\def \pl { Phys. Lett. }
\def \rmp { Rev. Mod. Phys. } 
\def \nc { Nuovo Cimento }
\def \jnp { Sov. Jour. Nucl. Phys. }

\newcommand{\Zn}      {\mbox{$ {\mathrm Z}^0                               $}}
\newcommand{\Wp}      {\mbox{$ {\mathrm W}^+                               $}}
\newcommand{\Wm}      {\mbox{$ {\mathrm W}^-                               $}}
\newcommand{\Hp}      {\mbox{$ {\mathrm H}^+                               $}}
\newcommand{\WW}      {\mbox{$ {\mathrm W}^+{\mathrm W}^-                  $}}
\newcommand{\ZZ}      {\mbox{$ {\mathrm Z}^0{\mathrm Z}^0                  $}}
\newcommand{\GG}      {\mbox{$ {\mathrm \gamma}{\mathrm \gamma}            $}}
\newcommand{\HZ}      {\mbox{$ {\mathrm H}^0 {\mathrm Z}^0                 $}}
\newcommand{\GW}      {\mbox{$ \Gamma_{\mathrm W}                          $}}
\newcommand{\gs}      {\mbox{$ G_{\mathrm s}                          $}}
\newcommand{\gl}      {\mbox{$ G_{\mathrm L}                          $}}
\newcommand{\gr}      {\mbox{$ G_{\mathrm R}                          $}}
\newcommand{\gn}      {\mbox{$ G_{\mathrm \nu}                        $}}
\newcommand{\gmk}     {\mbox{$ G_{\mathrm k}                          $}}
\newcommand{\gml}     {\mbox{$ G_{\mathrm l}                          $}}
\newcommand{\gsg}      {\mbox{$ \Gamma_{\mathrm s}                        $}}
\newcommand{\glg}      {\mbox{$ \Gamma_{\mathrm L}                        $}}
\newcommand{\grg}      {\mbox{$ \Gamma_{\mathrm R}                        $}}
\newcommand{\gng}      {\mbox{$ \Gamma_{\mathrm \nu}                      $}}
\newcommand{\gmkg}     {\mbox{$ \Gamma_{\mathrm k}                        $}}
\newcommand{\gmlg}     {\mbox{$ \Gamma_{\mathrm l}                        $}}
\newcommand{\XO}      {$\widetilde{\chi}^0$}
\newcommand{\XOI}     {$\widetilde{\chi}_1^0$}
\newcommand{\XOII}    {$\widetilde{\chi}_2^0$}
\newcommand{\XOJ}     {$\widetilde{\chi}_i^0$}
\newcommand{\XPI}{$\widetilde{\chi}_1^+$}
\newcommand{\XMI}{$\widetilde{\chi}_1^-$}
\newcommand{\XPM}{$\widetilde{\chi}^{\pm}$}      
 
\newcommand{\nue}      {\mbox{$ \nu_ e                           $}}
\newcommand{\num}      {\mbox{$ \nu_\mu                          $}}
\newcommand{\nut}      {\mbox{$ \nu_\tau                         $}}
 
\newcommand{\Zg}      {\mbox{$ \Zn \gamma                                  $}}
\newcommand{\ecms}     {\mbox{$ \sqrt{s}                                    $}}
\newcommand{\ee}      {\mbox{$ {\mathrm e}^+ {\mathrm e}^-                 $}}
\newcommand{\eeWW}    {\mbox{$ \ee \rightarrow \WW                         $}}
\newcommand{\MeV}     {\mbox{$ {\mathrm{MeV}}                              $}}
\newcommand{\MeVc}    {\mbox{$ {\mathrm{MeV}}/c                            $}}
\newcommand{\MeVcc}   {\mbox{$ {\mathrm{MeV}}/c^2                          $}}
\newcommand{\GeV}     {\mbox{$ {\mathrm{GeV}}                              $}}
\newcommand{\GeVc}    {\mbox{$ {\mathrm{GeV}}/c                            $}}
\newcommand{\GeVcc}   {\mbox{$ {\mathrm{GeV}}/c^2                          $}}
\newcommand{\TeV}     {\mbox{$ {\mathrm{TeV}}                              $}}
\newcommand{\TeVc}    {\mbox{$ {\mathrm{TeV}}/c                            $}}
\newcommand{\TeVcc}   {\mbox{$ {\mathrm{TeV}}/c^2                          $}}
\newcommand{\MZ}      {\mbox{$ m_{{\mathrm Z}^0}                           $}}
\newcommand{\MW}      {\mbox{$ m_{\mathrm W}                               $}}
\newcommand{\GF}      {\mbox{$ {\mathrm G}_{\mathrm F}                     $}}
\newcommand{\MH}      {\mbox{$ m_{{\mathrm H}^0}                           $}}
\newcommand{\MT}      {\mbox{$ m_{\mathrm t}                               $}}
\newcommand{\GZ}      {\mbox{$ \Gamma_{{\mathrm Z}^0}                      $}}
\newcommand{\TT}      {\mbox{$ \mathrm T                                   $}}
\newcommand{\UU}      {\mbox{$ \mathrm U                                   $}}
\newcommand{\alphmz}  {\mbox{$ \alpha (m_{{\mathrm Z}^0})                  $}}
\newcommand{\alphas}  {\mbox{$ \alpha_{\mathrm s}                          $}}
\newcommand{\alphmsb} {\mbox{$ \alphas (m_{\mathrm Z})
                               _{\overline{\mathrm{MS}}}                   $}}
\newcommand{\alphbar} {\mbox{$ \overline{\alpha}_{\mathrm s}               $}}
\newcommand{\qqg}     {\mbox{$ {\mathrm q}\bar{\mathrm q}\gamma            $}}
\newcommand{\Wev}     {\mbox{$ {\mathrm{W e}} \nu_{\mathrm e}              $}}
\newcommand{\Zvv}     {\mbox{$ \Zn \nu \bar{\nu}                           $}}
\newcommand{\Zee}     {\mbox{$ \Zn \ee                                     $}}
\newcommand{\ctw}     {\mbox{$ \cos\theta_{\mathrm W}                    $}}
\newcommand{\cwd}     {\mbox{$ \cos^2\theta_{\mathrm W}                  $}}
\newcommand{\cwq}     {\mbox{$ \cos^4\theta_{\mathrm W}                  $}}
\newcommand{\thw}     {\mbox{$ \theta_{\mathrm W}                          $}}
\newcommand{\gamgam}  {\mbox{$ \gamma \gamma                               $}}
\newcommand{\qaqb}    {\mbox{$ {\mathrm q}_1 \bar{\mathrm q}_2             $}}
\newcommand{\qcqd}    {\mbox{$ {\mathrm q}_3 \bar{\mathrm q}_4             $}}
\newcommand{\bbar}    {\mbox{$ {\mathrm b}\bar{\mathrm b}                  $}}
\newcommand{\djoin}   {\mbox{$ d_{\mathrm{join}}                           $}}
\newcommand{\Erad}    {\mbox{$ E_{\mathrm{rad}}                            $}}
\newcommand{\mErad}   {\mbox{$ \left\langle E_{\mathrm{rad}} \right\rangle $}}
 
\newcommand{\sel}     {\mbox{$ \tilde{e}                                $}}
\newcommand{\sell}     {\mbox{$ \tilde{e_L}                                $}}
\newcommand{\selr}     {\mbox{$ \tilde{e_R}                                $}}
\newcommand{\smu}     {\mbox{$ \tilde{\mu}                              $}}
\newcommand{\smul}     {\mbox{$ \tilde{\mu_L}                              $}}
\newcommand{\smur}     {\mbox{$ \tilde{\mu_R}                              $}}
\newcommand{\stau}    {\mbox{$ \tilde{\tau}                             $}}
\newcommand{\staul}    {\mbox{$ \tilde{\tau_L}                             $}}
\newcommand{\staur}    {\mbox{$ \tilde{\tau_R}                             $}}
\newcommand{\snu}      {\mbox{$ \tilde{\nu}                                $}}
\newcommand{\ssup}     {\mbox{$ \tilde{u}                                  $}}
\newcommand{\sdn}     {\mbox{$ \tilde{d}                                   $}}
\newcommand{\sch}     {\mbox{$ \tilde{c}                                   $}}
\newcommand{\sst}     {\mbox{$ \tilde{s}                                   $}}
\newcommand{\stp}     {\mbox{$ \tilde{t}                                   $}}
\newcommand{\sbt}     {\mbox{$ \tilde{b}                                   $}}
\newcommand{\squ}     {\mbox{$ \tilde{q}                                   $}}
\newcommand{\sge}     {\mbox{$ \tilde{g}                                   $}}
\newcommand{\snue}     {\mbox{$ \tilde{\nu_e}                              $}}
\newcommand{\snum}     {\mbox{$ \tilde{\nu_{\mu}}                          $}}
\newcommand{\schi}    {\mbox{$ \tilde{\chi}                               $}}
\newcommand{\sph}    {\mbox{$ \tilde{p}                               $}}
\newcommand{\sm}       {\mbox{$ \tilde{m}                              $}}
\newcommand{\sfe}       {\mbox{$ \tilde{f}                              $}}
\newcommand{\mxo}       {\mbox{$ m_{\chi_o}                              $}}
 
\newcommand{\achi}    {\mbox{$ \tilde{\chi}                               $}}
\newcommand{\achiip}  {\mbox{$ \tilde{\chi^{+}_i}                         $}}
\newcommand{\achijp}  {\mbox{$ \tilde{\chi^{+}_j}                         $}}
\newcommand{\achijn}  {\mbox{$ \tilde{\chi^{-}_j}                         $}}
\newcommand{\achii}   {\mbox{$ \tilde{\chi^{0}_i}                         $}}
\newcommand{\achij}   {\mbox{$ \tilde{\chi^{0}_j}                         $}}
\newcommand{\achiap}    {\mbox{$ \tilde{\chi}^{+}_{1}                    $}}
\newcommand{\achibp}    {\mbox{$ \tilde{\chi}^{+}_{2}                       $}}
\newcommand{\achian}    {\mbox{$ \tilde{\chi}^{-}_{1}                       $}}
\newcommand{\achibn}    {\mbox{$ \tilde{\chi}^{-}_{2}                       $}}
\newcommand{\achia}    {\mbox{$ \tilde{\chi}^{0}_{1}                        $}}
\newcommand{\achib}    {\mbox{$ \tilde{\chi}^{0}_{2}                        $}}
\newcommand{\achic}    {\mbox{$ \tilde{\chi}^{0}_{3}                        $}}
\newcommand{\achid}    {\mbox{$ \tilde{\chi}^{0}_{4}                        $}}
 
\newcommand{\Ptau}    {\mbox{$ P_{\tau}                                    $}}
\newcommand{\mean}[1] {\mbox{$ \left\langle #1 \right\rangle               $}}
\newcommand{\mydeg}   {\mbox{$ ^\circ                                      $}}
\newcommand{\thetabar}{\mbox{$ \theta^*                                    $}}
\newcommand{\phibar}  {\mbox{$ \phi^*                                      $}}
\newcommand{\thetapl} {\mbox{$ \theta_+                                    $}}
\newcommand{\phipl}   {\mbox{$ \phi_+                                      $}}
\newcommand{\thetamin}{\mbox{$ \theta_-                                    $}}
\newcommand{\phimin}  {\mbox{$ \phi_-                                      $}}
\newcommand{\ds}      {\mbox{$ {\mathrm d} \sigma                          $}}
\newcommand{\emis}    {\mbox{$ E_{miss}                                  $}}
\newcommand{\ptr}     {\mbox{$ p_{\perp}                                   $}}
\newcommand{\ptrjet}  {\mbox{$ p_{\perp {\mathrm{jet}}}                    $}}
\newcommand{\Wvis}    {\mbox{$ {\mathrm W}_{\mathrm{vis}}                  $}}
 
\newcommand{\jjlv}    {\mbox{$ j j \ell                         $}}
\newcommand{\jj}      {\mbox{$ j j                              $}}
\newcommand{\lele}    {\mbox{$ \ell \bar{\ell}                  $}}
\newcommand{\eemis}   {\mbox{$ e \mu                            $}}
\newcommand{\emmis}   {\mbox{$ e e                               $}}
\newcommand{\mmmis}   {\mbox{$ \mu \mu                           $}}
\newcommand{\ttmis}   {\mbox{$ \tau \tau                         $}}
\newcommand{\ginv}    {\mbox{$ \gamma + "nothing"                        $}}
\newcommand{\jjll}    {\mbox{$ j j \ell \bar{\ell}                         $}}
\newcommand{\jjjj}    {\mbox{$ j j j j                                     $}}
\newcommand{\jjvv}    {\mbox{$ j j \nu \bar{\nu}                           $}}
\newcommand{\lvlv}    {\mbox{$ \ell \nu \ell \nu                           $}}
\newcommand{\llll}    {\mbox{$ \ell \ell \ell \ell                         $}}

\newcommand{\barre}[1]{%
        \setbox1=\hbox{$#1$} \dimen2=\ht1 \dimen3=\dp1 \dimen4=\wd1
        \setbox2=\hbox{\sl /}
        \dimen1=\wd1 \advance\dimen1 by -\wd2 \divide\dimen1 by 2
        \advance\dimen1 by \wd2 \advance\dimen1 by 0.4pt
        \setbox3=\hbox to \wd1{\hss \box1 \kern -\dimen1 \box2\hss}
        \ht3=\dimen2 \dp3=\dimen3 \wd3=\dimen4
        \box3
        }

\def\tht{\theta_{t}}
\def\st{\widetilde{t}}
\def\stl{\st_{1}}
\def\sth{\st_{2}}
 
\newcommand {\sutry} {supersymmetry}
\newcommand {\rp } {${R}_{p}$\ }
\newcommand {\rpv} {\( \not \! {R}_{p} \) }
\newcommand {\Gv} {$GeV/c^{2}$}
\newcommand {\M} {$MeV/c^{2}$}
\newcommand {\Gc} {$GeV/c$}
\newcommand {\Mc} {$MeV/c$}
\newcommand {\mtwo} {$M_{2}$}
\newcommand {\tb} {$\tan \beta$}
\newcommand {\mzero} {$m_{0}$}
\newcommand {\nump}{  \[ \sum_{n=0}^{n_{0}} \frac{(b+N)^{n}}{n!}\]  }
\newcommand {\denp}{  \[ \sum_{n=0}^{n_{0}} \frac{b^{n}}{n!} \] }

\newcommand{\Labb}{$\lambda_{122}$}
\newcommand{\Lacc}{$\lambda_{133}$}
\newcommand{\epem}{$e^{+} e^{-}$}
\newcommand{\lum}{$\cal L$}
\newcommand {\Mmuplane} {($\mu$, $M_2$) plane}
\newcommand {\Mmuplanes} {($\mu$, $M_2$) planes}
\newcommand{\Emiss}{\( \not \! {E} \) }
\newcommand{\siminf}  {\mbox{$_{\sim}$ {\small {\hspace{-1.em}{$<$}}}       }}
\newcommand{\simsup}  {\mbox{$_{\sim}$ {\small {\hspace{-1.em}{$>$}}}       }}

\newcommand{\Rp}{\mbox{$\not \hspace{-0.15cm} R_p$}}
\newcommand{\lsim}{\raisebox{-1.5mm}{$\:\stackrel{\textstyle{<}}{\textstyle{\sim
}}\:$}}
\newcommand{\gsim}{\raisebox{-0.5mm}{$\stackrel{>}{\scriptstyle{\sim}}$}}
\newcommand{\cm}{\mbox{\rm ~cm}}
\def\GeV{\hbox{$\;\hbox{\rm GeV}$}}
\newcommand{\picob}{\mbox{{\rm ~pb}~}}
\def\figurename{{\bf Figure}}
\def\tablename{{\bf Table}}
 
\newcommand{\dfrac}[2]{\frac{\displaystyle #1}{\displaystyle #2}}
\newcommand{\df}      {$\delta_F^f$}
\newcommand{\db}      {$\delta_B^f$}
\newcommand{\epsif}   {$\varepsilon_F^f$}
\newcommand{\epsib}   {$\varepsilon_B^f$}
\newcommand{\swd}     {$\sin^2\theta$}
\newcommand{\xx}      {\rule[-3mm]{0mm}{8mm}}
\newcommand{\Z}       {$Z^0$}
\newcommand{\qfb}     {$<Q_{FB}>$}
\newcommand{\qtot}    {$<Q_{TOT}>$}
\newcommand{\afb}     {$A_{FB}$}
\newcommand{\afbraw}  {$A_{FB}^{raw}$}
\newcommand{\cmax}{cos\theta_{MAX}}
\newcommand{\ccmax}{cos^2\theta_{MAX}}
\newcommand{\cccmax}{cos^3\theta_{MAX}}
\newcommand{\cmin}{cos\theta_{MIN}}
\newcommand{\ccmin}{cos^2\theta_{MIN}}
\newcommand{\cccmin}{cos^3\theta_{MIN}}
\newcommand{\micron}{\mbox{$\mu {\rm m}$}}
\newcommand{\Zz}{\mbox{${\rm Z}^0$}}
\newcommand{\rphi}{\mbox{$r\phi$}}
\newcommand{\GeVe}{\mbox{${\rm GeV}$}}
\newcommand{\Pt}{\mbox{$p_t$}}
\newcommand{\Gbb}{\mbox{$\Gamma_{\rm b\bar b}$}}
\newcommand{\Gbbpart}{\mbox{$\Gamma_{\rm b\bar b}/\Gamma_{\rm h}$}}
\newcommand{\taub}{\mbox{$\tau_{\rm B}$}}
\def\sss{\scriptscriptstyle}
\def\ssstr{\scriptscriptstyle\rm}
\def\LQCD{\Lambda_{\ssstr QCD}}
\def\ce{CERN-EP/}
\def\cep{CERN-PPE/}
\def\eett{e^+e^-\to \tau^+\tau^-}
   \def\LMSB{\Lambda_{\overline{\ssstr MS}}}
\def\br{branching ratio}

\newcommand {\lb} {\lambda}
\newcommand {\lbp} {\lambda'}
\newcommand {\lbpp} {\lambda''}
\newcommand {\chio} {\tilde{\chi}^0}
\newcommand {\chip} {\tilde{\chi}^+}
\newcommand {\chim} {\tilde{\chi}^-}
\newcommand {\chipm} {\tilde{\chi}^{\pm}}
\newcommand {\sq} {\tilde{q}}
\newcommand {\sqb} {\overline{\tilde{q}}}
\newcommand {\slep} {\tilde{l}}
\newcommand {\slepb} {\overline{\tilde{l}}}
\newcommand {\qL} {\tilde{q}_L}
\newcommand {\qR} {\tilde{q}_R}
\newcommand {\uL} {\tilde{u}_L}
\newcommand {\uR} {\tilde{u}_R}
\newcommand {\dL} {\tilde{d}_L}
\newcommand {\dR} {\tilde{d}_R}
\newcommand {\cL} {\tilde{c}_L}
\newcommand {\cR} {\tilde{c}_R}
\newcommand {\sL} {\tilde{s}_L}
\newcommand {\sR} {\tilde{s}_R}
\newcommand {\tL} {\tilde{t}_L}
\newcommand {\tR} {\tilde{t}_R}
\newcommand {\stb} {\overline{\tilde{t}}_1}
\newcommand {\sbot} {\tilde{b}_1}
\newcommand {\sbotb} {\overline{\tilde{b}}_1}
\newcommand {\bL} {\tilde{b}_L}
\newcommand {\bR} {\tilde{b}_R}
\newcommand {\nuL} {\tilde{\nu}_L}
\newcommand {\nuR} {\tilde{\nu}_R}
\newcommand {\snub} {\bar{\tilde{\nu}}}
\newcommand {\eL} {\tilde{e}_L}
\newcommand {\eR} {\tilde{e}_R}
\newcommand {\et} {{E\hspace{-0.5em}/}_T}

\newcommand{\vev}[1]{%
	\langle #1 \rangle
	}

\newcommand{\svev}[1]{%
	|\langle #1 \rangle|^2
	}


\vspace*{1.3cm}
\begin{center}
{\Huge Report of the group on the $R$-parity violation}
\end{center}

\vspace*{1.cm}
\begin{center}
\normalsize {\bf
   {R. Barbier$^5$, C. B\'erat$^6$, M. Besan\c con$^{12}$, P. Binetruy$^{1,10}$,\\
    G.Bordes$^2$, 
    F. Brochu$^9$, P. Bruel$^8$, 
    F.Charles$^{14}$, C. Charlot$^8$,\\ M. Chemtob$^{13}$, P. Coyle$^3$,
    M. David$^{12}$,
    E. Dudas$^{1,10}$,\\
    D. Fouchez$^3$, C. Grojean$^{13}$, M. Jacquet$^7$, S. Katsanevas$^5$,
    S. Lavignac$^{4,10,11}$,\\
    F. Ledroit$^6$, R. Lopez$^6$, A. Mirea$^3$, G. Moreau$^{13}$, 
    C. Mulet-Marquis$^6$,\\ 
    E. Nagy$^3$, F. Naraghi$^6$, R. Nicolaidou$^{6,12}$, P. Paganini$^8$,
    E. Perez$^{12}$,\\
    G. Sajot$^6$, C.A. Savoy$^{1,13}$, Y. Sirois$^8$, C. Vall\'ee$^3$}
}
\end{center}
\vspace*{1.cm}
\footnotesize { \it {
$^1$ CERN, theory division, CH-1211 Geneva 23, Switzerland \\
$^2$ Coll\`ege de France, Lab. de Physique Corpusculaire,
IN2P3-CNRS; FR-75231, Paris Cedex 05, France\\
$^3$ CPPM, Universit\'e d'Aix-Marseille 2, IN2P3-CNRS, FR-13288
Marseille Cedex 09, France\\
$^4$ Institute for Fundamental Theory, Dept. of Physics,
Univ. of Florida, Gainesville FL 32611, USA\\
$^5$ IPNL, Universit\'e Claude Bernard de Lyon, IN2P3-CNRS,
FR-69622 Villeurbanne Cedex, France\\
$^6$ Institut des Sciences Nucl\'eaires, IN2P3-CNRS,
Universit\'e de Grenoble 1, FR-38026 Grenoble Cedex, France\\
$^7$ Laboratoire de l'Acc\'el\'erateur Lin\'eaire,
Universit\'e de Paris-Sud, IN2P3-CNRS, B\^at 200, FR-91405 Orsay Cedex, France\\
$^8$ Laboratoire de Physique Nucl\'eaire et des Hautes Energies,
Ecole Polytechnique, IN2P3-CNRS, 91128 Palaiseau Cedex, France\\
$^9$ Laboratoire de Physique des Particules - LAPP - IN2P3-CNRS,
74019 Annecy-le-Vieux Cedex, France\\
$^{10}$ Laboratoire de Physique Th\'eorique et Hautes \'Energies,
Universit\'e de Paris-Sud,
B\^at 210, FR-91405 Orsay Cedex, France\\
$^{11}$ Physikalisches Institut, Universit\"at Bonn, Nussallee 12, 
D-53115 Bonn, Germany\\
$^{12}$ DAPNIA/Service de Physique des Particules, CEA-Saclay,
FR-91191 Gif-sur-Yvette Cedex, France\\
$^{13}$ Service de Physique Th\'eorique, CEA-Saclay,
FR-91191 Gif-sur-Yvette Cedex, France\\
$^{14}$ Universit\'e de Haute Alsace, Mulhouse, France
}}

\vspace*{1.cm}
\begin{center}
\footnotesize{
\begin{tabular}{lcr}
Contact persons: & Marc Besan\c con (Marc.Besancon@cern.ch) \\
                 & Emilian Dudas (Emilian.Dudas@cern.ch)
\end{tabular}
}
\end{center}
\newpage
\tableofcontents
\newpage
\section{Introduction}
 It is a well-known fact that the conservation of the baryon and lepton
 number is an automatic consequence of the gauge invariance and 
 renormalizability in the Standard Model. Their non-conservation is
 generally considered in the context of grand unified theories. In this
 case, their effects (like proton decay, for example) are suppressed by
 powers of the
 grand unified scale, which is supposed to be of the order of $2 \times
 10^{16} GeV$. Therefore these effects are difficult to observe experimentally.
 
 In supersymmetric extensions of the Standard Model, gauge invariance and
 renormalizability no longer assures baryon and lepton number
 conservation. We will consider in what follows the MSSM as the minimal 
 supersymmetric extension, but the following considerations are easy to
 generalize. By renormalizability and gauge invariance, we can write two 
 different types of Yukawa-type interactions, described by a
 superpotential $W=W_1+W_2$, where
 \begin{equation}
 W_1 = \mu (H_u H_d) + (\lambda_u)_{ij} (Q_i H_u) D^c_j+
 (\lambda_d)_{ij} ( Q_i H_d) D^c_j+ (\lambda_e)_{ij}(L_iH_d) E^c_j 
 \  \label{1}
 \end{equation}
 and
 \begin{equation}   
 W_2 = \mu (H_u L_i) + \lambda_{ijk}( L_iL_j )E^c_k+
 \lambda'_{ijk}( Q_i L_j) D^c_k+ \lambda^{''}_{ijk} (U_i^cD_j^cD_k^c) 
 \ , \label{2}
 \end{equation}
 where we have exhibited the dependence  on the quarks and lepton family 
 indices, $i,j, \cdots $ and the parentheses enclosing fields products are meant
 to remind that one is to take overall singlet   contractions with respect to 
 the $SU(3)_c \times SU(2)_L$  gauge group  indices.   
 The superpotential $W_1$ contains the usual Yukawa interactions of the
 quarks and leptons and, in addition, the Higgs supersymmetric mass term
 $\mu$. The usual gauge interactions supplemented by $W_1$
 give a theory where the baryon and the lepton numbers are automatically
 conserved.
 
 On the other hand, even if renormalizable and gauge invariant, the terms
 in $W_2$ \footnote{the expression of $W_2$ developped in terms of fields
 is given in appendix B} 

 do violate the baryon ($B$ by the 9 
 $\lambda^{\prime \prime}_{ijk}$ couplings ) and the lepton number ($L$
 by the 9 $\lambda_{ijk}$ couplings and the 27 $\lambda^{\prime}_{ijk}$
 couplings) and
 they are not
 suppressed by any large mass scale. They may for example induce proton
 decay through product of couplings $\lambda' \times \lambda^{''}$ and,
 if these couplings are of order one, this is certainly
 unacceptable. Different combinations of couplings induce different
 baryon and lepton non-conserving transitions and, as we will see in
 detail in the next sections, are severely constrained experimentally.
 That's why, in 1978 Farrar and Fayet \cite{fayet} proposed a discrete symmetry $R$
 such that $RW_1=W_1$ and $RW_2=-W_2$ and therefore automatically
 guarantees the $B$ and $L$ conservation.  
  
 This symmetry, called R-parity, acts as $1$ on all known particles and
 as $-1$ on all the superpartners and can be written 
 \begin{equation} 
 R = (-1)^{3B+L+2S} \ , \label{3}
 \end{equation}
 where $S$ is the spin of the particle. The physics of the MSSM with a
 conserved R-parity is very peculiar, since the lightest supersymmetric
 particle (LSP) cannot desintegrate into ordinary particles and is
 therefore stable. In this case, the superpartners can be produced only
 in pairs and their direct search must typically wait for high energy
 colliders, LHC or NLC.   
  
 On the other hand, even if rather elegant, the ad-hoc imposition of the
 R-parity is not theoretically very-well motivated and neither sufficient
 for suppressing all the dangerous $B$ and $L$ violating terms. For
 example, if we consider MSSM as an effective theory, which is certainly
 the case and search for gauge-invariant higher-dimension operators, we can 
 immediately write down the terms
 \begin{eqnarray}
 W_3 &=&  {(\kappa_1)_{ijkl}\over \Lambda } (Q_iQ_j) (Q_kL_l)+
{(\kappa_2)_{ijkl}\over \Lambda }
 (U^c_iU^c_jD^c_k)E^c_l+ {(\kappa_3)_{ijk}\over \Lambda } (Q_iQ_j) (Q_kH_d)+ \cr
 &+& {(\kappa_4)_{ijk}\over \Lambda }
 (Q_iH_d)(U^c_jE^c_k)+{(\kappa_5)_{ij} \over \Lambda } (L_iL_j) (H_uH_u)+
 {(\kappa_6)_i\over \Lambda } (L_iH_d) (H_uH_u) \ , \label{4}
 \end{eqnarray}
 where $\Lambda$ can be viewed here as the scale of new physics, beyond MSSM.
 It is easy to check that the operators $\kappa_1$, $\kappa_2$ and $\kappa_5$
 do respect R-parity but still violate $B$ and $L$ and are experimentally
 constrained to be rather small. 
 
 On the other hand, in models without R-parity, the experimental
 signatures are spectacular: single production of supersymmetric
 particles accompanied by missing energy, which could be observed at
 lower energies compared to the R-parity conserving case, sizable effects
 in the flavour physics, etc. The study of these effects in the near
 future in the accelerators is the main purpose of our report.
  
  The plan of this report is the following. In Section 2, an updated and
 improved analysis on $R$-violating couplings coming from the various existing
 data is made. Then, in Section 3 we discuss theoretically motivated 
 alternatives to R-parity based on abelian family symmetries, relating
 the couplings $\lambda$, $\lambda'$ and $\lambda^{''}$ to the ordinary
 Yukawa couplings. In Sections 4 the single production of supersymmetric
 particles at LEP, Tevatron and LHC are discussed and in Sections
 5,6,7 we study in more detail the physics of R-parity violating
 couplings at HERA, LEP and LHC. In Section 8 we discuss the implications
 of R-parity violation on neutrino masses, which seems to find a more
 solid evidence in view of the last results at SuperKamiokande.
  We end with our projects and perspectives for the next two years.

\newpage

\section{Indirect bounds on $R$-parity odd interactions}
\label{chap2}
The  indirect bounds  concern  essentially the constraints deduced  from    
low and intermediate energy  particle,  nuclear,
atomic  physics  or astrophysics phenomenology, 
where the   superpartners of ordinary particles 
propagate off-shell in (tree or loop)  Feynman diagrams. 
The  interest in  this subject dates back to the  early period of
 the $R$-parity  litterature,  
see \cite{fayet} to \cite{moha},
 and  still continues to  motivate a strong activity \cite{reviews,reviews2}.
By contrast, the subject of direct bounds or measurements rather deals with
the high energy  colliders physics phenomenology
 (single production, LSP decays, etc...)     
with superpartners produced on the mass shell.
To extract  an experimental information on the 3 dimensionfull coupling constants,
 $\mu_i$, which mix the up Higgs boson $H_u$ with leptons,
 and the   45 dimensionless  Yukawa coupling constants, 
$\l_{ijk} = -\l_{jik}, \ \l '_{ijk} , \ \l ''_{ijk} =- \l ''_{ikj}$, 
 one must   define   some search  strategy  and look for reasonably 
motivated assumptions. It is important  to note first that an
independent discussion of  the bilinear interactions 
is necessary only for the case where the left-handed sneutrinos acquire, at
the stage of electroweak gauge symmetry breaking, non-vanishing VEVs,
$<\tilde \nu_i > =v_i$. In the alternate explicit breaking
 case,  characterized by $v_i=0$,  one can  by a field transformation  remove away   the
bilinear  interactions in favor of the trilinear and higher dimension   interactions.
The  case of a spontaneously  broken  discrete symmetry  may be characterized
 rather by, 
 $\mu_i =0 , \ v_i\ne 0$ or by the hypothetical  situation where  right-handed
 sneutrino raises a VEV. Strong bounds on these parameters have been deduced
in both  the explicit and spontaneous breaking cases.  

Concerning the  trilinear coupling constants,  the major part of  the 
existing  experimental  indirect  bounds  has been derived on
the basis of the  so-called  single coupling  hypothesis,
where a single     coupling constant is assumed to dominate
 over all the others, so that each  of the  coupling constants  
 contributes once at a time \cite{dimohall,bargerg}.
Apart from a few  isolated  cases, the typical bounds derived under this hypothesis,
 assuming  a linear dependence  on the superpartner masses, are of order,
$[ \l ,\  \l' ,\  \l '' ] <  ( 10^{-1} - 10^{-2}) { \tilde  m\over 100 GeV }$.

An important variant of the single operator dominance  hypothesis 
  can be defined by applying this  at the level of the gauge (current)
basis fields rather than the mass eigenstate fields.
This  appears as a more natural  assumption in models where 
the presumed hierarchies in coupling constants originate from physics at higher scales.
As an illustration, we quote   two useful   representations of the $\l '$ 
interactions, obtained by performing the linear  transformations on 
quarks  fields from current to  mass eigenstates bases,   
\begin{eqnarray}
  W(\l ') &=& \l '_{ijk} (\nu_i d_j -e_i u_j)d^c_k=
\l^{'A}_{ijk} (\nu_i d'_j -e_i u_j)d^c_k=
\l^{'B}_{ijk} (\nu_i d_j -e_i u'_j)d^c_k \ ,  \cr
\l^{'A}_{ijk}&=& \l '_{imn}(V_L^{u\dagger })_{mj}
(V_R^{dT })_{nk}  , \ \ 
 \ d'_i =V_{il} d_l;\quad  
\l^{'B}_{ijk}= \l '_{imn}(V_L^{d\dagger })_{mj}
(V_R^{dT })_{nk}  , \ \  
\ u'_i= (V^\dagger )_{il} u_l , 
\label{eqm0}
\end{eqnarray}
where $ V= V_{CKM}$ is the familiar  quarks  unitary  CKM matrix. The representations
 with the 
coupling constants,  $\l ^{'A}_{ijk} $ or $\l ^{'B }_{ijk}$,  
 allow for the presence of  flavour changing contributions in the d-quark or u-quark
  sectors, respectively,    even when
a single  $R$-parity odd coupling constant is assumed to  dominate \cite{agashe}.
 
To the extent that there is no preferred basis for fields,
it is useful to look  for   basis independent statements.
Thus, the two sets of mass basis   coupling constants, $\l ^{'A,B} $ and  the 
current basis  coupling constants, $\l ' $, obey the unitarity (sum rule)  type
 relations, 
$\sum_{jk} \vert \l ^{'A, B }_{ijk}\vert ^2 = \sum_{jk} \vert \l ^{'}_{ijk}\vert ^2 $.
 A classification of  all possible invariant products 
of the \rpv coupling constants has been  examined in  \cite{davidson}. 
Assuming that the linear  transformation matrices, $V_{L,R}^{q,l}$,
were known,  and that one is  given a  bound on some  interaction operator, then 
by applying  the single dominance hypothesis to the  current basis  coupling constants,
one could derive a string of bounds  associated to the
operators  which mix with it by the  current-mass fields transformations.
 For example, assuming $(V_L^u)_{1i}=(V_R^u)_{1i}=
(1, \ \e ,\  \e ' )$, starting  from the bound on $\l '_{111}$, 
one can deduce the following related   bounds: 
$\l '_{121}< \l'_{111}/\e , \ \   \l '_{131}< \l'_{111}/\e ' $, \cite{ellis98}.

At the next  level of complexity, one may apply an extended hypothesis 
where the  dominance is postulated for pair, triple,  etc...   products
 of coupling constants.
Several analyses  dealing with hadron  flavour changing effects 
(mixing parameters for the neutral light and heavy  flavoured  mesons,  mesons   decays,
$K\to \pi +\nu +\bar \nu $, ...  \cite{agashe,roy,jang});
lepton flavour changing  effects
 (leptons decays, $l_l^\pm \to l^\pm+l_n^-+l^+_p, $ \cite{roy}
conversion  processes, $ \quad \mu ^-\to e^-$  \cite{kim}, neutrinos
 Majorana mass \cite{dawson,godbole}, ...); lepton
number violating effects (neutrinoless double beta decay \cite{hirsch,babu,hirsch1});
 or  baryon number violating  effects (proton decay partial branchings \cite{smirnov},
rare non-leptonic decays of heavy mesons \cite{carlson},
nuclei desintegration \cite{goity},...)   have   led
to bounds on a large number of quadratic  products  of the coupling constants.

 Our purpose in this 
work is to present an encapsulated review of the litterature on indirect bounds, 
which complements the existing reviews \cite{reviews,reviews2}.
Our  main objective  is to  identify certain  important
unsettled problems where  effort is needed.    
The  contents  of this chapter  are  organized into 4 sections.
 Subsection \ref{secxxx1} is about the  bilinear interactions and spontaneously 
broken realization of $R$-parity.
Subsection \ref{secxxx2} is about the trilinear interactions. 
Subsection \ref{secxxx3}  reviews a variety of scattering and decay processes 
associated with lepton and baryon number and lepton and quark flavour
violations. Section \ref{secxxx5}  presents the main conclusions.
Some important
notations used in the sequel are summarized in appendix A.
\subsection{Bilinear  interactions and spontaneously broken $R$-parity}
\label{secxxx1}
The bilinear interactions, $W=\mu_iH_uL_i$, break $R$-parity and $L_i$ numbers. The 
physical effects  of these interactions bear on the parameters, $\mu_i$, and
the sneutrinos VEVs, $< \tilde \nu_i> =v_i$.  It is important to distinguish the
 cases of 
 a spontaneously broken $R$-parity, $\mu_i=0,\  v_i\ne 0, $ from the explicit
 breaking one, 
  $\mu_i\ne 0, \  v_i $ vanishing or not.  The viable models constructed so far, 
employ either  the explicit breaking option  or  a specific
spontaneous breakdown option where $R$-parity and lepton  numbers are broken by 
a right-handed neutrino VEV, $<\tilde \nu_R >\ne 0$, so as to have a  gauge
 singlet Goldstone
boson  (so-called majoron)  which  is decoupled from gauge interactions
 \cite{ellisvalle,rossvalle,valle1,barbieri}.  One expects the two sets of parameters,
 $\mu_i, \ v_i $, 
to be strongly correlated, since $\mu_i\ne 0$ 
may by themselves lead,  through minimization of  the scalar potential, to 
non vanishing VEVs,  $v_i \ne 0$,
at electroweak symmetry breaking \cite{nilles,banksnir}. As long as one
makes no commitment regarding 
the structure of the effective potential for the scalar fields, the four-vector
 of VEVs, $v_\a $,  must be regarded as free parameters.

In the limit of vanishing  superpotential, the  minimal supersymmetric
 Standard Model possesses an $SU(4)$ global symmetry 
which  transforms the column vector of down Higgs boson and leptons chiral
 supermultiplet fields, 
  $ L_\a =(H_d\ \  L_i),$   as, $ (H_d\ \  L_i ) \to  U (H_d \ \  L_i), \ U\in SU(4)$,
  where the indices, $ \a =[d, i],\   [i=(1,2,3) =( e,\mu ,\tau  )],  $ 
 label the down-type  Higgs bosons and the three  lepton
 families. The symmetry group $SU(4)$  reduces to   $SU(3)$  by  switching on the 
bilinear (d=3)  $R$-parity odd superpotential,  $ W= \mu_\a H_u L_\a , $ 
and is completely broken down by switching on the matter Higgs bosons  trilinear
  interactions. For vanishing $v_i$, one can apply  the superfields transformation,
 $ \ U\approx  \pmatrix{1 &  -\d_m \cr  \d_m  & I_3}$, with $ \d_m={\mu_m\over \mu }$,
  so as to rotate away  the lepton-Higgs boson  mixing terms,   leaving behind  the
  Higgs bosons mixing  superpotential,   $W_2= \mu H_uH_d$, along with trilinear
 $R$-parity odd   interactions  of  specific  structure, $\l_{ijk} = -(\l_e)_{ik} \d_j, 
\ \l'_{ijk} = -(\l_d)_{ik} \d_j. $ 
 
For non-vanishing  three-vector of  VEVs, 
$< \tilde \nu_i> =v_i$,  the remnant $SU(3)$  symmetry is spontaneously
  broken down to $SU(2)$.   This induces bilinear mass terms which mix 
neutrinos with neutralinos and charged leptons with charginos. 
The condition that the  six eigenvalues of the 
 neutralinos mass  matrix  can be assigned to the two massless, $\nu_e, \nu_\mu $,
 neutrinos,  the 
$\nu_\tau $ neutrino of mass $m_{\nu_\tau }$, and  the  three massive 
neutralinos of mass  $ O( M_Z)$,   imposes \cite{banksnir} the order of magnitude
  bounds,
  $  v_i <    O(\sqrt {M_Z m_{\nu_\tau }}) , \  \mu_i  < O(\sqrt {M_Z m_{\nu_\tau }}),$
 along with   the order of magnitude  alignment condition,
  $  \sin^2 \psi < O({m_{\nu_\tau } \over M_Z} )$. From the experimental bounds on
 neutrinos masses: 
$m_{\nu _{[e,\mu , \tau ]}  }< [5.1 eV,\  160 keV, \  24 MeV]$,  we conclude 
 that these strong bounds  affect not only  the \rpv coupling constants, $\mu_i $,
 but also 
place restrictive  conditions on the soft susy breaking parameters via
the sneutrinos VEVs vector,  $v_i$.  In the simpler case
 involving a single generation of leptons,
say the third, the mass bound on  $\nu_\tau $  yields, $v_\tau < 5 $
GeV \cite{barbieri}. For the general three generation case, in a suitable
 approximation, there arises a  single massive Majorana neutrino given by the
 fields linear combination,
 $ \nu_M(x)= {1\over (\sum_j v_j^2)^\ud  } \sum_i v_i \nu_i (x)$. 
If this is  identified with $\nu_\tau $, then the  associated mass bound  
(using, $ m_{\nu_ M} < 143 $ MeV, rather than the  stronger current experimental bound)
yields a bound on the quadratic form, $(\sum_i v_i^2 )^\ud < 12\  - \ 24 $ GeV
 \cite{hallsuz,lee,dawson}. If it is identified instead with $\nu_e$, 
the stronger bound   $(\sum_i v_i^2 )^\ud <  2 - 5  $ MeV  results.
The mixing  of neutrinos   with  neutralinos may also  induce new desintegration 
channels for the Z-boson consisting of single production of superpartners. 
 The Z-boson  current coupling to neutralinos pairs,  $Z^0\to \tilde \chi_l^0  + 
\tilde \chi_l^0  $,  leads through the  $\nu_\tau - \chi ^0_l $ mixing to 
the  decay  channels, $ Z\to \bar \nu_\tau + \tilde \chi_l^0. $
Similarly, the mixing of $\tau^\pm $ with charginos,  induces the decay channels, 
$Z^0\to \tau^++ \tilde \chi ^- , ...$  The associated  Z-boson decay   BF range 
inside the interval, $ 10^{-5} \ - \ 10^{-7}$  \cite{barbieri}.  

Neutrinos Majorana  masses and mixing parameters  can also be induced via one-loop
 mechanisms involving the $\l '$ interactions in combination with tadpoles of
 sneutrinos \cite{hallsuz,lee,dawson}.
 The experimental bounds on masses lead to bounds on the following products: 
$\l _{ilm} ({v_i \over 10 MeV }) <  
({\tilde m \over 250 GeV}) [ 10^2, 1.5\ 10^4, 1.6\ 10^5 ]  , \ [i=1,2,3] $
 \cite{dawson}.    One-loop mechanisms  may also  contribute 
to the rare forbidden processes, $\mu ^- \to e^- +\g , \ \mu ^- \to e^-+e^++e^- $,  
(where the current experimental bounds are,
 $B(\mu\to e+\g )< 4.9\ 10^{-11} , \ B(\mu \to e+e+e)<1.0  \ 10^{-12}$)  or to 
 neutralino LSP decays, $\chi^0 \to \nu +\g , \cdots $, which may have implications on 
cosmology. For a  very light  neutralino  LSP case,  exotic  pions decay reactions
 such as, 
 $ \ \pi \to e+\tilde \chi^0$, etc... \cite{hallsuz} are possible.      
Implications on $\tilde \nu \tilde {\bar \nu } $ oscillations  and 
bounds on sneutrinos Majorana-like masses,
 $  L=- \ud (\tilde m_M \tilde \nu \tilde \nu + c.c.)$,    have also been examined 
\cite{haber1}.
\subsection{Trilinear interactions}
\label{secxxx2}
 \subsubsection{Four-fermions contact interactions} 
Under the single   dominant  coupling constant hypothesis,
the  neutral current  four-fermion (dimension-6)  interactions
 induced by decoupling of exchanged  scalar superpartners at tree level, can be 
represented by the effective Lagrangian: 
\begin{eqnarray}
L_{EFF}&=&\sum_{ijk}\ud \vert \l_{ijk}\vert^2 \bigg [
{1\over m^2_{\tilde e_{kR}} } (\bar \nu_{iL}
\g_\mu \nu_{iL}) (\bar e_{jL} \g^\mu e_{jL})
-{1\over m^2_{\tilde e_{kR}} } (\bar \nu_{jL}
\g_\mu e_{jL}) (\bar e_{iL} \g^\mu \nu_{iL})\cr
&-&{1\over m^2_{\tilde \nu_{iL}} } (\bar e_{jL}
\g_\mu e_{jL}) (\bar e_{kR} \g^\mu e_{kR})
-{1\over m^2_{\tilde e_{iL}} } (\bar \nu_{jL}
\g_\mu \nu_{jL}) (\bar e_{kR} \g^\mu e_{kR}) +(i\to j) \bigg ] +h. c. \  .
\label{eq3}
\end{eqnarray}
An analogous formula holds for the $\l '_{ijk}$ interactions with the 
substitutions, $\nu_{[i,j]L}\to u_{[i,j]L},\  e_{jL} \to d_{jL}, \ e_{kR} \to d_{kR}$.
 The neutral  current (NC)   contact interactions can include scalar, vector or tensor
 Lorentz covariants. The least strongly  constrained of these three couplings, so far, 
are the vector  interactions. The conventional parametrization  for  leptons-quarks 
flavour diagonal couplings reads,  
$$ L_{NC}= \sum_{ij=L,R } {4\pi \eta_{ij}^q\over
 \L _{ij}^{\eta 2} } (\bar e_i\g_\mu e_i)( \bar q_j 
\g_\mu q_j) ,$$ 
where a sum over light flavours of leptons and quarks is 
understood and  $\eta^q_{ij}  =\pm 1$ are sign factors. 
  The analyses  of these interactions 
at high energy colliders are directed towards  tests of non-resonant continuum 
contributions associated to composite (technicolor, ...) 
  models of quarks and leptons, leptoquarks,  ...\cite{eichten}. 
Bounds of magnitude, $\L_{[LR,RL]}^{[-,+]d}> [1.4 ,\  1.6] $ TeV,
are  reported  by   ALEPH, DELPHI  and OPAL Collaborations at LEP \cite{lepcol}
 (based on 
the reactions, $e^-e^+\to s\bar s, ...$) , and  
$\L_{[LR,RL]}^{[+,+]u}>[ 2.5 ,\  2.5] $ TeV, by CDF Collaboration at
 the Tevatron \cite{abe}
(based on Drell-Yan processes or large $p_T$ jets production). The recent anomalous 
events observed by the H1 and ZEUS Collaborations at
  HERA seem to favor a small scale,  $\L \approx 1 $ TeV \cite{alta1,bargerz}. 

 The  charged current (CC) four-fermions contact  interactions
 have a Lorentz   vector component, which is conventionally   parametrized as, 
$$ L_{CC}= {4\pi \eta \over \L _{CC}^{\eta 2} } 
(\bar e_L\g_\mu \nu_L)( \bar  u_L
\g_\mu d'_L) . $$
While the bounds  obtained by the  Collaborations at the 
LEP or  Tevatron  colliders  lie typically  at, 
$\L^-_{CC}  > 1.5 TeV,$ 
the  fit to the recent deep inelastic scattering events observed by the Collaborations
 at
the   HERA  collider $(\sqrt s= 300 GeV, \  Q^2> 15,000 GeV^2$) favor again  lower
 values,
   $\L _{CC}= 0.8-1 TeV$ \cite{altarelli}. Let us note here that
the  bounds from  leptons and hadrons universality decays, APV etc..., to be
 discussed below, generally point to larger cut-off scales,
 $\L \approx  10 \ - \ 30 $ TeV,
 and $\L_{CC}  \approx  10 \ - \ 80 $ TeV.

 Under  the   hypothesis of   dominant pairs of coupling constants, 
 there arise  mixed leptons-quarks  four-fermion  \rpv induced 
interactions, of which a subset reads: 
\begin{eqnarray}
L_{EFF}&=& -{1\over m^2_{\tilde \nu_{iL}}}  \l'_{ijk} \l ^\star _{imn}
(\bar d_{kR}d'_{jL})(\bar e_{mL}e_{nR}) 
-{1\over  2m^2_{\tilde u_{jL}}}  \l'_{ijk} \l ^{'\star }_{ljn}
(\bar d_{kR}\g_\mu d_{nR})(\bar e_{lL}\g^\mu e_{iL}) \cr
&+&{1\over  2m^2_{\tilde d_{kR}}}  \l'_{ijk} \l ^{'\star }_{lmk}
(\bar e_{lL}\g_\mu e_{iL})(\bar u_{mL}\g^\mu u_{jL}) 
-{1\over m^2_{\tilde \nu_{iL}}} \l'_{ijk} \l ^{'\star } _{imn}
(\bar d_{kR}d'_{jL})(\bar d'_{mL}d_{nR} ) .
\label{eq3p}
\end{eqnarray}
These interactions can induce contributions to  rare  leptonic decay 
processes of mesons.  The bounds   on the \rpv coupling constants obtained from 
 the leptonic decays of light quarks mesons are: 

$(\l'_{l11} \l'_{l12} \pm \l '_{l11} \l _{l21} ) < 0.14, \ [ \pi^0\to e^-+\mu^+ ] $
 \cite{kim};  
  $\l_{122} \l '_{112,121}  < 3.8 \ 10^{-7},\  [K_L\to \mu^+ \mu^-], \ 
\l_{121} \l '_{212,221} < 2.5 \ 10^{-8},\  [K_L\to e^+ e^-], \ 
\l_{122} \l '_{212,221} < 2.3 \ 10^{-8},\   [K_L\to e^+ \mu^-] \ \  \cite{roy} $.

For leptonic decays  of  heavy quarks mesons, some bounds  reported in the literature, 
all in units of, $({\tilde m \over 100 GeV})^2 $, are: 

(1) $\l_{131} \l'_{333}< 0.075 e_{3L}^2, \  [  B\to e ^- +\bar   \nu ]  $;  
\ $\l'_{333} < 0.32  \tilde m^2, \ [ B^- \to  \tau ^-+\bar \nu  ] $; \   
 $\l_{131}\l'_{333} < 0.075  \tilde m^2, \  [B^- \to e ^-+\bar \nu ]; $ \cite{erler} 

(2) $\l_{121} \l'_{131}< 4.5 \ 10^{-5} \tilde m^2 , \  [  B\to e^+ +\mu^- ];$  \  
 $\l_{131} \l'_{131}< 5. \ 10^{-4} \tilde m^2 , \  [  B\to e^+ + \tau^- ];$ \  
 $\l_{123} \l'_{131}< 6. \ 10^{-4}  \tilde m^2 , \  [  B\to   \mu^+ +\tau^- ]$ \   
 \cite {jang}.

\subsubsection{Charged current interactions}
  
$ \bullet \bullet $ {\bf Lepton  families universality.} 
 Corrections  to  the leptons charged current universality in the $\mu $ -decay
 process, $\mu ^- \to e^-+\nu_\mu +\bar \nu_e$, 
 arise  at tree level from  the  \rpv  interactions. These redefine 
  the Fermi weak interactions  constant as \cite{bargerg},
$ {G_\mu \over \sqrt 2} = {g^2\over 8M_W^2} (1+r_{12k}(\tilde e_{kR})).$ 

 The redefinition, $G_\mu \to  G_\mu /  (1+r_{12k}(\tilde e_{kR}))$, can be tested
at the quantum  level of the Standard Model by testing the exact  relations,
  linking  the different basic coupling constants,   which incorporate  the
  one-loop renormalization corrections \cite{langarad}. The following two relevant
 relationships: 
$$ m_W^2={ \pi \a \over  \sqrt 2 G_\mu 
\sin^2\t_W (m_Z)\vert_ {\overline {MS} } 
(1-\D r(m_Z) \vert_{\overline {MS} }  )} , \quad 
({ m_W \over m_Z} ) ^2=1-{ \pi \a \over  \sqrt 2 G_\mu 
m_W^2 
(1-\D r(m_Z) \vert_{\overline {MS} }  )} ,$$
 link the renormalized  W-boson mass  and coupling constant parameters,
 $m_W,  \a , G_\mu ,  \ \sin^2 \t_W $, with the  combination  of radiative
 corrections,  
$\D r = 2\d e/e -2\d g /g -\tan ^{-2}\t_W (\d m_W^2/m_W^2- \d m_Z^2/m_Z^2) $. 
Recall that the input parameters employed in high precision tests 
of the Standard Model are chosen  as the subset of  
best experimentally  determined  parameters among the  following  basic  set:
$\a ^{-1} = 137.036, \
\a _s= 0.122 \pm 0.003 ,\ m_Z= 91.186 (2),\  G_\mu = 1.16639(1) \ 10^{5} GeV^{-2} , \ 
m_{top} (pole) = 175.6 \pm 5.0 , m_{Higgs}$.  The remaining parameters  are then 
deduced by means of  fits to  the familiar basic 
data (Z-boson lineshape and decay widths, $\tau $ polarization, forward-backward
 (FB)  or polarization asymmetries, atomic  parity violation (APV), beta decays,
 masses,  ...)
Unfortunately, at the  present level of 
precision for the fitted Standard Model parameters,  ($m_Z, \ m_W, \cdots $) 
no useful bound can be deduced on $\l_{12k}$.

The same interactions also govern  \rpv corrections in the  $\tau $-lepton  
decay process, $\tau^- \to e^-+\nu_\tau +\bar \nu_e,  $ and   related three-body
   beta decay  processes.   The  corrected BFs  read \cite{bargerg},
\begin{eqnarray}
R_\tau  &= &{\G (\tau  \to  e +\bar \nu_e  +\nu_\tau ) \over 
\G (\tau  \to \mu  +\bar \nu_\mu  +\nu_\tau ) } \simeq R_\tau ^{SM} [1+2
(r_{13k}(\tilde e_{kR}) -r_{23k}(\tilde e_{kR}))].
\label{eq5}
\end{eqnarray}
\begin{eqnarray}
R_{\tau \mu }  &= &{\G (\tau  \to  \mu +\bar \nu_\mu  +\nu_\tau ) \over 
\G (\mu  \to e +\bar \nu_e +\nu_\mu ) } \simeq R_{\tau \mu } ^{SM} [1+2
(r_{23k}(\tilde e_{kR}) -r_{12k}(\tilde e_{kR}))].
\label{eq5p}
\end{eqnarray}
These interactions  can also contribute to 
the pseudoscalar  mesons two-body leptonic decays of charged pions,
  $\pi^- \to l_i^-+\bar \nu_j  $.  
The  \rpv  corrections   lead to the  corrected  BFs \cite{bargerg}:
\begin{eqnarray}
R_\pi & =&{\G (\pi ^-\to  e^- +\bar \nu_e ) \over 
\G (\pi \to \mu  +\bar \nu_\mu  ) } \simeq R^{SM} [1+{2\over V_{ud}}
(r'_{11k}(\tilde d_{kR}) -r'_{21k}(\tilde d_{kR}))]. 
\label{eq4}
\end{eqnarray}

$ \bullet \bullet $ {\bf  Light quarks and leptons universality.}
The experimental information on light quarks charged current interactions 
 is deduced from data on neutron and  nuclear beta decay reactions in terms of  the  
 Fermi  coupling constant,  $G_F $, or equivalently the CKM matrix element, $V_{ud}$. 
 The \rpv corrections to the  d-quark  decay subprocess,
 $ d\to u +W^- \to u+e^-+\bar \nu_e$, combined with the above  redefinition
  of $ G_\mu $,  yields a  redefined  CKM  matrix element:
$$\vert V_{ud}\vert^2 = {\vert V^0_{ud} +r'_{11k}(\tilde d_R) \vert ^2
\over \vert   1+r_{12k}(\tilde e_R) \vert ^2 }.$$
Application to the s- and b-quarks decays 
yields analogous formulas  for $V_{us} $ and $ V_{ub} $,  which can be deduced from the 
formula for $V_{ud} $ by the substitutions, 
$r'_{11k} \to r'_{12k}$ and $  r'_{13k}$, respectively  \cite{ledroit}. 
Improved bounds obtained from tree and one-loop contributions to D-mesons
 beta decays are 
\cite{ledroit}: 

$ \l'_{22k}< 0.30,\ [D^\star ]  \  0.49 ,\ [D^+]  \   0.13,\ [D^0] ;\ \   
\l'_{12k}< 0.10,\ [D^\star ]\  0.28,  \ [D^+]    \  0.21,\ [D^0] .$

$ \bullet \bullet $ {\bf  Universality in $\tau $- lepton and mesons  semi-leptonic
 decays}. 
The \rpv contributions to the decay processes into pseudoscalar and vector mesons, 
$ \tau  ^-\to l^-+ P^0 , \ $ and  $ \tau ^-\to l^-+ V^0 , \ [P=\pi^0, \eta , K; 
\ V=\rho^0, \omega , K^\star]$ arise through tree level exchange of sneutrinos, 
\cite{kim}. The   bounds deduced from upper limits on experimental 
 rates are: $\l _{k31} \l '_{k11} < 6.4 \ 10^{-2} \tilde \nu_{kL}^2  $.
Several other  analogous bounds are also quoted in \cite{kim}.
The \rpv induced   decay process,  $\tau  ^-\to \pi^- +\nu_\tau $, 
 yields the bound \cite{kim}: $  \l '_{31k}< 0.16  \ \tilde d_{kR}$. From the  formally 
related ratios of  decay widths  $\tau $-lepton hadronic  and $\pi $-meson
 leptonic decays,  
$\G(\tau ^-\to \pi ^- + \nu_\tau )/\G (\pi  ^-\to \mu ^- +\nu_\mu ),
$ one also deduces \cite{ledroit}:
$ \l '_{31k}< 0.10  \ \tilde d_{kR}, \  \l '_{21k}< 0.03  \ \tilde d_{kR}$. 

The decay  processes,   $D^+\to \bar K^0  (K^\star ) +\mu^++\nu_\mu , \ 
 D^0 \to K^- +\mu^++\nu $ and related processes  involving the other leptons, 
  are induced  through the \rpv interactions by tree level  $\tilde d_{kR}$ exchange,  
 The  current experimental upper limits on the  BF for these processes yield
 the bounds \cite{bhattachoud}: 

$ \  \l'_{121, 123} < 0.29; \  [D^+\to \bar K^0] \  
\l '_{22k}< 0.18;\    [D^+\to \bar K^{0 \star }] \   \ 
\l'_{121, 123}< 0.34 \   [D^0\to K^-].  $ 

The  analysis of  tree level \rpv contributions to the 
 D-mesons three-body decay, $D\to K+l +\nu , \  
D\to K^\star +l +\nu  $  \cite{bhattachoud}, yields 
the bounds, 

$\l '_{12k=[1,3]}< 0.34, \ \l '_{22k}< 0.18, \ \l '_{31k}< 0.16 $.

For the $B-$ meson decay processes, one finds the bound: 

 $ \l '_{333} < 0.12  ( { \tilde  m_d \over 100 GeV   } ), 
[B^- \to  X_q +\tau ^-+\bar \nu ]     $,  \cite{grossman}.

$ \bullet \bullet $ {\bf Summary of charged current  experimental bounds.}   
Building on the initial analysis \cite{bargerg} where the tree level
  Standard Model predictions were used, the  analysis in \cite{ledroit}
  combined  both tree and one-loop 
level contributions.  The combined  list, including the refined 
 updated $1\s $  bounds from \cite{ledroit},  is given below.

$\l_{12k}:   0.04 e_{kR}  \ [V_{ud}]; \ 0.14  \pm 0.05  e_{kR}\  0.05  e_{kR}
[R_{\tau \mu  }];$ 

$\l_{13k}:      0.05 e_{kR}\ 
[R_{\tau  }]; $

$\l_{23k}:  \  0.05   e_{kR},\  [R_{\tau }]; \   0.05 e_{kR}\  [R_{\tau \mu  }]; $

$ \l'_{11k}: 0.01 d_{kR} \ [ V_{ud} ]$.
 
$ \l'_{12k}: 0.04 d_{kR} \ [ V_{us} ]$.

$ \l'_{13k}: 0.37 d_{kR} \ [ V_{ub} ]$.

 $ \l'_{21k}: 0.05 d_{kR} \ [R_\pi ]$. 

\subsubsection{Neutral current  interactions} 
 $ \bullet \bullet $ {\bf Neutrinos-leptons and quarks-leptons elastic scattering.}
 The  elastic  $\nu_\mu $ scattering processes, 
 $\nu_\mu +e_i \to \nu_\mu +e_j, \  \nu_\mu + q_i\to \nu_\mu +q_j$, 
   at energies well below $m_Z$, are described at tree 
level  by Z-boson exchange contributions in terms of the effective Lagrangian, 
\begin{eqnarray}
L&=&- {4G_F\over \sqrt 2} (\bar \nu_L\g_\mu \nu_L)(g^f _L\bar f_L\g^\mu f_L+
g^f _R \bar f_R\g^\mu f_R)  .
\label{eq7}
\end{eqnarray}
The related  $\nu_e$  scattering processes  include an additional $t$-channel
 contribution. 
The \rpv corrections read,   \cite{bargerg}
\begin{eqnarray}
g^e_L&=&(-\ud +x_W)(1-r_{12k}(\tilde e_{kR}) )-r_{12k}(\tilde e_{kR}), \
g^e_R=x_W(1-r_{12k}(\tilde e_{kR}) )+r_{211}(\tilde e_{1L}) +r_{231}(\tilde 
e_{3L}), \cr 
g_L^d&=&(-\ud +{1\over 3} x_W)(1-r_{12k}(\tilde e_{kR}) )-r'_{21k}(\tilde d_{kR}), \ 
g^d_R={x_W\over 3}(1-r_{12k}(\tilde e_{kR}) )+r'_{2j1}(\tilde d_{jL}).
\label{eqs7}
\end{eqnarray}

$ \bullet \bullet $ {\bf  Fermion-antifermion  pair production.} The
  forward-backward augular asymmetries (FB) in the differential cross sections 
 for the  reactions, 
 $e^++e^-\to f +\bar f,  \ [f=l,q] $ can be  parametrized  in terms 
 of the axial vector  coupling in the effective Lagrangian density,
$L= -{4G_F\over \sqrt 2} A^eA^f (\bar e \g_\mu \g_5 e) (\bar f \g^\mu \g_5 f), $
where, $   A^f= -T_{3L}^f .$ 
The off Z-boson pole asymmetry is defined as, $A_{FB}= -{3 G_F s m_Z^2\over 
16\sqrt 2 \pi \a (m_Z^2  -s ) } $, and  the Z-boson pole asymmetry 
as, $ A_{FB} = {3\over 4} A^e A^{l,q} $. 
The  formulas for the \rpv corrections  to the Z-pole asymmetries
in terms of the  products of parameters $ A^eA^f$   read \cite{bargerg},
\begin{eqnarray}
A^eA^\mu &=&{1\over 4} -\ud r_{ijk}(\tilde \nu_{kL} ) , \ \bigg [(ijk)= 
(122), (132), (121), (321) \bigg ]\cr
A^eA^\tau  &=& {1\over 4} -\ud r_{ijk}(\tilde \nu_{kL} ) , \ \bigg  [(ijk)= 
(213), (313), (131), (231)\bigg ]\cr
A^eA^{u_j} &=&-{1\over 4} -\ud r'_{1jk}(\tilde d_{kL} );\ 
A^eA^{d_k} ={1\over 4} -\ud r'_{1jk}(\tilde u_{jL} ).
\label{eq9}
\end{eqnarray}

$ \bullet \bullet $ {\bf  Atomic parity violation (APV).} The 
conventional parametrization for the effective flavour-diagonal 
interaction between leptons and quarks is,
\begin{eqnarray}
L= {G_F\over \sqrt 2} \sum_{i=u,d} C_1(i)  (\bar e \g_\mu \g_5 e)
(\bar q_i \g^\mu q_i) 
+C_2(i)  (\bar e \g_\mu e)
(\bar q_i \g^\mu  \g_5q_i). 
\label{eq9p}
\end{eqnarray}
Combining the Standard Model Z-boson pole  contributions with those of the \rpv
  interactions yields
 \cite{bargerg}: 
 \begin{eqnarray}
C_1(u)&=&(-\ud +{4\over 3} x_W )(1-r_{12k}(\tilde e_{kR})) 
-r'_{11k}(\tilde d_{kR}),  \ 
C_2(u)=(-\ud +2x_W )(1-r_{12k}(\tilde e_{kR}))-r'_{11k}(\tilde d_{kR}); \cr
C_1(d)&=&(\ud -{2\over 3} x_W )(1-r_{12k}(\tilde e_{kR}))
+r'_{1j1}(\tilde q_{jL}),\  
C_2(d)=(\ud -2x_W )(1-r_{12k}(\tilde e_{kR}))
-r'_{1j1}(\tilde q_{jL}).
\label{eq10}
\end{eqnarray}
An important experimental parameter  in the  APV phenomenology \cite{wood} 
is the weak charge, $ Q_W= -2[(A+Z) C_1(u)+(2A-Z) C_1(d)]$. For the reference
case of the $ \ ^{133}  _{55} Cs $ atom, the discrepancy between  experimental
and Standard Model fitted values  is: $\d (Q_W)= Q_W^{exp}-Q_W^{SM} =
 (-72.41\pm 0.84)+ (73.12\pm 0.09)=0.71 \pm 0.84 $.
 A refined analysis in \cite{alta1} yields: $\l '_{1j1} < 0.028  \tilde q_j$. 

$ \bullet \bullet $ {\bf Z-boson pole observables.}
 The  corrections to  the  Standard Model predictions to the  leptonic  BF 
(averaged over families)  and the  b-quarks Z-boson decays  BF, 
 $R_l^Z= \G_h/\G_l, \ R_b ^Z=\G_b/\G_h $, 
can be expressed as: $ \d R_l\equiv {R_l\over R_l^{SM} } -1 
= -R_l^{SM} \D_l +R_l^{SM} R_b^{SM} \D_l,\ 
\d R_b =  R_b^{SM} \D_b (1-R_b^{SM}) ,$ where $ \D_f ={\G (Z\to f+\bar f) \over
\G_{SM}  (Z\to f+\bar f) } -1 .$  The fitted Standard Model values \cite{pdg} are: 
$ R_l= 20.786, \   R_b= 0.2158,\ 
 \  R_c= 0.172, $  while the  experimental values  \cite{pdg} are: 
$ R_l=  20.795\pm 0.04,  \ R_b= 0.2202\pm 0.0020, \  R_c=0.1583\pm 0.0098. $ 
 The \rpv corrections to these BF may be   induced at one-loop  level via
 fermions-sfermions
intermediate states  \cite{ellishar}.

$ \bullet \bullet $  {\bf Summary of neutral current  experimental bounds.} The list of
bounds from a  tree  level analysis is given below. 

$\l_{12k}: \  0.34 e_{kR}, \  0.29 e_{k=1L}\  [\nu_\mu + e]; \     
   [0.10, 0.10,0.24]  
\  \nu_{kL}  \ [A_{FB}].$

$\l_{13k}: \   [0.10, 0.10,0.24]\ 
\nu_{kL}  \ [A_{FB}].$

$\l_{23k}:  \   0.26 e_{k=3L} \ [\nu_\mu + e];\   [0.10, 0.24]\ 
\nu_{k=1,2L}  \ [A_{FB}].$

$\l'_{11k}:  0.26 q_{k=3L} \ [A_{FB}]; \  0.30 d_{kR} ,  0.26 q_{k=1L} \ 
[APV]; $ 

$\l'_{12k}: 0.45 d_{kR} ,  \  0.26 q_{k=3L}\  [A_{FB}]; \  0.26 q_{k=1L}\  [APV]; \ 
0.29 \ \tilde d_{kR} \ [ D^\star \to \bar K^\star ]$

$\l'_{13k}:  0.45 q_{k=3L}, \  [A_{FB}] ; \ 0.26 q_{k=1L}\  [APV]; 0.63\  [R^Z_{l,b}] $

$\l'_{21k}: 0.11 d_{kR} ,  \  0.22 d_{k=1L}\  [\nu + q]; $

$\l'_{22k}:   0.22 d_{k=2L}\  [\nu_\mu + q]; 0.18 \ \tilde d_{kR} \ [ D \to \bar K ]$

$\l'_{23k}:   0.22 d_{k=1L} \  [\nu_\mu  +q]; \  0.44  \ [R_\mu ];  0.56 \ 
 [ R^Z_{l,b}]$

$\l'_{31k}: 0.16 \ \tilde d_{kR} \ [\tau  \to \pi +\nu ] \ .$   

$\l'_{32k}:   0.36  $ \ \cite{ledroit}.

 $\l'_{33k}: 0.26 \  [R_\tau  ^Z];\  0.45\   [R^Z_{l,b}] ;\  0.6 \ - 1.3 \ [k=3]$

$ \l ''_{312}: 0.097  \   [ R_l^Z]$. 

$ \l ''_{313}: 0.097  \   [ R_l^Z]$. 

$ \l ''_{323}: 0.097  \   [ R_l^Z]$.

 \subsection{Scattering and decay processes}
\label{secxxx3}
A multitude of bounds for  the \rpv coupling constants can be  deduced 
from  analyses of   low and intermediate energy processes. 
To present  the  results available from the current litterature,  we shall 
organize  the  discussion according
to  the  four  main themes  associated with
 violations of  leptons  and  quark flavours  and  violations of leptonic and
 baryonic numbers.  
\subsubsection{Lepton flavour violation ($\d L=0$)}
$ \bullet \bullet $ {\bf    Radiative  decays of leptons.}
The flavour non-diagonal, chirality-flip  photon emission processes,
 $l_J  \to l_{J'}  +\g  (q),\ [J\ne J']$,    acquire 
 \rpv contributions   at one-loop order from  fermions-sfermions
exchanges. The  fit to experimental bounds leads to the bounds, \cite{chen,decarlos}: 
 $ \l_{i1k} \l_{i2k} < 4.6\  10^{-4} \tilde \nu_{iL} ^2 $  or $\tilde e_{kR}^2 , \ \ 
\l_{ij1}  \l_{ij2} < 2.3 \  10^{-4}   \tilde \nu_{iL} ^2 $  or $\tilde e_{jL}^2 .$ 
           The virtual (time-like)  photon decay case,  which is associated to the
 physical 
processes, $l_J\to l_{J'}+e^++e^-$,  depends  on  vectorial type couplings 
 in addition to
the above tensorial couplings.

$ \bullet \bullet $ {\bf Electric dipole moments (EDM).}
 In one-loop  diagrams  propagating sfermion-fermion internal lines and  
incorporating mass insertions for  both fermions and sfermions lines, the \rpv 
interactions   can induce a contribution to  the  leptons EDM,  where a CP-odd phase  
is introduced through 
 the $A$  soft supersymmetry parameter  describing the 
$\tilde d_L\tilde d_R $ mixing \cite{hamidian}.  The strongest  bounds,  found by 
assuming 
a CP odd phase, $\psi ={\pi \over 4} $, are: 
$\l '_{1jk} <  5 \ 10^{-5} - 10^{-6} $, $\l '_{2jk} <    3 \ 10^{-1} - 10^{-2} $. 
A contribution to the neutron EDM, $d_n$,  \cite{masiero} from the \rpv interactions 
arises 
at two-loop order  through $W , \ \tilde d $ exchange. This  involves a relative 
complex phase between \rpv coupling constants described by  the  formula:
 $Im (\l ''_{32k} \l _{12k}^{''\star } ) = 
10^{-5} {d_n\over 10^{-34} e \times cm } ({ m_{\tilde q }\over 1TeV })^2.$

$ \bullet \bullet $ {\bf Anomalous magnetic dipole (M1) moments of leptons.} 
The discrepancies in  the anomalous 
 magnetic moments, $a=g/2-1$,  of the electron and muon, 
$\d a_l =a_l^{exp}-a_l^{SM},$  of 
Standard Model predictions  (including higher loop orders of 
 electroweak  corrections and hadronic corrections)
with respect  to the measured  values are
determined with  high precision. For the electron,  the $a_e$ observable   serves  
mainly as a 
measurer of the hyperfine constant, $\a $. Still, in the comparison with other 
determinations of $\a $, there  arises a finite discrepancy,
 $\d a_e \equiv  a_e^{exp}-a^{SM}_e)= 1 \ 10^{-11}, $ The discrepancy for the muon, 
 $ \d  a_\mu = 11659230(84) \ 10^{-10} - 11659172(15.4) \ 10^{-10} < 2.6 \ 10^{-8}$, 
should 
serve as a sensitive test for  new physics \cite{marciano}. 
 The \rpv interactions  contribute to the 
  $M1$ moments through the  same type of  one-loop diagrams as for the  EDM.
  These    contributions    scale  with the lepton mass as $m_l$.
 Related   observables, which should be accessible at  LEP,  are the  Z-boson 
current magnetic moment   of  the $\tau $-lepton or of  heavy $b $ or $ t $ quarks, 
$a_{\tau , b}(m_Z^2)$.   These  
calculations are described  by  the same  complex valued amplitude through an 
 $s\ -\  t$ crossing  transformation.   
 
 $ \bullet \bullet $ {\bf  Charged leptons conversion.} 
The   $\mu^-\to e^-$ transition can be  observed
 in the  muonium -antimuonium atoms
conversion process, $M(\mu^-e^+)\to \bar M( \mu^+e^-)$.  The associated    
bounds \cite{kim} read:  $\l_{231} \l^\star _{132} < 6.3 \ 10^{-3} 
{\tilde \nu_{3L} } ^2 $.  The other important 
 atomic transition conversion process, 
$\mu^- + \ ^{45} Ti\to e^-+  \ ^{45}Ti $, 
 gives a strong bound on a rather peculiar  linear combination  
of coupling constants \cite{kim},
$[\sum_k \l '_{2k1} \l '_{1k1} m^{-2}_{\tilde u_{kL} } 
-2 \l '_{k11} \l '_{k12}m^{-2}_{\tilde \nu _{kL} }
 \mp 2 \l '_{k11} \l '_{k21}m^{-2}_{\tilde \nu_{kL} } -
{70\over 14}  \l '_{21k} \l '_{11k}m^{-2}_{\tilde d_{kR} } ] < 1.6 \  10^{-11} $.

 \subsubsection{Lepton number violation, ($\vert \d L\vert >0$)}
$ \bullet \bullet $ {\bf Three-body leptons decays.}
 The analysis of the  flavour non-diagonal decay processes, 
 $l^\pm _m\to l^\pm_i +l^-_j+l^+_k$,   yields  several 
bounds for pair products of coupling constants \cite{roy}.  Among   the 
strongest ones are, 
$F_{1112}^2 + F^2_{2111}< 4.3 \ 10^{-13},\   [\mu \to 3e] \ 
F_{1113}^2 + F^2_{3111}< 3.1 \ 10^{-5}, \ [\tau  \to 3e]$. 
 If one excludes accidental cancellations
these bounds on  sums can be converted to equivalent  bounds for fixed
family indices \cite{roy}.

$ \bullet \bullet $ {\bf Neutrinos Majorana masses.} 
 The general structure of the  mass  Lagrangian of charge  neutral fermions
 allows $\d L=2$   Majorana  mass terms, $ \bar \nu_L A \nu_R^c +\bar \nu_L^cS \nu_R $,
  along with the $\d L=0$ Dirac mass  terms,
 $ \bar \nu_L D \nu_R +\bar \nu_L^c D \nu_R^c. $
The \rpv contributions may  occur  at one-loop level, via exchange  of
$l - \tilde l_{kH}$,  with a mass insertion on   fermions and a $LR$ insertion on 
sfermions 
\cite{dimohall}.
These yield: 
$\d m_{\nu _e} = {\l '_{1jk}\over 8\pi^2} { M_{susy } m_{q_j} m_{q_k} \over 
m_{\tilde q} ^2 } $  \cite{godbole}. 
Based on the empirical bound for the neutrino  $\nu_e$ mass,
 $m_\nu <   5 \ eV $  as deduced from a fit to 
$0\nu \b \b $ (neutrinoless double beta decay)  data,  one   infers 
the bounds \cite{godbole}:

$\l '_{133} < 3.5\ 10^{-3} ({ m_{\tilde q} \over 100 GeV } )^\ud , \ 
\l '_{122} < 7.\ 10^{-2} ({ m_{\tilde q} \over 100 GeV } )^\ud  .$ 

For the other coupling constants,  especially those involving light families indices, 
such as,  $ \l'_{111,112,121}$,  one obtains uninterestingly weak bounds.  
The neutrino mass bounds also imply bounds on the sneutrinos Majorana masses, which are 
defined as, $L= -\ud (\tilde m_M^2 \tilde \nu_L \tilde \nu_L +h.c. )$ \cite{haber1}. 

$ \bullet \bullet $ {\bf Neutrinoless double beta decay.}
The nuclear desintegration processes, $(Z,N)\to (Z+2,N-2)+l^-_i+l^-_j$, 
are measured  through geochemical or laboratory experiments. 
The $\ ^{76} Ge$ target (of half-life $T_\ud > 1.1 \ 10^{25} yrs $) 
stands as one of the most favorite test case. 
A  list of experimental  data is provided in the review \cite{expbb}.
The tree level contributions from $R$-parity odd interactions can be 
described by Feynman diagrams  where  t-channel exchanged   pairs of 
scalars, $\tilde e_L ,\ e_L^\star $ or $  \tilde u_{L}, \  u_L^\star $,
annihilate by emission of the final  leptons pair via 
an intermediate neutralino or gluino  t-channel exchange \cite{hirsch}. 
It can also be described
by the reaction scheme, 
$ d+d \to {\tilde \chi , \tilde g} \to
  \tilde d+\tilde d \to (u+e)+(u+e).$
The  stringent bound  deduced from this analysis is  \cite{hirsch}:
$\l'_{111}/[ ({m_{\tilde q} \over 100 GeV })^2 
({m_{\tilde g} \over 100 GeV })^{\ud } ]
< 3.3\times 10^{-4} . $
An order  of magnitude  stronger  bound, replacing 
the right hand side of the above inequality 
by $ 3.2 \ 10^{-5} $ was recently
obtained in \cite{wodecki}, using an analysis based on a
gauge mediated supersymmetry breaking scenario.

Another  class of contributions involves the t-channel 
exchange of a charged gauge boson $W^\pm $ and  a $d-$squark    
 according to the reaction scheme, $d+d\to u+W^-+d \to {\nu }
 \to u +e+\tilde d \to 
 (u+e)+(u+e),$
 \cite{babu}. This mechanism requires an $L-R$ mixing vertex for the produced 
down-squark, 
 $\tilde b_L -\tilde b_R$. 
The strongest bounds occur  for the following configurations of  flavour indices
(using the reference value ${\tilde m = 100 GeV} $): 

$\l'_{113}\l'_{131}< 7.9 \times 10^{-8}, \ 
\l'_{112}\l'_{121}< 2.3 \times 10^{-6}, \ \l ^{'2}_{111}< 4.6 \times 10^{-5} $, quoting 
from \cite{hirsch1} where
the  initial analysis of \cite{babu} was updated. 

\subsubsection{Hadrons flavour  violation}
 $ \bullet \bullet $ {\bf Semi-leptonic decays of pseudoscalar mesons.} 
The decay  process,  $K^+\to \pi^++\nu +\bar \nu  $, is viewed as one of the 
 the most  favorite  test  case for new physics beyond the  Standard Model \cite{bigi}.
  The \rpv
 interactions contribute at tree level  by $\tilde d_L $ and    $\tilde d_R$ exchange. 
 Based on the  experimental bound,  $ BF_{exp}< 5.2 \times 10^{-9} , $ one  deduces 
\cite{agashe} the  upper bounds,  
$\l '_{imk}< 0.012 {\tilde d_{kR}} , \ 
\l '_{i3k}< 0.52 {\tilde d_{kR}} .$
For the $B-$ meson decay processes, one finds the bound: 

 $ \l '_{ijk}\l '_{[l3k,lj3]} < 1.1 \ 10^{-3} {\tilde d_{k[R,L]} }^2 ,
 \  [B\to X_q +\nu +\bar \nu  ]$ \cite{grossman}.

$ \bullet \bullet $ {\bf Mixing  of  light and heavy quarks neutral mesons.}
In the single  dominant \rpv  coupling constant  hypothesis \cite{agashe},
the one-loop box diagrams, involving internal  sfermions and fermions lines, 
can contribute  to the transition matrix elements of neutral mesons charge conjugate 
pairs, 
$K-\bar K, \ D-\bar D, \ B-\bar B$ \cite{masiero}. 
 The  deduced bounds,  involving  the multiplicative  mass scaling dependence,
 $ [({100 GeV \over m_{\tilde \nu_{iL} } ^2  } )^2 +
({100 GeV \over m_{\tilde d_{kR} } ^2  } )^2  ]^{-1/4},$ are  \cite{agashe}:  

 $ \l '_{imk} < 0.11, \ [K\bar K]; \ \l '_{ijk} < 0.16 , \ [D\bar D],\ 
 \l '_{i3k} < 1.1, \ [B\bar B] $,

Under the  hypothesis of two  dominant \rpv coupling constants  \cite{roy},  
tree level contributions can occur via scalar exchange diagrams. 
 The four-fermion couplings are  then controlled by 
the quadratic products,  $F_{abcd}$, where 
the various entries span the sets: $ (ab)= [13, 23, 31, 32], \ 
 (cd)= [11, 12, 21, 22]$. Some of   the  strongest bounds  are \cite{roy,kaplan}: 

$ F'_{1311}< 2\ 10^{-5}, F'_{1331}< 3.3\ 10^{-8}, \ 
F'_{1221}< 4.5\ 10^{-9}.$

 The CP violation  asymmetries in the  $B-$ mesons decays, say, to CP-eigenstates,  
$$ a_{f(CP)}= {\G (B^0(t)\to f(CP) ) -\G (\bar B^0(t) \to f(CP)) \over 
 \G (B^0(t)\to f(CP) ) +\G (\bar B^0(t) \to f(CP))  }= 
{(1-\vert r_{f(CP)} \vert ^2 )\cos \D mt -2Im r_{f(CP)} \sin \D m t  \over
  (1+\vert r_{f(CP)}  \vert ^2  )   }, $$
are controlled by the 
ratio of amplitudes, $ r_{f(CP)} =q A (\bar B\to f(CP) )/(p A (B\to f(CP) )) $, 
where the ratio 
of $B-\bar B$ mixing parameters  is numerically, $q/p \approx 1 $. 
The tree level  \rpv interactions to the b-quarks decay  subprocesses,
$\bar b \to \bar c \bar c \bar d_i , \  \bar c  u_i  \bar d, \cdots $, can lead
to  significant contributions to  $r_{f(CP)}$, \cite{kaplan}. 
In particular,  these could contribute to subprocesses such as,
$ \bar b \to \bar d_i  d \bar d_i, \ [i=1,2,3]$, inducing decay channels such as,
$B \to K^0\bar K^0, \ \phi \pi^0, $  which are  tree level forbidden in the standard 
model. 
 
$ \bullet \bullet $ {\bf Non-leptonic decays of    heavy quarks mesons.}
 To the    flavour changing rare decay processes,  $B^+\to \bar K^0 +K^+ $
 and their  charge conjugate partners,  are assigned the 
 experimental upper bounds, $ BF <  5 \ 10^{-5} $.  
Fitting these with the \rpv contributions  provides   the bounds  \cite{carlson}: 
$\l ''_{i32} \l ''_{i21}< 5 \ 10^{-3} ({m_{\tilde q } \over m_W })^2 $. The 
 BF,   $\G ( B^+\to \bar K^0  +\pi^+) /\G (B^+\to J/\psi +K^+ )$, 
implies  $ \l ''_{i31} \l ''_{i21}< 4.1 \ 10^{-3} ({m_{\tilde q } \over m_W })^2 $.

  $ \bullet \bullet $ {\bf   Top-quark decay channels.} The \rpv induced  two-body 
decay channels, 
$t\to  \tilde l^+_i + \tilde d_k $, if kinematically allowed,  can compete with the  
electroweak decay channels,  $t\to b+W^+ $.
In reference to the weak interaction decay channel, the decay schemes, $\tilde l^+ \to
\tilde \chi^0+l, \ \tilde \chi^0\to \nu_i +b +\bar d_k$ cause violation of $e - \mu $
universality and a surplus of $b-$quarks events through the interactions, $\l '_{i3k}$. 
This possibility  can be probed on 
$p+\bar p\to t+ \bar t$ production  events recorded at the Tevatron,
by comparison of final states having $ e$ or $\mu $ accompanied by hadronic  jets.
  Fitting the \rpv contributions to the ratio of single  $e$ to $\mu $ BFs 
to the  ratio determined from the  CDF Collaboration
 top-quark-antiquark production  events,  
yields the  bounds,  \cite{agashe} $\l'_{13k}< 0.41 $.

\subsubsection{Baryon number violation}
$ \bullet \bullet $ {\bf Proton decay channels,   $\d B=1, \d L=\pm 1$. }  The  
effective Lagrangian 
description of the elementary baryons decays involves  dimension-6 
 operators built with quarks and lepton fields. 
 The \rpv interactions can  induce $B-L$  conserving contributions 
to the  two decay processes, $P\to \pi^0+e^+ $ and $ \  \pi^+ +\bar \nu $, through
tree level $\tilde d_{kR}$ squarks s-channel exchange. 
 Also at tree level,  there can occur   $B+L$ conserving 
interactions, through   the insertion of mass 
 mixing terms coupling the left and right chirality squarks. These contribute to the
chirality-flip,   $ \d B=-\d L = 1$,  decay process, $P\to \pi^++\nu $. 
  Borrowing  the  familiar dimensional analysis  argument from GUT physics,
\cite{langacker,masiero1,weinberg82},  one derives,
 based on  the naive rescaling, $ m_X^2 \to  \tilde m^2/\l '' \l '$,   the bounds, 
 $ \l '_{l1k} \l ^{''\star }_{11k} <10^{-25} - 10^{-27}\  {\tilde d_{kR} } ^2 $ 
for the first two processes,  $P\to \pi^0+e^+ ,  \  \pi^+ +\bar \nu $) 
and   $ \l '_{11k} \l ^{''\star }_{m1k} < 10^{-25} - 10^{-27}\ 
{\tilde d_{kR} } ^2  ({m^2_{\tilde d_{kR} } \over \d \tilde m^d_{LR} } ) $,
for the third process  ($P\to \pi^++\nu $). 

The analysis of  vertex  loop diagrams associated with the Higgs boson 
dressing of the vertex $\tilde d u d$ and the box loop diagrams, 
$ u+d\to {h^+}\to  d+u \to {\tilde d}\to  \bar \nu + d$,   having the same 
configurations 
of external lines as for tree level diagrams, 
 and propagating   charged and neutral 
Higgs bosons internal lines \cite{smirnov}, indicates that  these could 
provide competitive bounds on the \rpv coupling constants.  This gives
strong bounds for all combinations of pair products,  
$\l'   \l ''  < 10^{-7}-\ 10^{-9} $.
Stronger bounds, $\l' \l ''  < 10^{-11} $, hold if ones takes CKM  flavour mixing 
into account. Some representative examples are: 
$\l'_{3j3} \l ''_{121} < 10^{-7}, \  $ (no matching case) 
$\l'_{2j2} \l ''_{131} < 10^{-9} \  $ (matching case), 
where matching (no matching) refers to the case in which
the generation index of $d $ or $ \ d^c$ fields  
in $\l '$ coincides (differs)  from that of the $d^c$ field in $\l ''$.

Another mechanism for proton decay, involving a sequential
 tree  level exchange of $\tilde b ,  \ \tilde \chi^\pm $,  \cite{carlson}  
gives bounds 
 for the following three product combinations, 

   $ \l '_{ijk}   \l ''_{m21}< 10^{-9}, \ 
  \l '_{ijk} \l ''_{m31} < 10^{-9} , \   \l '_{ijk} \l ''_{m32} < 10^{-9}$.

However, there remains in this analysis 
certain weakly constrained products, such as,

$\l '_{12m} \l '' _{33m} < 10^{-2}, 
\l '_{112} \l '' _{331} < 10^{-2},\ \l '_{33l} \l '' _{221} < 10^{-1}$.

 The contributions  to the $\d B=2$  B-meson decay 
 processes, $B\to \L +\L  $ or $ B\to \S^++\S^-$,  
at  tree level with sequential $\tilde q $ and $\tilde \chi ^+$ exchanges  
\cite{carlson},   give bounds on several products,

 $\l_{ijk} \l ''_{131} < 10^{-13}, \ 
\l_{ijk} \l ''_{132} < 10^{-12}, \ \l_{ijk} \l ''_{221} < 10^{-13},
\ \l_{ijk} \l ''_{321} < 10^{-13}$. 

There remain, however, certain weakly constrained products, such as,
$ \l_{ijk} \l ''_{33[m=1,2]} < 10^{-3}, \   10^{-2},
\l_{ijk} \l ''_{23[m=1,2]} < 10^{-3}  , \ 10^{-2}.$

$ \bullet \bullet $ {\bf   Decays of scalar and neutralino LSPs
(Lightest Supersymmetric Particles).} 
The \rpv interactions can contribute at tree level to desintegrations
 of scalar neutrinos, $\tilde f \to f' +\nu $,  or of  
 neutralinos, $\tilde \chi^0 \to f+f'+\nu $. In order for these  processes to    
 occur  inside a  detector of  length $l>1 $  meter,  (corresponding to
proper  lifetimes   above $3\ 10^{-9} s$)  based on the tree-level decay mechanisms, one
must require the lower   bounds on coupling constants \cite{dawson,dreiner1},
 $$ [\sqrt 3 \l ' , \l ]_{sneutrinos} >  { 10^{-7}\sqrt {\b \g } \over 
 ( \tilde m/GeV )^\ud  } ,\ \
 [\sqrt 3 \l '' , \sqrt 3 \l ' , \l ]_{gauginos} >  5 \ 10^{-2}\sqrt {\b \g }  
({ m_{\tilde f} \over 100 GeV} )^2
({1 GeV \over m_{\tilde \chi }  })^{5/2},  $$
where $\b =v/c $ is the decaying particle velocity, and $\g = (1-\b ^2)^{-\ud }.$ 

The      stability conditions  of sneutrinos or neutralinos
 against decays occurring within the age of the  Universe today, 
$\tau_U\approx 10^{9} $ yrs, 
would place   bounds on all the coupling constants $\l , \ \l' , \ \l ''$,
which are smaller than the detectors bounds  given above by a factor of $10^{12}$.  
 One concludes  from this that cold dark matter  candidates from supersymmetry   
are practically dismissed, unless $R$-parity is broken at  infinitesimal levels.

The one-loop level decay modes, $\tilde \chi ^0\to \nu + X, \ \g + X $, place 
bounds on products of  $\l $ with the sneutrino VEVs $v_i$ \cite{salati}. 
Weaker bounds on the \rpv coupling constants 
would be imposed if one allowed the LSPs to start decaying 
after the  nucleosynthesis period   or after the protons
 and  electrons-positrons recombination period.  
The cosmological bounds on  the LSPs masses  obtained from the 
familiar constraints on the age of the Universe and the energy density abundance, 
$\Omega_0 h^2 < 1$,  depend  principally
on the LSPs annihilation rates. These  are practically unaffected by the \rpv 
interactions
except for the implications  derived  from the  effects of 
the LSPs decays following their thermal decoupling 
from the plasma.  The physics here is similar to that of heavy neutrinos \cite{kolb}.

$ \bullet \bullet $ {\bf Cosmological  baryon  and lepton number asymmetries.}
 The phenomenology of   baryogenesis (ratio of baryon  
number to entropy densities of the Universe  set today at the small value,
 $B=n_B/s \approx 10^{-10 } $) faces three basic problems \cite{kolb}: 
 (i) Generation  of a baryon asymmetry at some temperature, $T_{BA}$.
Minds are still unsettled concerning the relevant mechanism  and 
the scale of $T_{BA}$,  for which  
a variety of possibilities are still envisaged (high energy  GUTs,  
$T_{BA} \approx m_X/10 $;
low energy Standard Model,  $T_{BA} \approx T_C= m_W/\a_W $;    or intermediate  
non-perturbative approach as 
in the  Dine-Affleck squarks condensate  mechanism). 
(ii) Erasure of  the prexisting  baryon asymmetry  via 
B and/or  L violating interactions  inducing reactions among quarks and 
 leptons or gauge and  Higgs bosons,  which might  be in thermal  equilibrium at some 
temperature,
$T< T_{BA}$ during the cosmic expansion. This is formulated in terms 
of the  reaction rate $\G _D $  and the Universe 
 expansion rate, $H \approx 20 T^2/M_P$, by  the out-of-equilibrium  
condition, $\G_D/H < 1$. The erasure takes place for all linear combinations, $B+aL$,
 except for the (non-thermalizing modes) which   remain conserved by 
the interactions.   
(iii) The non-perturbative contributions associated to  
 the electroweak sphalerons, which induce vacuum transition processes,
 $ 0\to  \prod_i (u_{iL} u_{iL} d_{iL} l^-_{iL}), \quad  0\to 
\prod_i (u_{iL} d_{iL} d_{iL} \nu _{iL}), $  violating $B, \ L$ 
via the anomalous combination, $\d (B+L)= 2N_{gen}$, 
while conserving $B_i-L_i, \ [i=1,2,3]$.  Accounting for the flavour changing 
interactions of quarks, 
the effectively conserved combinations are in fact, 
 $(B/3 -L_i)$.  Since the sphaleron induced rates, over the  wide period, 
$m_X < T < T_C$,  are very much faster than the expansion rate, 
 $\G _{sphal}/H  \approx  ( T/H )e^{-2m_W/(2\a _W T) }  \approx 10^{17 } $,
 this will damp the  $(B+L)$ component of the asymmetry, while 
 leaving the components  
$(B/3-L_i)$ constant. A necessary condition for baryon asymmetry 
 erasure in the presence of sphalerons is  then that this must have been produced
via $(B-L)$ or $(B-L_i)$ (for some fixed $i$) violating interactions 
\cite{campbell,fischler}. 

The  out-of-equilibrium conditions, taking into account the 
set of $ 2 \to 2 $ processes, $u+d \to \tilde d^\star \to \bar d+\tilde \chi^0,
\  u+e \to \tilde d \to d+\tilde \chi^0, \ \nu +e \to \tilde \mu 
\to \mu +\tilde \chi^0,$ and $ 2 \to 1  $ processes,  
$ d +\nu \to  \tilde  d , \   u +e \to  \tilde { d },\  \nu +e\to \tilde \mu $, 
give on all \rpv coupling constants the strong bounds, 
$\l , \l ' , \ \l ''\  < 5\times 10^{-7} 
({\tilde m\over 1 TeV })^\ud  , $  
corresponding to an  updated version \cite{dreiross} of  previous 
analyses \cite{salati,campbell}.   

 A more refined analysis in \cite{dreiross},  accounting for all 
the  relevant symmetries of the Standard Model, through the equations
on the particles chemical potentials expressing  chemical 
equilibrium  constraints,  turns out to lead
to milder constraints. Thus, it is found that  the bounds on the 
B-violating  $\l ''$ interactions  are removed in the absence of sphalerons, 
but remain in force when these are included.  
For the L-violating interactions,    only a subset of the coupling constants,
 $\l ,\ \l '$, remains bounded. The reason is that 
one need impose the out-of-equilibrium  conditions
only for one lepton family, say $J$, corresponding to one conserved combnination, 
$(B/3 -L_J)$. The above bounds would then hold only for the subsets, $\l_{Jjk}, \ 
\l '_{Jjk}$.   An indicative analysis of  
the fields  basis dependence of these  bounds  is  made in \cite{davidson}. 

For the dimension $ D> 5$ non-renormalizable 
operators, the out-of-equilibrium conditions, as  formulated by 
the inequalities: 
$ \G_D\approx  T ({T\over \L } )^{2(D-4) }  < H \approx {20 T^2\over M_P} $, 
lead to  the bounds:
 $ \L > [{ T^{2(D-4) -1} M_P \over 20 }]^{1/2(D-4)}$ \cite{campbell}. 
The strongest bound, associated with $ T= T_{GUT} \approx  10^{14} $ GeV,  is:  
 $ \L > 10^{ 14 +2/(D-4) }$ GeV.

It is important, however,  to note that these baryogenesis erasure constraints 
 are really sufficient conditions  and do not constitute strict bounds.
 They  could be   evaded  if    baryogenesis occurred 
at the electroweak scale or in non-perturbative  models 
in case of an  insufficient reheat temperature. 

$ \bullet \bullet $ {\bf Nucleon-antinucleon oscillations.} 
 The $N\to \bar N$ transition is described by the effective Lagrangian,
 $L= \d m \bar N^c N +h.c.$,  such that the  oscillation time for free neutrons reads, 
$\tau ^{-1} _{oscill}  = \G = 1/\d m  $.  Recall \cite{masiero1} that this is linked 
to the nuclear lifetime
against decays, $ NN \to X$, denoted as, $\tau_{NN}$,  by the relationship:
  $\tau_{NN} =a \d m ^2 /m_N ,$ where 
$a \approx 10^{-2}$ is a nuclear wave function factor and $m_N$ the nucleon mass. 
 The present  experimental bound  on oscillation time is, 
$\tau_{oscill} > 1.2 \ 10^{8}$ s. This is to be compared with 
the bound deduced from, $ \tau_{NN} > 10^{32}$ years, which yields: 
$ \d m< 10^{-28} $, hence $\tau_{oscill} > 10^6 $ s.   

The initialy proposed \rpv induced  mechanism  \cite{zwirner}
 involved an intermediate three-scalars annihilation coupling.
A more competitive  mechanisms  has been proposed 
which involve the  three-body mechanism,
$udd_R\to \tilde d_R +d_R\to {\tilde g}\to  d^c \tilde d'_R \to d^c u^cd^c$.
   For a simple estimate, one can borrow 
the result from GUT physics, $L= e^4 O/M^5$, with $ M= 4\ 10^5 - 10^6$ GeV.
The  bound resulting  from an analysis of the oscillation amplitude reads,  
 $ \l''_{112}< 3.3 \ 10^{-10} - 10^{-11 }$ \cite{zwirner} .

Another three-body mechanism uses the reactions scheme,
 $udd_R\to \tilde d_R +d_R \to \tilde d'_L +d_R \to { q\tilde q}\to 
 d^c \tilde d''_R \to d^c u^cd^c$,  which involves a $W^\pm $  gauge boson box
 diagram \cite{goity}.
This is described by the Lagrangian, 
$ L_{\tilde w}\propto  Col (\bar u^cd)(\bar d^cd)(\bar u^cd),  $
where $ Col$ stands for a color factor.  The resulting bound is:   
$ \l''_{132} <  10^{-3}   $ \cite{goity}.

$ \bullet \bullet $ {\bf Double nucleon decay processes, $\d B=2$.}
The nuclear decay processes,  $ ^{16} O\to^{14}C+ 2\pi^0, \   +\cdots $, are 
described by dimension-9 six-fermion operators, 
$ O= d_Rd_R d_Ru_Rq_Lq_L,\ d_Rd_R q_Lq_Lq_Lq_L, \cdots $.
Using  a naive  rescaling  from the GUT-like Lagrangian,
$ L= {e^4\over M^5} O$,  the associated inverse
lifetime formula reads, $(\tau /yrs)^{-1} \approx { e^8 10^{29}  \over (M /GeV)^{10} }
 $ where, $ e = (4\pi \a )^\ud \approx 0.3  $,  is the electron charge in  Heaviside 
units. From the experimental
bound, $\tau > 10^{30,33} $ yrs, one deduces the bounds: 
$\l ''_{131} < 5 \ 10^{-3} ,\ \l ''_{121} < 10^{-6}$ \cite{zwirner}.
The peculiar  double nucleon, $\d B= \d S=-2$, 
 decay process,  $ N+N \to ^{14}C+ K^++K^+$ could provide a 
competitive channel for  nuclear decay reactions. The analysis in \cite{goity} gives 
the   
 bound, $ \l ''_{121}< 10^{-15} {\cal R}^{-5/2},$
  where  the parameter $\tilde \L $ in  ${\cal R} =
 {\tilde \L  \over (m_{\tilde g} m_{\tilde q}^4)^{1/5}  } $,  
represents an estimate for the nuclear matrix element. Varying 
${\cal R}$ in the interval, $10^{-3} - 10^{-6} $, one finds:
 $\l ''_{121}< 10^{-7} - 10^{0}$.

 $ \bullet \bullet $ {\bf Dimension-5 operators contributions to proton decay.}    
Except for few isolated works,  little attention  was 
devoted  so far to the dimension-5  dangerous operators.
The analysis in \cite{murakap} has focussed on the baryon number violating  F-term
 operator, $(QQQL)_F$, which can be induced by tree level  exchange of a 
massive  color triplet Higgs bosons.  This superpotential term contributes 
two-fermions two scalars interactions,
which  induce via a one-loop  gaugino dressing  mechanism  
\cite{hisano,hisano1,hisano2}  contributions described by dimension-6 four-fermion 
operators, $qqqq$. 
There arise a set of several such operators, $O_{ni}$,
 which could cause proton or neutron  decays in peculiar  channels, such as  
$P\to K^0+l^+_i,\ N \to K^0 +\bar \nu_i $. Restricting to the dominant contribution 
from  wino dressing, one deduces the effective Lagrangian as, 
$ L= {g^2\over 4\pi ^2} \sum_{n,i} {g_{1i}\over \L } a_{ni} O_{ni} +
{g_{2i}\over \L } b_{ni} O_{ni} $, where $a_{ni} , b_{ni} $ are calculable
 loop amplitudes factors, $n $ labels the 
independent operators and $i$ the   emitted leptons flavour. The experimental bounds, 
based on the choice of gravitational Planck mass scale, 
$\L =M^\star =M_P/\sqrt {8\pi }$, yield:
$(\sum_i\vert g_{[1,2] i} \vert^2 )^\ud < [3.6 \ 10^{-8} , 1.0 \ 10^{-7}] .$

$ \bullet \bullet $ {\bf Infrared fixed points.}  In direct analogy  with the familiar  
 estimate for the top-quark mass, $m_t  (pole) \equiv (200 GeV)\sin \b $, which is
derived by assuming the existence 
of an infrared  fixed point in the Yukawa coupling constant, $\l^u_{33}$, one can 
deduce similar fixed point bounds for  the third generation \rpv coupling constants.  
The argument  is 
again based  on  the vanishing of the beta function in the renormalization group flow, 
via the competition between Yukawa and gauge interactions, as displayed schematically 
by the equation,
$ (4\pi )^2 {\dh \ln \l_{ijk} \over \dh t} = {8\over 5} g_1^2 
+3 g_2^2 -(\d _{j3} +2\d _{k3} )
\l^{u2} _{33}, \ [t =\ln m_X^2/Q^2]$, 
 where the  c-number  coefficients in front of the coupling constants 
represent the fields anomalous dimensions.     Equivalently, this reflects on  
 the assumption of perturbative unitarity 
(absence of Landau poles) for the \rpv coupling constants at high energies scales.
 The  predicted fixed point 
 bounds  \cite{bargerfp,goity,fp1,fp2} are: 
$\l _{233} < 0.90, \ \l '_{333} < 1.01, \  \l ''_{323} < 1.02$.

\subsection{Conclusions}
\label{secxxx5}
The \rpv interactions represent one among several  
sources of physics beyond the  Standard Model.  
Other  possibilities in the context of non-minimal supersymmetry 
(leptoquarks, fourth family of quarks and leptons,
left right symmetric gauge 
groups,  mirror fermions, extra gauge bosons,  etc...) may be realized.
 It is  likely, however, as  has been assumed, 
 that  one can exclude interference effects between these various possibilities.

It is  clear that the low energy phenomenology is a rich and  valuable  source of 
information on the \rpv interactions.
 Perhaps  the strongest and most robust  bounds are those derived  from the rare 
forbidden 
neutrinoless double beta decay, proton decay and $l\to eee $ decay processes.
Some general trends here are that: (1)  Most  
of the  B-violating coupling constants $\l ''_{ijk}$ are below
$10^{-6}$ or so,  except for $\l ''_{332}$;  (2)  The  L-violating
coupling constants, 
$\l_{lmn} ,\ \l '_{lmn} $, associated with the first and second families 
and  also those  involving  one third family index only, such as, 
$\l_{3mm} ,\ \l '_{3mn}, $ and permutations thereof,  
tend to be more suppressed. 
There still survives a number of weakly  constrained cases in the  specific   
family configurations, 
$ \l_{123}, \ \l_{l33}; \l '_{i13}, \ \l' _{i23},  $  etc... 
 This nourishes the  (foolish?) 
hope that  a few coupling constants may just happen to be of order $ 10^{-1} $
or so, enough to lead to directly observable effects  at high energy colliders.

Nevertheless, one must exercise 
a critical eye on the model dependent assumptions and not
treat all bounds indiscriminately.    The  apparently strong bounds deduced from 
the leptons EDMs,  the two-nucleon decay processes  or the cosmic  baryon asymmetry  
erasure
appear as  fragile  bounds relying on model-dependent assumptions.
One must also  keep in mind  the limitations in 
the basic hypotheses of single or pairs of dominant coupling constants.
These presume the absence of cancellations from different configurations 
and the existence of  strong flavour hierarchies.  Often this is 
taken as a  reflection of dynamics associated with  horizontal flavour 
symmetries.  However, to satisfy the  various
 constraints imposed on  supersymmetry models,  
it is possible that Nature  may  have chosen a different option. This 
could be  string theory  or gauge dynamics.  It could also   
be along the lines of the so-called effective supersymmetry  approach  \cite{cohen},  
implying  TeV scale  supersymmetry breaking parameters with lightest scalar 
superpartners to be found amongst the third   family quarks or leptons. 
 
The prospects on the long term are encouraging. Thanks to 
 the planned machines, experimental  measurements of rare  forbidden decay processes 
are expected to  gain several order of magnitude in  sensitivities 
\cite{marciano}. Factors 
of $10 - 100 $ improvements in accuracies are also  anticipated for  
high precision  measurements of magnetic or electric dipole moments.
 Some progress, although at a more modest  level, 
 is expected to take place for the high precision physics
observables. Our   theoretical understanding  of supersymmetry  and of  physics beyond
 the standard  model is likely also to deepen in the meantime. Efforts on all these
 fronts should be needed  in meeting with the future challenges of  high precision 
physics.

\newpage

\section{Alternatives to conserved $R$-parity}
\label{chap3}
On the theoretical side, one has a priori little knowledge on $R$-parity violating 
couplings, since they have the same structure as Yukawa couplings, which are not 
constrained by the symmetries of the MSSM. Turning the argument the other way, 
one expects any model of fermion masses to give predictions for broken $R$-parity 
\cite{BenHamo,Banks,BLR,Borzumati,BDLS,ELR}. In this note\footnote{Most of what 
follows is based on work done in collaboration 
with P. Bin\'etruy and C.A. Savoy \cite{BDLS}.}, 
we want to show that abelian family symmetries, which can explain the observed 
fermion mass spectrum, naturally generate a flavour hierarchy between $R$-parity 
violating couplings that can easily satisfy all present experimental bounds.
\vskip .3cm

Let us first explain how a family-dependent symmetry $U(1)_X$ constrains the Yukawa 
sector \cite{FN}. Consider a Yukawa coupling $Q_i \bar u_j H_u$; invariance under 
$U(1)_X$ implies that $Q_i \bar u_j H_u$ appears in the superpotential only if 
its $X$-charge vanishes, i.e. $q_i+u_j+h_u=0$ (we denote generically the charge 
of any superfield $\Phi_i$ by a small letter $\phi_i$). To account for the large 
top quark mass, we shall assume that this happens only for the Yukawa coupling 
$Q_3 \bar u_3 H_u$; thus all fermions but the top quark are massless before the 
breaking of the symmetry. One further assumes that the family symmetry is 
broken by the vacuum expectation value of a Standard Model singlet $\theta$ 
with $X$-charge $-1$, and that the other Yukawa couplings are generated from 
interactions of the form
\begin{equation}
  y^u_{ij}\ Q_i \bar u_j H_u\, \left( \frac{\theta}{M} \right)^{q_i+u_j+h_u}
\label{eq:nonrenormalizable}
\end{equation}
where $M$ is a mass scale, and $y^u_{ij}$ is an unconstrained coupling of order one. 
Such nonrenormalizable terms typically appear in the low-energy effective field 
theory of a fundamental theory with heavy fermions of mass $M$ - one may also 
think of a string theory, in which case $M=M_{Pl}$. If $U(1)_X$ is broken below 
the scale $M$, $\epsilon =\, <\theta> / M$ is a small parameter, 
and (\ref{eq:nonrenormalizable}) generates an effective Yukawa coupling
\begin{equation}
  Y^u_{ij}\ =\  y^u_{ij}\ \epsilon^{\, q_i+u_j+h_u}
\label{eq:Yu}
\end{equation}
whose order of magnitude is fixed by the $X$-charges. Similarly, one has, for 
down quarks and charged leptons:
\begin{eqnarray}
  Y^d_{ij} & \sim & \epsilon^{\, q_i+d_j+h_d}  \label{eq:Yd} \\
  Y^e_{ij} & \sim & \epsilon^{\, l_i+e_j+h_d}  \label{eq:Ye} 
\end{eqnarray}
A family-dependent symmetry thus naturally yields a hierarchy between Yukawa 
couplings. Notice that if a particular coupling $Y_{ij}^u$ has a negative charge, 
$q_i+u_j+h_u <0$, it is not possible to generate it from (\ref{eq:nonrenormalizable}), 
due to the property of holomorphicity of the superpotential $W$.
\vskip .3cm

An explicit example of a model which reproduces the observed masses of quarks and 
their mixing angles is the following. Consider the charge assignment
\begin{eqnarray}
&& q_1-q_3=3 \ , \ q_2-q_3=2 \ , \ u_1-u_3=5 \ , \ u_2-u_3=2 \ , \nonumber \\
&& d_1-d_3=1 \ , \ d_2-d_3=0 \ . 
\end{eqnarray}
The corresponding quark mass matrices are of the form
\begin{eqnarray} 
Y^u  
&=& 
\left(
\begin{array}{ccc} 
\epsilon^8 & \epsilon^5 & \epsilon^3 \\ 
\epsilon^7 & \epsilon^4 & \epsilon^2 \\
\epsilon^5 & \epsilon^2 & 1
\end{array}
\right) \ \ \
Y^d  = \epsilon^x 
\left(
\begin{array}{ccc} 
\epsilon^4 & \epsilon^3 & \epsilon^3 \\ 
\epsilon^3 & \epsilon^2 & \epsilon^2 \\
\epsilon & 1 & 1
\end{array}
\right)
\label{eq:FN_matrices}
\end{eqnarray}
where in fact all entries are only known up to factors of order one.  The small
number $\epsilon$ has been assumed here to be numerically equal to the
Cabibbo angle, $V_{us} \simeq 0.22$. The assignement above gives the following relations
\begin{eqnarray}
&& \frac{m_u}{m_t} \sim \epsilon^8 \ ,\ \frac{m_c}{m_t} \sim \epsilon^4 \ ,\
\frac{m_d}{m_b} \sim \epsilon^4 \ ,\ \frac{m_s}{m_b} \sim \epsilon^2 \ ,
\nonumber \\
&& V_{us} \sim V_{cd} \sim \epsilon \ ,\ V_{ub} \sim V_{td}
\sim \epsilon^3 \ ,\  V_{cb} \sim V_{ts} \sim \epsilon^2 \ , \nonumber \\
&& \frac {m_b}{m_t} \sim \epsilon^x \frac{1}{\tan \beta} \ ,   
\label{eq:FN_masses}
\end{eqnarray}
which hold at the scale where the abelian symmetry is broken, usually taken to be
close to the Planck scale. With renormalization group effects down to the weak scale 
taken into account, this charge assignment can accommodate the observed masses
and mixings. More generally, assuming that the charge carried by each Yukawa 
coupling is positive, there are only a few structures for $Y^u$ and $Y^d$ allowed
by the data, which differ from (\ref{eq:FN_matrices}) only by a $\pm 1$ change in 
the powers of $\epsilon$. In the lepton sector there is more freedom, as the leptonic
mixing angles (which are physical only if the neutrinos are massive) are not yet
measured. The number $x = q_3+d_3+h_d$, which is related through (\ref{eq:FN_masses})
to the value of $\tan \beta$, is actually constrained if one imposes gauge anomaly
cancellation conditions.
\vskip .3cm

$R$-parity violating couplings are constrained by $U(1)_X$ exactly in the same
way as Yukawa couplings. They are generated from the following nonrenormalizable
interactions:
\begin{equation}
L_i L_j \bar e_k \left( \frac{\theta}{M} \right)^{l_i+l_j+e_k}\ , \
  \hskip .5cm  L_i Q_j \bar d_k \left( \frac{\theta}{M} \right)^{l_i+q_j+d_k}
\label{eq:Rp_NR}
\end{equation}
To avoid unnaturally large values of the quark charges, we have assumed a baryon 
parity that forbids the appearance of $\bar u \bar d \bar d$ terms in
the superpotential\footnote{This baryon parity can be a residual discrete 
symmetry resulting from the breaking of $U(1)_X$. Another possibility is that
the couplings are forbidden by holomorphy, which happens when all combinations
of charges $u_i+d_j+d_k$, $i,j,k=1,2,3$, are negative \cite{BDLS}.}, thus
preventing proton decay. One can see from (\ref{eq:Rp_NR}) that abelian
family symmetries yield a hierarchy between $R$-parity violating couplings 
that mimics (in order of magnitude) the down quark and charged lepton mass
hierarchies. Indeed, one has \cite{BDLS}:
\begin{eqnarray}
  \lambda_{ijk} & \sim & \epsilon^{\, l_i - h_d}\ Y^e_{jk} \label{eq:lambda} \\
  \lambda'_{ijk} & \sim & \epsilon^{\, l_i - h_d}\ Y^d_{jk} \label{eq:lambda'}
\end{eqnarray}
Provided that the Yukawa matrices $Y^d$ and $Y^e$ are known, experimental
limits on $\lambda$ and $\lambda'$ can be translated into a constraint
on $l_i - h_d$. We shall assume here that the charge carried by each
operator is positive, and take for $Y^d$ the structure (\ref{eq:FN_matrices}).
In the lepton sector, there is not enough data to determine completely
the $Y^e_{ij}$; however, it is possible to derive upper bounds on the 
couplings (\ref{eq:lambda}) from the three charged lepton masses.
Assuming a small value of $\tan \beta$ (corresponding to $x=3$), 
one finds that the experimental bounds on product couplings including
\cite{babu,roy,barbieri}
\begin{eqnarray}
\mbox{Im} \left( \lambda'_{i12}\ \lambda^{\prime *}_{i21} \right)
 \leq 8.10^{-12} & & (\epsilon_K)  \label{eq:epsilon_K}  \\
\end{eqnarray}
 are satisfied as soon as:
\begin{equation}
  l_i-h_d\ \geq\ 2 - 3
\label{eq:condition}
\end{equation}
For moderate or large values of $\tan \beta$, larger charges would be required.

Now the condition (\ref{eq:condition}) can be used, together with (\ref{eq:lambda})
and (\ref{eq:lambda'}), to derive upper bounds\footnote{It should be stressed here 
that, while (\ref{eq:condition}) strongly depends on $\tan \beta$, this is not the
case for the couplings $\lambda$ and $\lambda'$ themselves.} on the individual
couplings $\lambda$ and $\lambda'$. We find that all of them are (well) below
the experimental limits - in particular, there is no explanation of the possible
HERA large-$Q^2$ anomaly. Thus, if abelian family symmetries are responsible for
the observed fermion mass spectrum, we expect the first signals for broken $R$-parity
to come from FCNC processes. Let us stress, however, that these conclusions are not
completely generic for abelian family symmetries: they would be modified if $U(1)_X$
were broken by a vector-like pair of singlets \cite{ELR}, or if we gave up the
assumption that the $X$-charge carried by each operator is positive.
\vskip .5cm

In addition, the inclusion of the bilinear $R$-parity violating 
terms $\mu_i\, L_i H_u$ can modify the previous picture. 
In the presence of these terms, the $L_i$ fields assume
a vacuum expectation value together with the Higgs fields.
The low-energy $H_d$ and $L_i$ fields have then to be redefined in 
such a way that only $H_d$ has a nonzero vev. This may modify significantly
the order of magnitude relations (\ref{eq:lambda}) and (\ref{eq:lambda'}), as
we show below. For convenience, we write the superpotential as
\begin{equation}
  W\ =\  \lambda^e_{\alpha \beta k}\, \hat{L}_{\alpha} \hat{L}_{\beta}
        \bar e_k\ +\ \lambda^d_{\alpha j k}\, \hat{L}_{\alpha} Q_j
        \bar d_k\ +\ \mu_{\alpha}\, \hat{L}_{\alpha} H_u
\end{equation}
where $\hat{L}_{\alpha}$, $\alpha = 0,1,2,3$ denote four $SU(2)_L$ doublets 
with hypercharge $Y=-1$ and well-defined $X$-charges $l_{\alpha}$.
After supersymmetry breaking, each $\hat{L}_{\alpha}$ acquires a vev,
$v_{\alpha} \equiv < L^0_{\alpha} >$. The standard Higgs field $H_d$ 
is defined as the combination of the $\hat{L}_{\alpha}$ whose vev breaks
the hypercharge:
\begin{equation}
  H_d\ =\ \frac{1}{v_d} \sum_{\alpha}\, v_{\alpha} \hat{L}_{\alpha}
\end{equation}
where $v_d \equiv \left( \sum_{\alpha} v^2_{\alpha} \right)^{1/2}$.
The orthogonal combinations $L_i$, $i=1,2,3$ are the usual lepton fields:
\begin{equation}
  \hat{L}_{\alpha}\ =\ \frac{v_{\alpha}}{v_d}\, H_d\ +\
        \sum_i\, e_{\alpha i}\, L_i
\label{eq:redefinition}
\end{equation}
The ambiguity in the rotation $e_{\alpha i}$ is partially lifted by requiring 
that $L_1$ and $L_2$ do not couple to $H_u$. After this redefinition, 
the superpotential reads:
\begin{eqnarray}
  W & = & Y^e_{ik}\, L_i \bar e_k H_d\ +\ Y^d_{ik}\, Q_i \bar d_k H_d\
        +\ \lambda_{ijk}\, L_i L_j \bar e_k\
        +\ \lambda'_{ijk}\, L_i Q_j \bar d_k\  \nonumber  \\
        & & +\ \mu \cos \xi\, H_d H_u\ +\ \mu \sin \xi\, L_3 H_u\
\end{eqnarray}
where $\mu \equiv \left( \sum_{\alpha} \mu^2_{\alpha} \right)^{1/2}$, $\xi$ is
the angle between the vectors 
$\vec{\mu}$ and $\vec{v}$, $\cos \xi \equiv \sum_{\alpha} \mu_{\alpha} v_{\alpha}
 / \mu v_d$, and the physical Yukawa and $R$-parity violating couplings are given by:
\begin{eqnarray}
  Y^e_{ik}\ =\ 2 \sum_{\alpha,\, \beta}\, e_{\alpha i}\,
        \frac{v_{\beta}}{v_d}\: \lambda^e_{\alpha \beta k} & & Y^d_{ik}\
        =\ - \sum_{\alpha}\, \frac{v_{\alpha}}{v_d}\: \lambda^d_{\alpha i k}
        \label{eq:redefined_lambda}  \\
  \lambda_{ijk}\ =\ \sum_{\alpha,\, \beta}\, e_{\alpha i}\, e_{\beta j}\,
        \lambda^e_{\alpha \beta k} & & \lambda'_{ijk}\ =\ \sum_{\alpha}\,
        e_{\alpha i}\, \lambda^d_{\alpha j k}
\label{eq:redefined_lambda'}
\end{eqnarray}
Due to the residual term $L_3 H_u$, the tau neutrino acquires
 a mass through mixing with the neutralinos \cite{Hempfling}:
\begin{equation}
 m_{\nu_3}\ =\ m_0\, \tan^2 \xi  \hskip 1cm 
 m_0\ \sim\ (100\, GeV)\, \cos^2 \beta\, \left( \frac{500\, GeV}{\widetilde{m}} \right)
\label{eq:m_nunu}
\end{equation}
where $\tilde{m}$ is a typical supersymmetry breaking scale, 
and the exact value of $m_0$ depends on the gaugino masses,
$\mu$ and $\tan \beta$. To be compatible with the LEP limit
on $m_{\nu_\tau}$, and with the even stronger cosmological
bound on neutrino masses ($m_{\nu} \leq {\cal O}\, (10\, eV)$
for a stable doublet neutrino), one needs a strong alignment 
($\sin \xi \ll 1$) of the $v_{\alpha}$ along the $\mu_{\alpha}$.
\vskip .3cm

Let us specify formulae (\ref{eq:redefinition}), (\ref{eq:redefined_lambda}),
(\ref{eq:redefined_lambda'}) and (\ref{eq:m_nunu}) in the presence of an abelian 
family symmetry. Assuming that the bilinear terms are generated through 
supersymmetry breaking \cite{GM} (which ensures that the $\mu_{\alpha}$ 
are of the order of the weak scale, as required by electroweak symmetry breaking), 
one finds:
\begin{equation}
  \mu_{\alpha}\ \sim\ \tilde{m}\, \epsilon^{\, \tilde{l}_{\alpha}}
        \hskip 1cm  v_{\alpha}\ \sim\ v_d\, \epsilon^{\,
        \tilde{l}_{\alpha} - \tilde{l}_0}
\end{equation}
where $\tilde{l}_{\alpha}\ \equiv\ |l_{\alpha}+h_u|$, and the above estimates are 
valid for $0 \leq \tilde{l}_0 < \tilde{l}_i$, $i=1,2,3$. Thus the $v_{\alpha}$ are 
approximatively aligned along the $\mu_{\alpha}$ by the family symmetry \cite{Banks}, 
which implies (assuming with no loss of generality $\tilde{l}_3 \leq \tilde{l}_{1,2}$):
\begin{equation}
  \sin^2 \xi\ \sim\ \epsilon^{\, 2\, (\tilde{l}_3 - \tilde{l}_0)}
\label{eq:sin_xi}
\end{equation}
Furthermore, the redefinition (\ref{eq:redefinition}) is completely fixed by requiring 
$L_1 \simeq \hat{L}_1$ and $L_2 \simeq \hat{L}_2$, with
\begin{equation}
  \frac{v_{\alpha}}{v_d}\ \sim\ \epsilon^{\, \tilde{l}_{\alpha} -
        \tilde{l}_0}\  \hskip 1cm\  e_{\alpha i}\ \sim\ \epsilon^{\,
        |\tilde{l}_{\alpha} - \tilde{l}_i|}
\end{equation}
Note that $H_d \simeq \hat{L_0}$, which allows us to define $h_d \equiv l_0$.

The low-energy $R$-parity violating couplings depend on the signs of the charges 
$l_{\alpha}+h_u$. In all phenomenologically viable cases, the order of magnitude 
relations (\ref{eq:lambda}) and (\ref{eq:lambda'}) are modified to:
\begin{eqnarray}
  \lambda_{ijk} & \sim & \epsilon^{\, \tilde{l}_i - \tilde{l}_0}\ Y^e_{jk}
        \label{eq:lambda_bis} \\
  \lambda'_{ijk} & \sim & \epsilon^{\, \tilde{l}_i - \tilde{l}_0}\ Y^d_{jk}
        \label{eq:lambda'_bis}
\end{eqnarray}
By combining the eqs. (\ref{eq:m_nunu}), (\ref{eq:sin_xi}), (\ref{eq:lambda_bis}) 
and (\ref{eq:lambda'_bis}), we can write down a relation between the
mass of the tau neutrino, $R$-parity violating couplings $\lambda'$ and
down-quark Yukawa couplings
\begin{eqnarray}
m_{\nu_3} \sim m_0 (\frac{\lambda'_{3jk}}{\lambda_{jk}^d})^2
\end{eqnarray}
which is a generic prediction of this class of models.
 
For the sake of simplicity, we shall only describe two cases of interest.
The first one, in which all $l_{\alpha}+h_u$ are positive, yields
the standard Froggat and Nielsen structure:
\begin{eqnarray}
  Y^e_{ik}\ \sim\ \epsilon^{\, h_d+l_i+e_k}  &  &  \lambda_{ijk}\ \sim\
        \epsilon^{\, l_i+l_j+e_k}  \\
  Y^d_{ik}\ \sim\ \epsilon^{\, h_d+q_i+d_k}  &  &  \lambda'_{ijk}\ \sim\
        \epsilon^{\, l_i+q_j+d_k}
\end{eqnarray}
The second one, in which $l_i+h_u \geq 0 > l_0+h_u$, leads to an enhancement of
 flavour diagonal couplings relative to off-diagonal couplings. Indeed, the dominant
 terms in (\ref{eq:redefined_lambda}) and (\ref{eq:redefined_lambda'}) correspond
 to $\alpha=0$ or $\beta=0$, which provides an alignment of the $R$-parity violating
 couplings along the Yukawa couplings:
\begin{eqnarray}
  \lambda_{ijk} & \simeq & \frac{1}{2} \left( e_{0j}\, Y^e_{ik}\, -\,
        e_{0i}\, Y^e_{jk} \right)  \\
  \lambda'_{ijk} & \simeq & -\, e_{0i}\, Y^d_{jk}
\end{eqnarray}
As a consequence, $R$-parity violating couplings are almost diagonal in the basis of
 fermion mass eigenstates. Furthermore, they undergo an enhancement relative to the
 naive power counting, since e.g.
\begin{equation}
  \lambda'_{ijk}\ \sim\ \epsilon^{\, \tilde{l_i}-\tilde{l_0}}\ Y^d_{jk}\
        \sim\ \epsilon^{\, -2\, \tilde{l_0}}\ \epsilon^{\, l_i+q_j+d_k}
\end{equation}

This opens the phenomenologically interesting possibility that $R$-parity violation be
 sizeable while its contribution to FCNC processes is suppressed, as required
 by experimental data. Let us stress, however, that if the cosmological bound on
 neutrino mass is to be taken seriously, (\ref{eq:sin_xi}) indicates
 that $R$-parity violation should be very suppressed - unless some other 
mechanism provides the required alignment between the $v_{\alpha}$ and 
the $\mu_{\alpha}$.

\newpage

\section{Single production of supersymmetric particles}
\label{chap4}
\subsection{Indirect effects}

$\bullet$ {\bf fermion pair production.} The alternative that direct 
production rates would turn to be 
too small at LEP energies, is a possibility that might be envisaged \cite{Chou}.
The reason could be
either relatively heavy masses for \susyq particles or too weak couplings of 
the lighter particles with the $\tilde m$ particles. In this case, the virtual effects
of the \rpv interactions could lead to possible indirect signals. Sneutrino 
or squark t-channel exchange could contribute to processes 
$e^+e^- \to f \bar f $ with $f=e$,$\mu$,$\tau$,b,c if $\l$ or $\l'$ couplings 
were present, respectively, assuming a single dominant \cc. 
For $f=e$, s-channel exchange is also possible. Since the angular
distributions for the $\tilde m$ and the \rpv contribution are different, it is
proposed to divide the experimental angular width   
into bins, and to compare the observed number of events in each bin with the $\tilde m$
prediction. A contour of the detectability in the \rpv coupling constant-sfermion mass
plane gives some interesting bounds on some of the $\l_{ijk}$ and $\l_{ijk}'$
coupling constants. In \cite{Sridhar}, the contributions from 
t-channel exchange of squarks or sleptons
to the process $q \bar q \to t \bar t$ were studied. The comparison with
 the data from Tevatron on $t \bar t$ production is used to
 constrain the B-violating $\l''$ couplings and the L-violating
 $\l$ couplings.\\
$\bullet$ {\bf CP violation asymmetries.} The effects of  
$R$-parity interactions on flavor changing rates  and CP asymmetries  in the production
 of fermion-antifermion pairs at leptonic colliders
are examined in \cite{art1}. In the reactions, $e^-+e^+\to f_J +\bar f_{J'}, \ 
[J\ne J']$,  the  produced fermions may be  leptons, down-quarks or up-quarks,
and the center of mass energies may range  from
the Z-boson  pole up to $ 1000$ GeV.
Off the  Z-boson pole, the flavor changing rates  are controlled by tree  level 
amplitudes and the  CP asymmetries
by interference  terms between tree and loop level  amplitudes. At the 
Z-boson pole, both observables involve loop amplitudes.
The  lepton number violating interactions,
associated with the coupling constants, $\l_{ijk} , \ \l '_{ijk}$, are only taken
 into account.
The consideration of  loop amplitudes  is
restricted  to the Z-boson vertex corrections. The
 Z-boson decays  branching ratios, $B_{JJ'}= B(Z\to l^-_J+l^+_{J'})$, 
scale in order of magnitude as, 
$ B_{JJ'}\approx ({\l \over 0.1})^4 ({100 GeV \over \tilde m} )^{2.5}
 \ 10^{-9} $, and  
the   off Z-boson pole rates as,
 $ \s_{JJ'}\approx ({\l \over 0.1})^4 ({100 GeV \over \tilde m})^{3.5}
10^{2} fbarns$. 
 The corresponding results     for quarks have an extra color factor, $N_c=3$. 
The CP asymmetries at the Z-boson pole, 
$A_{JJ'}={B_{JJ'}-B_{J'J}\over B_{JJ'}+B_{J'J} } $, vary in the range, 
 $10^{0}, \ 10^{-1}\sin \psi $, where $\psi $ is the CP odd phase.
 The off Z-boson pole asymmetries, $A_{JJ'}={\sigma_{JJ'}-\sigma_{J'J}\over
 \sigma_{JJ'}+\sigma_{J'J} } $,  lie at $10^{-3} \sin \psi $ for leptons and
 d-quarks and reach  $ \sin \psi $  order of magnitude for 
 reactions (such as $t \bar  c +\bar t c)$  involving one top-quark
in the final state. 
 
\subsection{Single production}

\subsubsection{Resonant production at LEP}

$\bullet$ {\bf Bhabha scattering.}
The only single resonant production which is allowed at leptonic colliders
is the sneutrino production, via $\l_{ijk}$ couplings. This was first considered
in \cite{dimohall}, as a contribution to Bhabha scattering:
 $e^+e^- \to \tilde \nu \to e^+e^-$. The characteristic quantity describing
 the $\tilde \nu-Z^0$ interferences is: ${e^+e^- \ event \ rate \ at 
\ \tilde \nu \ peak
\over e^+e^- \ event \ rate \ at \ Z^0 \ peak} \approx 100 ({100GeV \over 
m_{\tilde \nu}})({250MeV \over \D E})({\l \over 0.2})^2$ where $\D E$ is the 
beam spread. If $\tilde \nu \to \nu \tchi^0$ dominates over
$\tilde \nu \to e^+e^-$, all is not lost since this would give new 
signals associated with the \rpv decay of $\tchi^0$.
Cross sections for reactions $e^+e^- \to e^+e^-$ and 
$e^+e^- \to \tchi^0 \nu, \ \tchi^{\pm} l^{\mp}$ via a resonant sneutrino 
have been computed in \cite{bargerg}. Bounds have been deduced on \rpv \ccs
 by comparing with experimental 
results, from TRISTAN data, on Bhabha scattering and events with 
two or more charged leptons plus missing energy.
Motivated by the interpretation of the very small x and high $Q^2$ events 
reported at HERA \cite{Coll}, based on charm squark production with a 
squark mass of order $m_{\tilde c} \simeq 200 GeV$ \cite{highq2},  
J. Kalinowski et al., \cite{Kal1} have considered the corrections to Bhabha
scattering for LEP II energies. Using
the indirect known bounds on the products $\l \l'$, they argue that if the HERA data
are interpreted as charm squark production (i.e. $\l_{121}'>0.05$), then 
$\l_{131}$ and $\l_{123}$ are weakly constrained. At  $\sqrt s=192GeV$, the 
relative correction effect from the sneutrino exchange,
${\sigma(SM + \tilde \nu) \over \sigma (SM)}-1$, lies between $3.10^{-1}$ and 
$4.10^{-3}$, for $200GeV<m_{\tilde \nu}<500GeV$, using $\l_{131}=0.1$.
 For the sneutrino $\tilde \nu_{\tau}$ resonance, cross sections
 values reach 300pb for $\sqrt s = m_{\tilde \nu_{\tau}}= 200GeV$ if $\l_{131}=0.1$.
With the same \cc $\l_{131}$, sneutrino exchange in 
t-channel could also contribute to $e^+e^- \to \tau^+ \tau^-$. The 
effect lies between $6.5 \ 10^{-3}$ and $1.5 \ 10^{-4}$ 
for the same choice of parameters.
A dominant $\l_{123}$ would affect $\mu^+ \mu^-$ and $\tau^+\tau^-$
pair production woul.
\\ $\bullet$ {\bf $b \bar b$ production.}\cite{Feng} 
The sneutrino $ \tilde  \nu_{\tau} $ exchange contribution to the process 
$ e^+e^- \to  \tilde  \nu_{\tau} \to b \bar b$ is especially promising 
because the Yukawa 
renormalisation of the scalar particle spectrum typically 
gives the third generation scalar 
field lighter than the first two. The authors have concentrated on 
$b \bar b$ production since this one has a factor $\l_{131}^2 \l_{333}'^2$,
giving sneutrino width, $\Gamma_{\tilde \nu_{\tau}} \approx 6GeV \l_{333}'^2 
({m_{\tilde \nu_{\tau}} \over 100GeV})$.
By calculating the required luminosity to get a $5 \sigma \ b
\bar b$ excess from sneutrino resonance with $\sqrt s \approx m_{\tilde \nu}
\approx 190GeV$, they concluded that values of the product $\l_{131}
\l_{333}'$ more than two order magnitude below actual bounds ($\l_{131}
\l_{333}' < 0.075({m_{\tilde \tau_L}\over 100GeV})^2$ from $B \to e \bar \nu$)
could be probed by the LEP experiments. In case where the sneutrino
peak is near the Z peak, sneutrino resonance could be still 
observable since its increases the
branching
ratio $R_b=B(Z \to b \bar b)$ and reduces the b quark forward-backward asymmetry
 $A_{FB}(b)$.
\\ $\bullet$  {\bf Experimental searches.}
A recent study by the DELPHI Collaboration \cite{Delphi} 
has analysed sample of events at $\sqrt s= 161$ and $172GeV$.  
 They account for sneutrino resonant production, followed by the \rpv decays,
$\tilde \nu \to b \bar b$ and $\tilde \nu \to \tchi^0 \nu \to e^+ e^- \nu \nu$.
Using the same cuts for the data and the simulated background and signal, 
they find bounds on the \cc $\l$ between $0.002$ and $0.04$, and 
on $\l'$ between $0.003$ and $0.014$, for $100GeV<m_{\tilde \nu}<200GeV$.

\subsubsection{Resonant production at Tevatron and LHC}

The first systematic study of final states for the \he hadron colliders 
was given in \cite{dreiner1}. H. Dreiner and G. G. Ross described all the different 
final state signatures, taking into account \rpv \ccs
in decays and both \rpv (single or resonant) and RPC (pair) supersymmetric
production mechanisms. Furthermore, analytic expressions of rates were given for 
each superpartner decay. The encouraging conclusion was that in all cases,
\rp violation leads to new visible signals
for physics at LHC or Tevatron. This is due largely to an important r\^{o}le
played by the RPC cascade decays into LSP.
 S. Dimopoulos et al., have presented in \cite{Esmail} 
cross sections for all resonant superparticle production at the Fermilab
Tevatron: $p \bar p \to \tilde l$ and $p \bar p \to \tilde \nu$ via 
$\l_{ijk}$ interactions or $p \bar p \to \tilde q$ via $\l_{ijk}''$ 
interactions.
For $ \sqrt s <2TeV$, the rates range between $10^{-1}nb $ and $ 10^{-4}nb$
in the interval $20GeV< \tilde m <250GeV$, if $\l=\l''=1$
If the produced sleptons decay to leptons via $\l$ couplings, a large 
 range of sleptons masses
and $\l$ \ccs can be explored. The slepton decays to pairs of jets via 
$\l'$ \ccs are not favorable because the QCD background is important.
 Cross sections for the single production reactions: $p \bar p \to \nu
\tilde \gamma, \ l \tilde \gamma, \ q \tilde g$ range between 
$10^{-5}$ and $10^{-1}nb$ for the same choice of parameters as above. Some 
of the final states have small background, as for exemple in the case where
 the photinos decays via $\l'$ into a lepton and two jets.
\\ $\bullet$ {\bf Single top quark production.}
The process $p \bar p \to q \bar q' \to t \bar b$ at the Tevatron
induced by couplings $\l''$ (via the exchange of a squark 
in the t-channel) and by couplings  $\l$ and $\l'$ (via the
exchange of a slepton in the s-channel) has been studied in \cite{Zhang1}.
It was found that the upgraded Tevatron can probe  efficiently the $\l''$ 
couplings, but less so  the $\l'$ couplings. In \cite{Zhang2}, the single top 
quark production via the processes, 
$q \bar q' \to slepton \to t \bar b$ and $q q' \to squark 
\to t b $ at Tevatron and LHC, respectively, were investigated. R. J. Oakes et al.,
 found that given the 
existing bounds on \rpv coupling constants, single top quark production by \rpv
may be greatly enhanced over the RPC contribution, and 
that both colliders can set strong
constraints on the relevant \rpv coupling constants. They further found that 
the LHC is more powerful than the Tevatron in probing the squark couplings, 
but the two colliders have comparable sensitivity for the slepton couplings.
\\ $\bullet$ {\bf Sneutrino and slepton production.}
The $\tilde \nu$ and $\tilde l$ resonant production for $p \bar p$
collisions (via $\l'$) in combination with their decays to leptons (via $\l$),
was studied in \cite{Kal2}. 
Coupling constants product $\l_{131} \l_{311}'$ was chosen to produce 
$\tilde \tau \ or \ \tilde \nu_{\tau}$. The cross sections for: $p \bar p \to
\tilde \nu_{\tau} \to e^+e^-$ and $p \bar p \to \tilde \tau \to e^+e^-$ 
range between 0.015 and 0.8pb for the set of parameters: 
$\sqrt s = 1.8TeV$, $\l_{131} \l_{311}'=(0.05)^2$ and 
$\Gamma_{\tilde \nu_{\tau}}=\Gamma_{\tilde \tau}=1GeV$. 
A study of the di-electron invariant mass distribution for the process 
$p \bar p \to e^+e^-$ gives a constraint on the product 
$\l_{131} \l_{311}'$, assuming
the sneutrino contribution to be smaller than the experimental error of the data
 points.
For $m_{\tilde \nu}=200GeV$ and at $\sqrt s=1.8TeV$, the constraint
obtained from Tevatron data was: $(\l_{131} \l_{311}')^{1/2}<0.08 
\Gamma_{\tilde \nu_{\tau}}^{1/4}$ . A bound can also be deduced from the
contribution of the process $e^+e^- \to \tilde \nu_{\tau} \to p \bar p$ 
to the inclusive reaction $e^+e^- \to hadrons$. The constraint is 
$(\l_{131}  \l_{311}')^{1/2}<0.072(0.045)$ from LEP data, at 
$\sqrt s=184(192)GeV$ and for $m_{\tilde \nu_{\tau}}=200GeV$.

\subsection{Systematic study of single production}

\begin{figure}
\begin{center}
\leavevmode
\psfig{figure=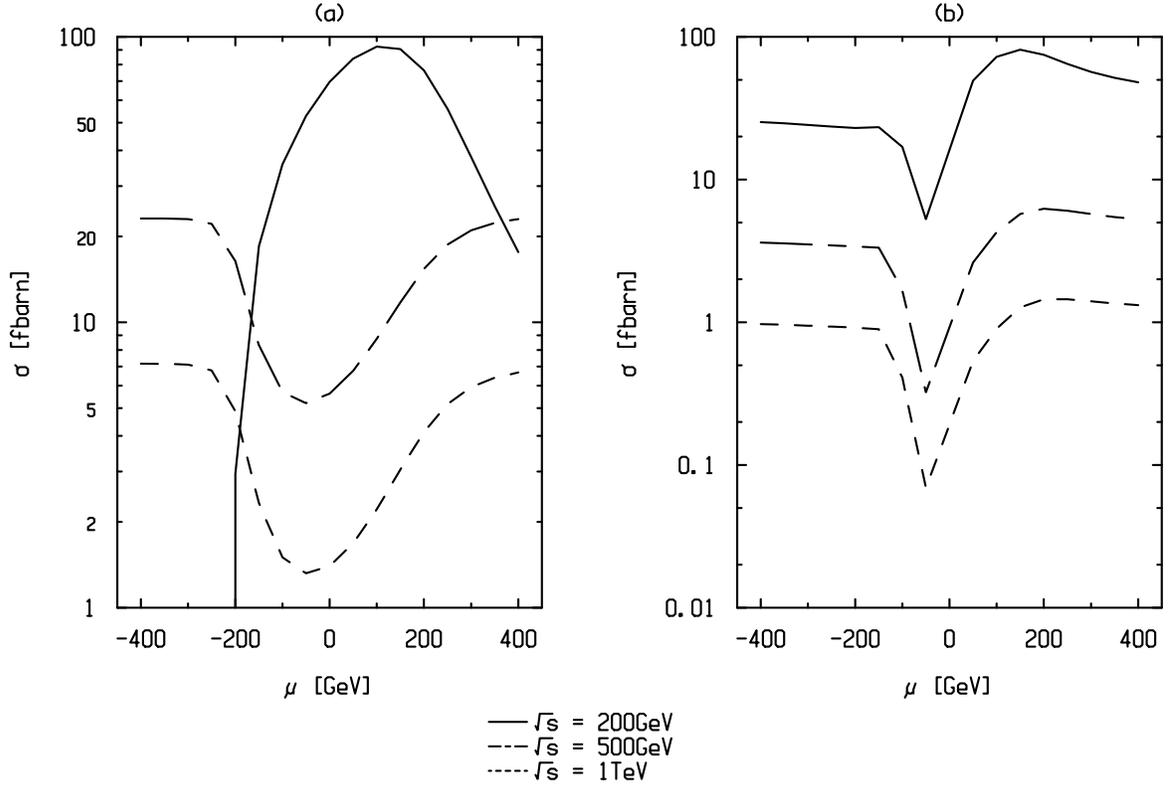}
\end{center}
\caption{\footnotesize  \it
The cross sections for the processes $e^+e^- \to \tchi^{\pm} l^{\mp}$ (a)
 and $e^+e^- \to \tchi^0 \nu$ (b), as a function of $\mu$, using the set of parameters:
 $M_2=200GeV$, $m_0=150GeV$, $\l_{m11}=0.05$ and $\tan \beta =2$. 
The different center of mass energies are indicated under the figures.
\rm \normalsize}
\label{f1}
\end{figure}

The studies of single resonant
production are restricted to the hypothetical
situation where the center of mass energy is chosen to be exactly the mass of 
a \susyq particle, which is not easy to achieve. The prospect in the distant future
of disposing of high precision measurements from high energy supercolliders (LHC, NLC)
makes it interesting to study single production for reactions such as 2 $\to$ 2 body
in a more systematic way. 
\\ $\bullet$ {\bf Lepton-photon collisions.}
B.C. Allanach et al., have examined in \cite{Alla} for LEP and NLC energies,
the processes: $e^{\pm} \gamma \to e^{\pm} \tilde \nu, \ \tilde e^{\pm} \nu$, 
where the photon is a tagged photon radiated by one of the colliding leptons.
These processes could test seven of the 9 $\l_{ijk}$ coupling constants. 
The cross section for $e^{\pm} \gamma \to \nu \tilde e^{\pm}$ is smaller than that for
$e^{\pm} \gamma\to \tilde \nu e^{\pm}$ because the t-channel exchange
amplitude involves a heavy slepton in the first reaction and a lepton in the second.
 For $\l=0.05$, the sneutrino production
cross section ranges between $30$ and $1000fb$ at $\sqrt s=192GeV$, and between 
$6$ and $1000 fb$ at $\sqrt s=500GeV$. A Monte Carlo analysis was performed to 
investigate the sensitivity 
to the sneutrino signal, and $5\sigma$ discovery contours in the $m_{\tilde \nu}$
versus $\l$ plane were presented. By comparing these contours with recent bounds 
on $\l$, B.C. Allanach et al., have concluded that sneutrinos with mass up 
to 170GeV could be discovered in the near future of LEPII.
\\ $\bullet$ {\bf Systematics of single production 
              at leptonic supercolliders}
Indicative results for the processes:
$e^+e^- \to \tchi^0 \nu, \ \tchi^{\pm} l^{\mp}$ are given in \cite{DESY}.
A systematic study of all the single production in $e^+e^- \to$ two-body reactions 
at leptonic colliders, $e^+e^- \to \tchi^0 \nu, 
\ \tchi^{\pm} l^{\mp}, \ \tilde l^{\pm} W^{\mp}, 
\ \tilde \nu Z^0,$ and $ \tilde \nu \gamma$ is performed in \cite{art2}.
A supergravity model is employed and a wide range of the parameter space is covered. 
As an illustration, we present in figure \ref{f1} some representative results.
For the chosen  values of $M_2$ and $\tan \beta$, the pair production of charginos 
and neutralinos is kinematically forbidden at LEP II, for $\vert \mu \vert > 100GeV$.
Branching ratios for the \susyq particles decays are calculated, assuming one
dominant  $\l_{ijk}$ coupling constant.

\newpage

\section{On the discovery potentiel of HERA for $R$-parity violating SUSY}

\subsection{Introduction}
\label{sec:intro}

The search for squarks, the scalar supersymmetric (SUSY) partners of
the quarks, is especially promising at the $ep$ collider HERA if they
possess a lepton number violating Yukawa coupling $\lambda'$ to
lepton--quark pairs.
Such squarks, present in the $R$-parity violating (\Rp) SUSY extension
of the Standard Model (SM), can be singly produced via the coupling
$\lambda'$ as $s$-channel resonances.
Masses up to the kinematic limit of $\sqrt{s} \simeq 300 \GeV$ are
accessible by the fusion of the $27.5 \GeV$ initial state positron
with a quark from the $820 \GeV$ incoming proton.

The interest in such new scalar bosons has been considerably renewed
recently following the observation by the H1~\cite{H1HIQ2} and 
ZEUS~\cite{ZEUSHIQ2} experiments of an excess of events at very high 
masses and $Q^2$, above expectations from Standard Model (SM) neutral 
current (NC) and charged current (CC) deep-inelastic scattering (DIS).
These early results were based on data samples collected in
1994 to 1996. 
Of particular interest was the apparent ``clustering'' of outstanding 
events at masses around $200 \GeV$ observed in H1 which, although not 
specifically supported by ZEUS observations~\cite{DREES,BERNARDI},  
have motivated considerable work on leptoquarks~\cite{HIGHXYLQ} and
$\Rp$ squarks~\cite{HIGHXYSQ} constraints and phenomenology.

In this report, the search at HERA for squarks through single production 
via a \Rp\ coupling, considering both \Rp\ decays and decays
via gauge couplings involving mixed states of gauginos and higgsinos 
is investigated and the discovery potential of HERA beyond other
existing colliders and indirect contraints from low energy processes
is established.

\subsection{Phenomenology}
\label{sec:pheno}

The general SUSY superpotential allows for gauge invariant terms with
Yukawa couplings between the scalar squarks ($\tilde{q}$) or sleptons
($\tilde{l}$) and the known SM fermions.
Such couplings violate the conservation of $R$-parity 
$R_p\,=\,(-1)^{3B+L+2S}$ where $S$ denotes the spin, $B$ the baryon 
number and $L$ the lepton number of the particles.
To minimize the number of free parameters (couplings) in the theory,
an exact conservation of $R_p$ has traditionally been assumed in the
context of the so-called Minimal Supersymmetric Standard Model (MSSM).
But provided that (e.g.) baryon number is exactly conserved, sizeable
lepton number violating Yukawa couplings are possible. 
%
%
%
The most general case, allowing for all such possible couplings,
would lead to a complicated phenomenology. There are however 
theoretical motivations for a strong hierearchy of the 
$\Rp$ couplings~\cite{BLR}, \cite{BDLS}, \cite{ELR}, 
analogous to that observed for the
known fermion masses, which simplify a lot the phenomenological implications
of the existence of such couplings. \\
%
Non vanishing \Rp\ couplings would have dramatic 
consequences in cosmology such as
the instability of the lightest SUSY particle which otherwise could 
contribute to the dark matter in the universe, and a possibly important
role in some baryogenesis models~\cite{dreiross} and~\cite{BARYOGEN}. 
But the most important consequence is that the discovery of SUSY matter
itself might be made through a sizeable $\Rp$ coupling.
This has to do with the trivial fact that sparticles can be singly 
produced in the presence of $\Rp$ couplings and this might provide 
the crucial mass reach increase for collider facilities.
The extension of the SUSY discovery potential in the presence of $\Rp$ 
couplings is particularly manifest at HERA. There, the search for R$_p$ 
conserving MSSM through slepton-squark associated production via 
$t$-channel gaugino exchange only marginally probes the parameter space 
beyond existing LEP collider constraints~\cite{MSSMHERA}. 

\begin{figure}[t]
\vspace{-0.3cm}

  \begin{center}
    \begin{tabular}{p{0.40\textwidth}p{0.60\textwidth}}
      \vspace{-4.0cm}
      \caption[]{ \label{fig:xsect}
                Squark production cross-sections in $ep$ collisions for
                a coupling $\lambda'=0.1$. } &
      \mbox{\epsfxsize=0.5\textwidth \epsffile{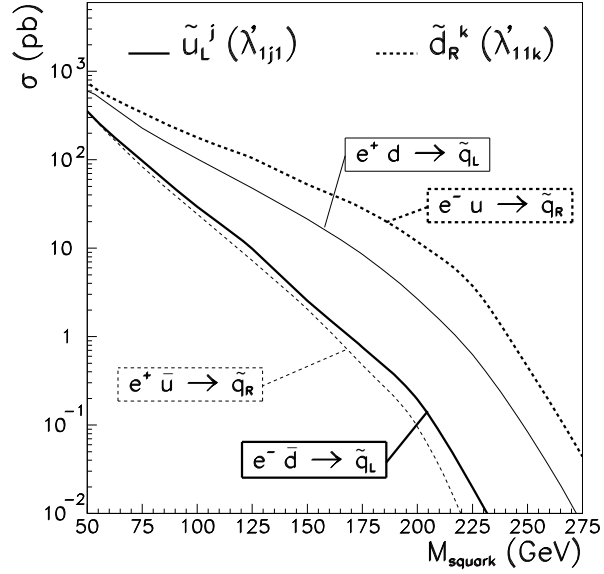}}
    \end{tabular}
  \end{center}
\vspace{-0.5cm}
\end{figure}

Of particular interest for HERA are the \Rp\ terms
$\lambda'_{ijk} L_{i}Q_{j}\bar{D}_k$ of the superpotential
which allow for lepton number violating processes.
By convention the $ijk$ indices correspond to the generations of
the superfields $L_{i}$, $Q_{j}$ and $\bar{D}_k$
containing respectively the left-handed lepton doublet, quark doublet
and the right handed quark singlet.
Expanded in terms of matter fields, the interaction Lagrangian
reads~\cite{RPVIOLATION} :
\begin{eqnarray}
{\cal{L}}_{L_{i}Q_{j}\bar{D_{k}}} &=
   & \lambda^{\prime}_{ijk}
              \left[ -\tilde{e}_{L}^{i} u^j_L \bar{d}_R^k
              - e^i_L \tilde{u}^j_L \bar{d}^k_R - (\bar{e}_L^i)^c u^j_L
     \tilde{d}^{k*}_R \right.           \nonumber \\
 \mbox{} &\mbox{}
 & \left. + \tilde{\nu}^i_L d^j_L \bar{d}^k_R + \nu_L \tilde{d}^j_L
    \bar{d}^k_R + (\bar{\nu}^i_L)^c d^j_L \tilde{d}^{k*}_R \right]
   +\mbox{h.c.}             \nonumber
 \nonumber
\end{eqnarray}
where the superscripts $^c$ denote the charge conjugate spinors
and the $^*$ the complex conjugate of scalar fields.
For the scalars the `R' and `L' indices distinguish independent
fields describing superpartners of right- and left-handed fermions.
Hence, with an $e^+$ in the initial state, the couplings
$\lambda'_{1jk}$ allow for resonant production of squarks through
positron-quark fusion.
The list of possible single production processes is given in
table~\ref{tab:sqprod}.
%
%
\begin{table*}[htb]
  \renewcommand{\doublerulesep}{0.4pt}
  \renewcommand{\arraystretch}{1.2}
 \begin{center}
 \begin{tabular}{p{0.40\textwidth}p{0.60\textwidth}}
         \caption
         {\small \label{tab:sqprod}
         Squark production processes at HERA ($e^+$ beam)
         via a $R$-parity violating
         $\lambda'_{1jk}$ coupling.} &
   \begin{tabular}{||c||c|c||}
   \hline \hline
   $\lambda'_{1jk}$ & \multicolumn{2}{c||}{production process} \\
   \hline
   111 & $e^+ +\bar{u} \rightarrow \bar{\tilde{d}_R}$
       &$e^+ +d \rightarrow \tilde{u}_L $\\
   112 & $e^+ +\bar{u} \rightarrow \bar{\tilde{s}_R}$
       &$e^+ +s \rightarrow \tilde{u}_L $\\
   113 & $e^+ +\bar{u} \rightarrow \bar{\tilde{b}_R}$
       &$e^+ +b \rightarrow \tilde{u}_L $\\
   121 & $e^+ +\bar{c} \rightarrow \bar{\tilde{d}_R}$
       &$e^+ +d \rightarrow \tilde{c}_L $\\
   122 & $e^+ +\bar{c} \rightarrow \bar{\tilde{s}_R}$
       &$e^+ +s \rightarrow \tilde{c}_L $\\
   123 & $e^+ +\bar{c} \rightarrow \bar{\tilde{b}_R}$
       &$e^+ +b \rightarrow \tilde{c}_L $\\
   131 & $e^+ +\bar{t} \rightarrow \bar{\tilde{d}_R}$
       &$e^+ +d \rightarrow \tilde{t}_L $\\
   132 & $e^+ +\bar{t} \rightarrow \bar{\tilde{s}_R}$
       &$e^+ +s \rightarrow \tilde{t}_L $\\
   133 & $e^+ +\bar{t} \rightarrow \bar{\tilde{b}_R}$
       &$e^+ +b \rightarrow \tilde{t}_L $\\
   \hline \hline
  \end{tabular}
  \end{tabular}
\end{center}
\end{table*}
With an $e^-$ beam, the corresponding charge conjugate processes are
$e^- u_j \rightarrow \tilde{d}_R^k$ 
($e^- \bar{d}_k \rightarrow \bar{\tilde{u}}_L^j$) for $u$-like 
($d$-like) quarks of the j$th$ ($k$th) generation.
Squark production via $\lambda'_{1j1}$ is especially interesting in
$e^+p$ collisions as it involves a valence $d$ quark,
whilst $\lambda'_{11k}$ are best probed with an $e^-$ beam since
squark production then involves a valence $u$ quark.
This is seen in Fig.~\ref{fig:xsect} which shows the production 
cross-sections $\sigma_{\tilde{q}}$ for ``up''-like squarks 
$\tilde{u}^j_L$ via $\lambda'_{1j1}$, and for ``down''-like 
squarks ${\tilde{d}}^{k*}_R$ via $\lambda'_{11k}$,
each plotted for coupling values of $\lambda'=0.1$.
In the narrow width approximation, these are simply
expressed as
\begin{equation}
   \sigma_{\tilde{q}} = \frac{\pi}{4 s}
                        \lambda'^2 q'(\frac{M^2}{ s })
\end{equation}
where $\sqrt{s} = \sqrt{ 4 E^0_e E^0_p } \simeq 300 \GeV$ is the
energy available in the CM frame for incident beam energies of
$E^0_e = 27.5 \GeV$ and $E^0_p = 820 \GeV$, and $q'(x)$ is the
probability to find the relevant quark (e.g. the $d$ for $\tilde{u}_L$
and the $\bar{u}$ for $\bar{\tilde{d}}_R$) with momentum fraction
$ x = M^2 / s \simeq M^2_{\tilde{q}} / s$ in the proton.
Hence the production cross-section approximately scales in $\lambda'^2$.

%
%
%
The squark search at HERA reported here has been carried with the
simplifying assumptions that:
 
 \begin{itemize}
 \item the lightest supersymmetric particle (LSP) is the lightest
       neutralino;
 \item gluinos are heavier than the squarks such that decays
       $\tilde{q} \rightarrow q + \tilde{g}$
       are kinematically forbidden.
 \item either one of the $\lambda'_{1jk}$ dominates, or 
       one product of couplings $\lambda'_{1jk} \times \lambda'_{3jl}$
       is non vanishing. This latter possibility, leading to lepton
       flavor violation processes will be adressed independently. 
\end{itemize}
%
The squarks decay either via their Yukawa coupling into fermions,
or via their gauge couplings into a quark and either a neutralino
$\chi_i^0$ ($i=1,4$) or a chargino $\chi_j^{+}$ ($j=1,2$).
The mass eigenstates $\chi_i^0$ and $\chi_j^{+}$ are mixed
states of gauginos and higgsinos and are in general unstable.
In contrast to the MSSM, this also holds in \Rp\ SUSY for the lightest
supersymmetric particle (LSP) which decays via $\lambda'_{1jk}$
into a quark, an antiquark and a lepton~\cite{RPVIOLATION}.
 
\begin{figure}[htb]
  \vspace{-1.5cm}
  
  \begin{center}
     \mbox{\epsfxsize=0.9\textwidth \epsffile{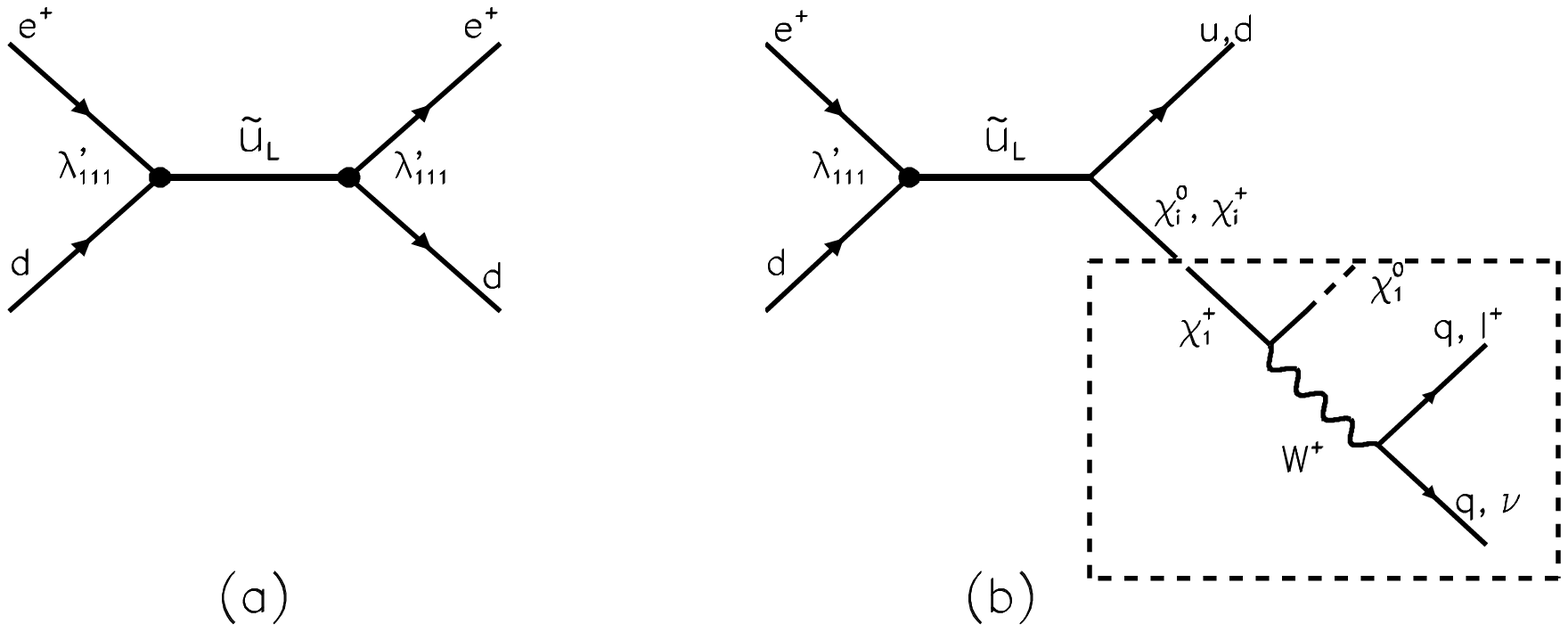}}
 
     \vspace{-2.1cm}
 
     \mbox{\epsfxsize=0.9\textwidth \epsffile{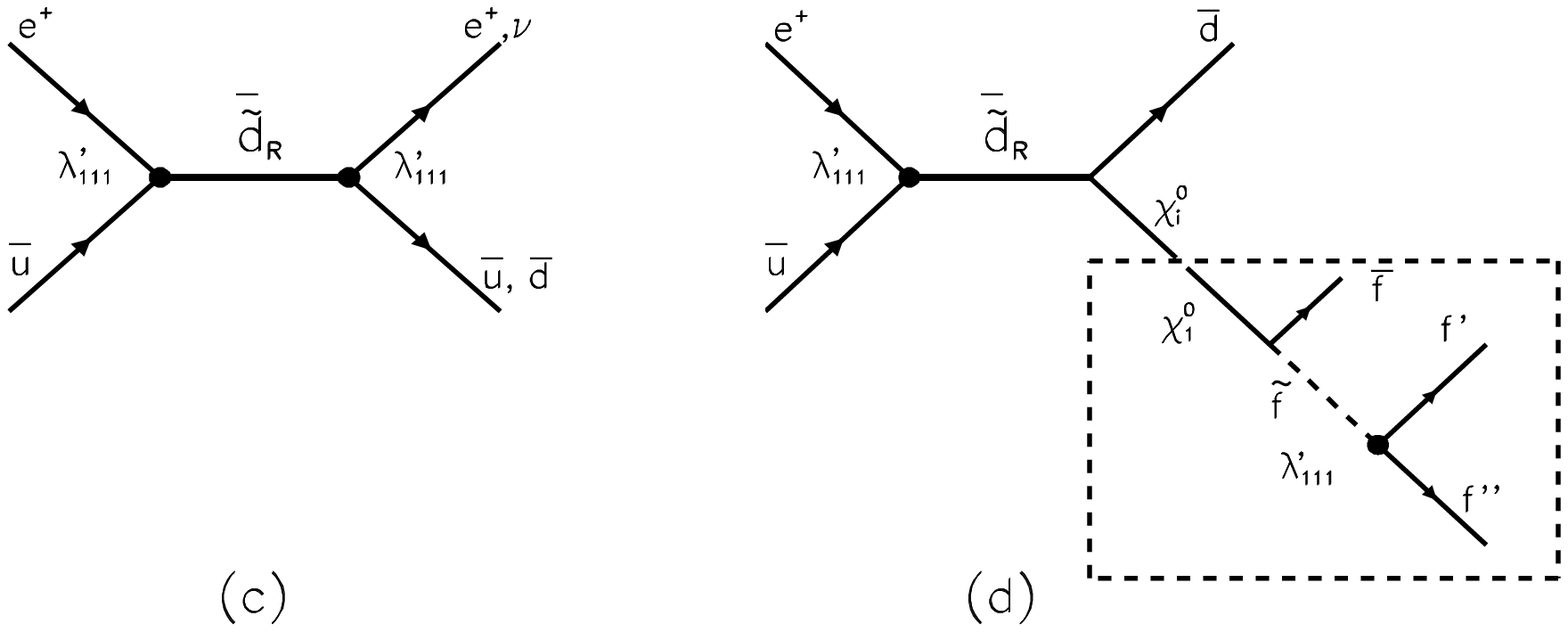}}
  \end{center}
 
\vspace{-1.5cm}

 \caption[]{ \label{fig:sqdiag}
    {\small Lowest order $s$-channel diagrams for first generation
      squark production at HERA followed by
      (a),(c) \Rp\ decays and (b),(d) gauge decays.
      In (b) and (d), the emerging neutralino or chargino might
      subsequently undergo \Rp\ decays of which examples are
      shown in the doted boxes for (b) the $\chi_1^{+}$ and
      (d) the $\chi_1^0$. }}
\end{figure}

Typical diagrams for the production of first generation squarks are
shown in Fig.~\ref{fig:sqdiag}.
%
By gauge symmetry only the $\bar{\tilde{d}}_R$ and $\tilde{u}_L$ are
produced via the $\lambda'$ couplings.
These have in general widely different allowed or dominant decay modes.
 
 
In cases where both production and decay occur through a
$\lambda'_{1jk}$ coupling (e.g. Fig.~\ref{fig:sqdiag}a and c for
$\lambda'_{111} \ne 0$), the squarks behave as scalar
leptoquarks~\cite{H1LQ95,BUCHMULL}.
For $\lambda'_{111} \ne 0$, the $\bar{\tilde{d}}_R$ resemble the
$\bar{S^0}$ leptoquark and decays in either $e^+ + \bar{u}$ or
$\nu_e + \bar{d}$  while the $\tilde{u}_L$ resemble the
$\bar{\tilde{S}}_{1/2}$ and only decays into $e^+ d$.
Hence, the final state signatures consist of a lepton and a jet and
are, event-by-event, indistinguishable from the SM neutral
and charged current DIS.
The strategy is then to look for resonances in DIS--like events
at high mass, exploiting the characteristic angular distribution
of the decay products expected for a scalar particle.
 
 
In cases where the squark decay occurs through gauge couplings
(e.g. Fig.~\ref{fig:sqdiag}b and d), one has to consider for the
$\tilde{u}_L$ the processes $\tilde{u}_L \rightarrow u \chi_i^0$ or
$d \chi_j^+$ while for the ${\tilde{d}}^*_R$
only $\bar{\tilde{d}}_R \rightarrow \bar{d} \chi_i^0$ is allowed.
This is because the $SU(2)_L$ symmetry which implies in the SM
that the right handed fermions do not couple to the $W$ boson
also forbids a coupling of $\bar{\tilde{d}}_R$ to the $\tilde{W}$.
Hence, the $\bar{\tilde{d}}_R$ can only weakly couple (in proportion
to the $d$ quark mass) to the $\chi_j^+$ through its higgsino component.
 
 
The possible decay modes of the chargino, when it is the lightest
chargino $\chi_1^+$, are the gauge decays
$\chi_1^+ \rightarrow \chi_1^0 l^+ \nu$ and
$\chi_1^+ \rightarrow \chi_1^0 q \bar{q}'$,
and the \Rp\ decays $\chi_1^+ \rightarrow \nu u \bar{d}$ and
$\chi_1^+ \rightarrow e^+ d \bar{d}$.
The fate of the $\chi_1^0$ depends on its gaugino-higgsino
composition.
The question of how this $\chi_1^0$ nature depends on free fundamental
parameters of the MSSM, as well as the corresponding $\tilde{q}$
branching fractions for various possible decay channels will be
discussed briefly in relation to our analysis in
section~\ref{sec:results} and was studied in more detail
in~\cite{H1RPV96,DRPEREZ,DARKMATTER}.
In general, the $\chi_1^0$ will undergo the decay
$\chi_1^0 \rightarrow e^{\pm} q \bar{q}'$ or
$\chi_1^0 \rightarrow \nu q \bar{q}$.
The former will be dominant if the $\chi_1^0$ is photino-like
(i.e. dominated by photino components) in which case both the
``right'' and the ``wrong'' sign lepton (compared to incident
beam) are equally probable leading to largely background free
striking signatures for lepton number violation.
The latter will dominate if the $\chi_1^0$ is zino-like.
A higgsino-like $\chi_1^0$ could
be long lived and escape detection since its coupling to
fermion-sfermion pairs (e.g. Fig.~\ref{fig:sqdiag}d) is proportional
to the fermion mass~\cite{GUNION}.
Hence processes involving a $\tilde{H}$-like $\chi_1^0$ can be
affected by an imbalance in transverse momenta.
 
 
Taking into account the dependence on the nature of the $\chi_1^0$,
the possible decay chains of the $\tilde{u}_L$ and $\bar{\tilde{d}}_R$
squarks has been classified in~\cite{H1RPV96} in eight 
distinct event topologies among which we shall mostly concentrate
here on the first four, namely: 
\begin{itemize}
 \item {\cal {S1}}, high $P_T$ $e^+$ + 1 jet, 
            e.g. $\tilde{q} \stackrel{\lambda'}{\longrightarrow} e^+ q'$; 
 \item {\cal {S2}}, high $P_{T,miss}$ + 1 jet, e.g.
  $\bar{\tilde{d}}_R \stackrel{\lambda'}{\longrightarrow} \nu_e \bar{d}$; 
 \item {\cal {S3}}, high $P_T$ $e^+$ + multiple jets,
            e.g. $\tilde{q} \longrightarrow q \chi_1^0$
                 followed by
	         $\chi_1^0 \stackrel{\lambda'}{\longrightarrow} 
		                                       e^+ \bar{q}' q''$; 
 \item {\cal {S4}}, High $P_T$ $e^-$ (i.e. wrong sign) + multiple jets,
            e.g. $\tilde{u}_L \longrightarrow d \chi_1^+$
                 followed by
	         $\chi_1^+ \longrightarrow W^+ \chi_1^0$ and
    $\chi_1^0 \stackrel{\lambda'} {\longrightarrow}e^- \bar{q}' q''$; 
\end{itemize}

 
For a squark decaying into a quark and the lightest neutralino,
the partial width can be written as
$$ \Gamma_{\tilde{q}\rightarrow \chi_1^0 q}
 = \frac{1}{8\pi} \left( A^2 + B^2 \right) M_{\tilde{q}}
   \left(1-\frac{M^2_{\chi_1^0}}{M^2_{\tilde{q}}}\right)^2
\Longrightarrow
   \Gamma_{\tilde{q}\rightarrow \tilde{\gamma}q}
 = \Gamma_{\tilde{q}\rightarrow eq'} \; \frac{2e^2e^2_q}{\lambda'^2} \;
   \left(1-\frac{M^2_{\tilde{\gamma}}}{M^2_{\tilde{q}}}\right)^2  $$
where $A$ and $B$ in the left expression are chiral couplings
depending on the mixing matrix of the neutralinos.
Detailed expressions for such couplings can be found in~\cite{GUNION}.
Under the simplifying assumption that the neutralino is a pure
photino $\tilde{\gamma}$, this gauge decay width reduces to the
expression on the right. Here we introduced the partial width
$ \Gamma_{\tilde{q}\rightarrow eq'} = \lambda'^2 M_{\tilde{q}}/16 \pi$
for squarks undergoing \Rp\ decays.
It is seen that, in general, gauge decays contribute strongly at low
$\chi_1^0$ masses and small Yukawa couplings.
 
%
The case $\lambda'_{131} \ne 0 $ (or $\lambda'_{132} \ne 0$) is of special 
interest~\cite{KONRP} since it allows for direct production of the stop 
via $e^+ d \rightarrow \tilde{t}$ ($e^+ s \rightarrow \tilde{t}$).
The stop is particular in the sense that a ``light'' stop mass eigenstate 
($\tilde t_{1}$) could (depending upon the free parameters of the model)
exist much lighter
than the top quark itself and lighter than other squarks.
This applies only for the stop since the off-diagonal terms which
appear in the mass matrix associated to the superpartners
of chiral fermions are proportional to the partner fermion mass.
 
\subsection{Results from HERA}
\label{sec:results} 

The search for squarks or $R$-parity violating SUSY was originally
carried~\cite{H1RPV94,H1RPV96} by the H1 Collaboration at HERA for the first 
time combining $\Rp$ decays and gauge decays of the squarks. 
It has been very recently extended~\cite{H1ICHEP98} to include the 
1995 $\to$ 1997 datasets which represent an increase of integrated 
luminosity of more than an order of magnitude.
In view of the excess observed in particular by H1~\cite{H1HIQ2} for 
NC-like (i.e. $\Rp$-decay like) event topologies for mass 
$M \sim 200 \GeV$, it became particularly important to analyse the full 
available datasets in gauge-decay topologies.
No deviations from Standard Model expectation was found and these
channels were used by H1 in combination with the NC-like channel to 
derive exclusion domains.  

\begin{figure}[htb]
  \begin{center}
  \begin{tabular}{cc}
   \mbox{\epsfxsize=0.5\textwidth
 \epsffile{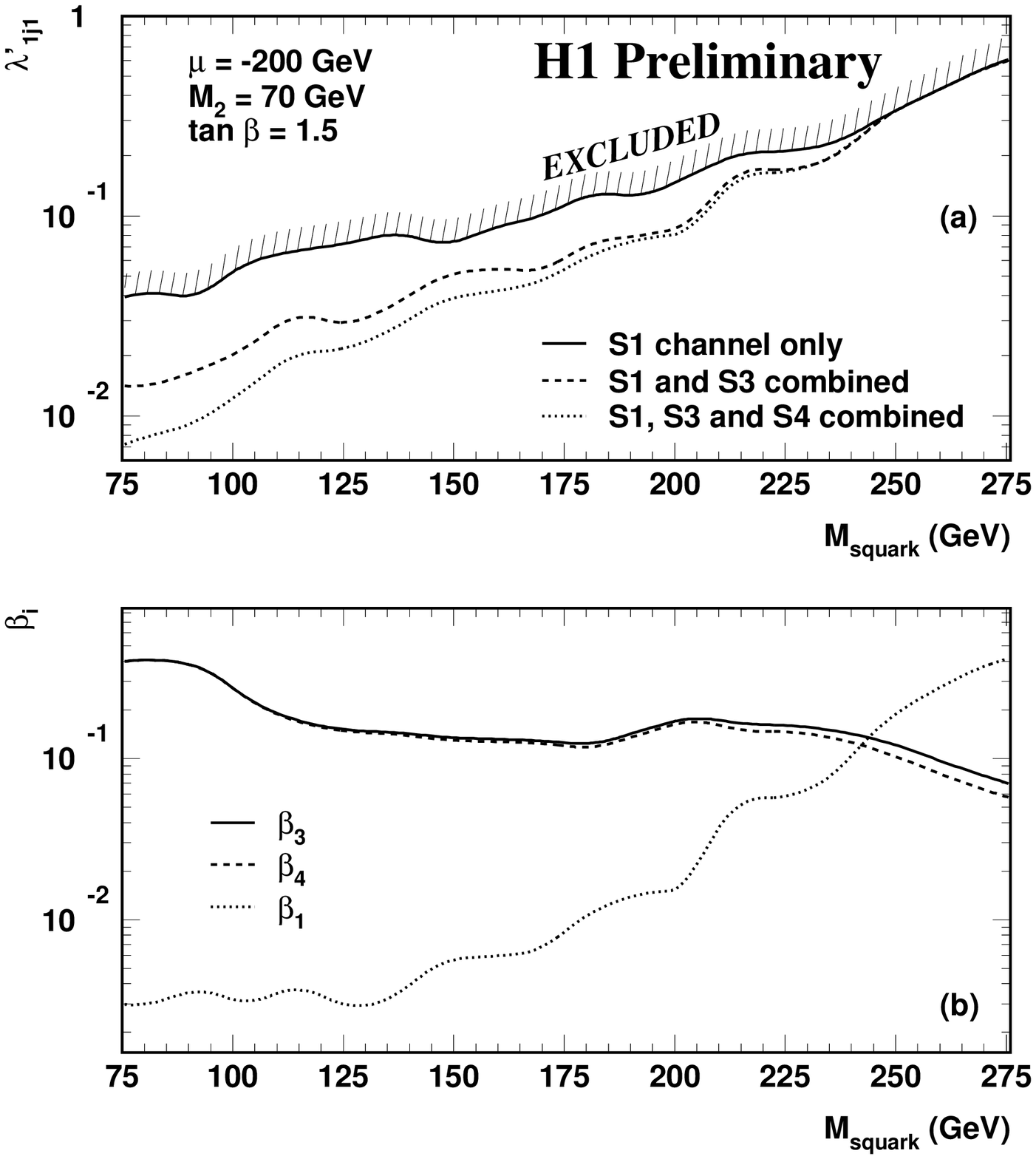}}
 &
   \mbox{\epsfxsize=0.5\textwidth
 \epsffile{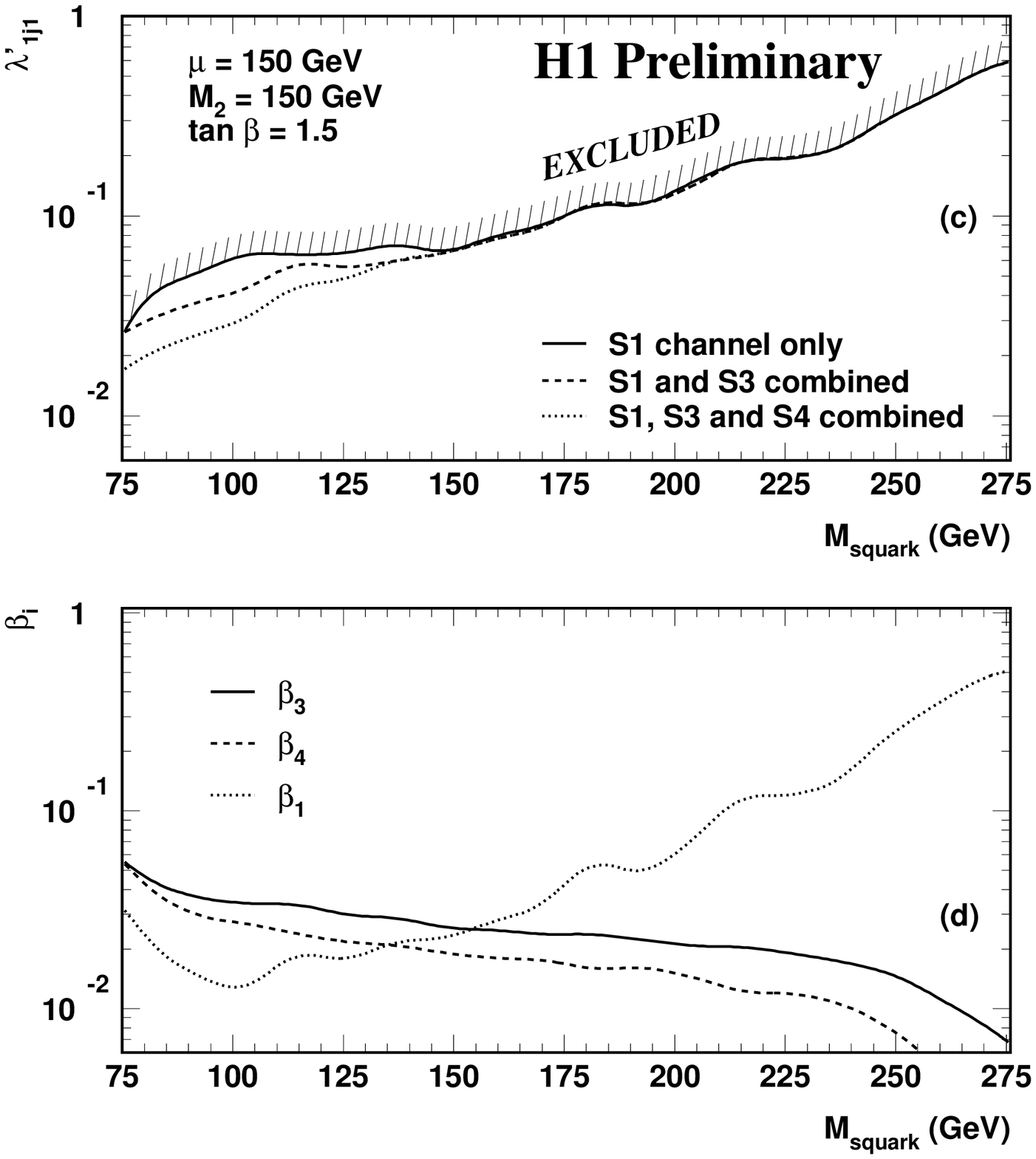}}
  \end{tabular}
  \end{center}
 \caption[]{ \label{fig:lim_combine}
 {\it  (a) Exclusion upper limits at $95 \%$ C.L. for the coupling
       $\lambda'_{1j1}$ as a function of the squark mass for a specific
       set of MSSM parameters ($M_{\chi^0_1} = 40 \GeV$, 
       $\chi^0_1 \sim \tilde{gamma}$, see text). Gauge and $\Rp$ decays
       of the squarks have been combined. 
       Regions above the curves are excluded;
       (b) The relative contributions of channels {\large{S1}},
       {\large{S3}} and {\large{S4}} versus the squark mass;
       (c) and (d) : as (a) and (b) but for a $40 \GeV$  
       $\chi^0_1$ dominated by its zino component. }}
\end{figure}

The rejection limit obtained obtained by H1 at $95 \%$ confidence level 
on $\lambda'_{1j1}$ as a function of the $\tilde{u}^j_L$ mass is shown 
in Fig.~\ref{fig:lim_combine}a for a specific choice of the MSSM 
parameters, $\mu = -200 \GeV$, $M_2 = 70 \GeV$ and $\tan \beta = 1.5$. 
These have been set such that the lightest neutralino $\chi^0_1$ 
is mainly dominated by its photino component and its mass is 
about $40 \GeV$. 
The $\chi^+_1$ and $\chi^0_2$ appear nearly degenerate around $90 \GeV$.
By combining three contributing channels {\cal {S1}}, {\cal {S3}} and
{\cal {S4}}, H1 improves the sensitivity for squarks considerably 
compared to an analysis which would rely solely on $\Rp$ two-body
decay of the squarks. 
For example, at masses $M \sim 100 \GeV$, an improvement of a factor 
$\simeq 5$ is obtained beyond an analysis relying on NC-like data  
(i.e. channel {\large{S1}}).
%
%
%
The branching ratios $\beta_1$, $\beta_3$ and $\beta_4$ in
channnels {\large{S1}}, {\large{S3}} and {\large{S4}} respectively
are shown on Fig.~\ref{fig:lim_combine}b
versus the squark mass, at the sensivity limit on the Yukawa coupling.
For masses up to $\simeq 230 \GeV$, channels {\large{S3}} and
{\large{S4}} dominate and contribute each at a similar level.
As soon as squark decays into $\chi^+_1$ and
$\chi^0_2$ become kinematically allowed,
$\beta_3$ and $\beta_4$ are hampered by the fact that the 
both $\chi^+_1$ and $\chi^0_2$ decay preferentially into $\nu q \bar{q}$
instead of $e^{\pm} q q'$ (because they are dominated
respectively by their wino and zino components~\cite{H1RPV96}).
In the very high mass domain, a large Yukawa coupling is necessary
to allow squark production, hence the relative contribution of
{\large{S1}} is largely enhanced.

In order to study the dependence of our rejection limits on the MSSM
parameters, another set of values for $(\mu, M_2, \tan \beta)$
is chosen, which leads to $40 \GeV$ $\chi^0_1$ dominated
by its zino component.
The masses of $\chi^+_1$ and $\chi^0_2$ are respectiveley
$\simeq 100 \GeV$ and $\simeq 72 \GeV$, and $\chi^0_2$ is mainly
a $\tilde{\gamma}$ state.
As before, $95 \%$ C.L. limits on $\lambda'_{1j1}$ versus the squark
mass are displayed in Fig.~\ref{fig:lim_combine}c.
The gain obtained by the combination of the three channels
is less substantial than in previous case.
Indeed, the $\chi^0_1$ being here dominated by its $\tilde{Z}$
component, it decays with a high branching ratio into 
$\nu q \bar{q}$ instead of $e^{\pm} q q'$.
The same holds for the lightest chargino.
Hence, total branchings $\beta_3$ and $\beta_4$ are quite small
though gauge decays of the squark are important.
On the contrary to the ``photino'' case, $\beta_3$ is here
substantially higher than $\beta_4$. This is mainly due
to the fact that the fraction of \Rp\ decays
into $\nu q \bar{q}$ is smaller for the $\chi^+_1$ than for
the $\chi^0_1$.
In fact, the dominant decay channel here would be the one labelled
{\large{S5}} in~\cite{H1RPV96}, leading to multijets +
$P_{T,miss}$ topology, which has not been adressed in
this analysis. 
The separation of the signal from the SM background
in this channel is however not easy, and, following the analysis
presented in~\cite{H1RPV96}, we expect that the limit obtained
when also combining {\large{S5}} will not improve too much
the result given here.
%
\begin{figure}[htb]
  \begin{center}
     \mbox{\epsfxsize=0.6\textwidth 
\epsffile{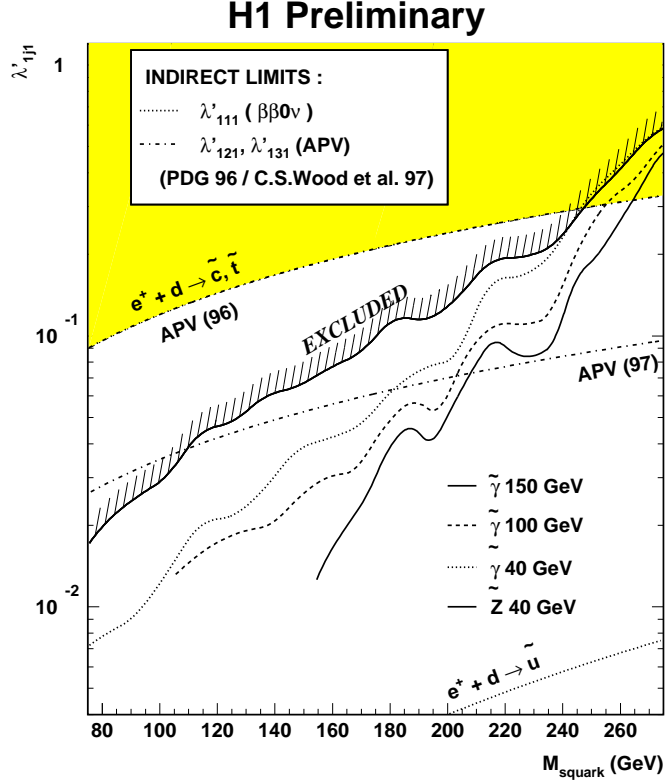}}

  \end{center}
 \caption[]{ \label{fig:lim_l1j1}
 {\it  Exclusion upper limits at $95 \%$ C.L. for the coupling
       $\lambda'_{1j1}$ as a function of squark mass, for 
       various masses of a $\tilde{\gamma}$-like $\chi^0_1$;
       the difference obtained between a $40 \GeV$ 
       $\tilde{\gamma}$-like and 
       $\tilde{Z}$-like $\chi^0_1$ is also shown; 
       also represented are the most stringent indirect
       limits on $\lambda'_{111}$ and $\lambda'_{1j1}$,
       $j=2,3$. }} 
\end{figure}
%
%
Limits obtained when combining {\large{S1}}, {\large{S3}} and
{\large{S4}} in the two cases detailed above are compared to each
other in Fig.~\ref{fig:lim_l1j1}.
Our sensitivity on $\lambda'_{1j1}$ is better by a factor
$\simeq 2$ for squark masses below $\simeq 200 \GeV$ 
for a $\tilde{\gamma}$-like $\chi^0_1$ than for a $\chi^0_1$
dominated by its zino component, due to the highest part of
total branching actually ``seen'' by our analysis.
One can infer from~\cite{H1RPV96}
that the two cases presented here are somewhat ``extreme'' and in
that sense quite representative of the sensitivity we can
achieve for any other choice of MSSM parameters leading
to a $40 \GeV$ $\chi^0_1$.
The same figure shows the $95 \%$ C.L. limits obtained
for a $100 \GeV$ and a $150 \GeV$ $\tilde{\gamma}$-like
$\chi^0_1$. 
For electromagnetic coupling strenghs 
$\lambda^{'2}_{1j1} / 4 \pi \simeq \alpha_{em}$, squark masses
up to $262 \GeV$ are excluded at $95 \%$ C.L. by this analysis,
and up to $175 \GeV$ for coupling strenghs $\gsim 0.01 \alpha_{em}$.
For low masses, these limits represent an improvement 
of a factor $\simeq 3$ compared to H1
previously published results. \\

\begin{figure}[htb]
  \begin{center}
     \mbox{\epsfxsize=0.6\textwidth 
\epsffile{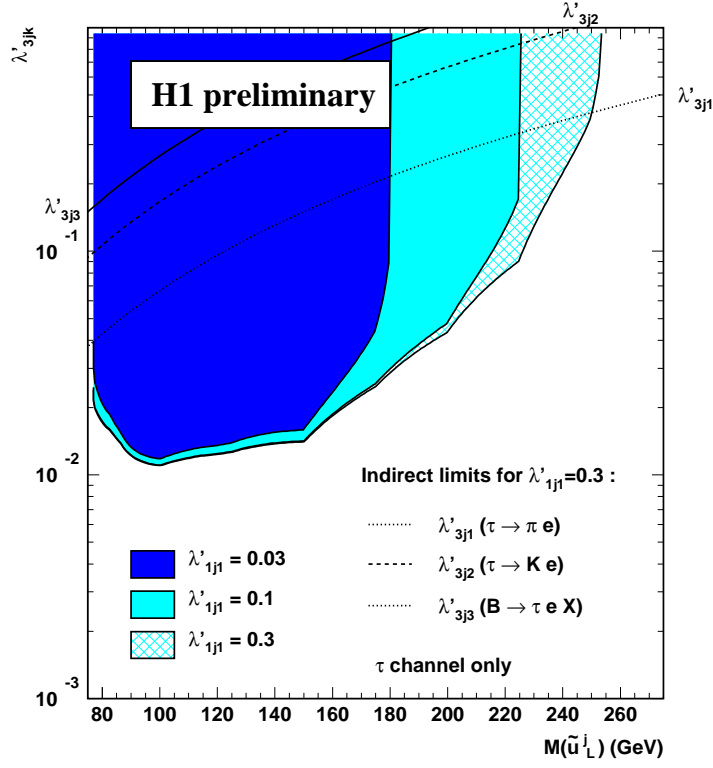}}
  \end{center}
 \caption[]{ \label{fig:lim_l3jk}
 {\it  Exclusion upper limits at $95 \%$ C.L. for the coupling
       $\lambda'_{3jk}$ as a function of squark mass, for 
       for several fixed values of $\lambda'_{1j1}$ (greyed
       domains). The regions above the full, dashed and dot-dashed
       curves correspond to the best relevant indirect
       limits. }}
\end{figure}

%
%
We now turn to the case where two couplings $\lambda'_{1j1}$ and
$\lambda'_{3jk}$ are non vanishing.
On the contrary to what has been done above, we assume
here that gauge decays of squarks are forbidden, so that the 
only squark decay modes are $\tilde{u}^j_L \rightarrow e^+ + d$
and $\tilde{u}^j_L \rightarrow \tau^+ + d^k$.
$\tau$ + jet final states have been searched for in the hadronic
decay mode of the $\tau$ and no signal has been observed.
Assuming a given value for the production coupling $\lambda'_{1j1}$
exclusion limits at $95 \%$ C.L. on $\lambda'_{3jk}$ have been derived
combining both $e$ + jet and $\tau$ + jet channels.
Results are shown in Fig.~\ref{fig:lim_l3jk} versus the
mass of $\tilde{u}^j_L$, for $\lambda'_{1j1}$ equal to 0.3
(i.e. an electromagnetic coupling strength), 0.1 and 0.03.
Greyed domains are excluded.
For $\lambda'_{1j1} = \lambda'_{3jk} = 0.03$, $\tilde{u}^j_L$
squarks lighter than $165 \GeV$ are excluded at $95 \%$ C.L.
%
%
A similar analysis has been carried out by ZEUS Collaboration,
with an integrated luminosity of $\simeq 3 \picob^{-1}$
using $e^+ p$ 1994 data~\cite{ZEUSLFV}. Instead of fixing $\lambda'_{1j1}$,
results were presented assuming $\lambda'_{1j1} = \lambda'_{3jk}$.
When these two couplings are both equal to 0.03, the analysis
presented here extends the excluded squark mass range by 
$\simeq 65 \GeV$. \\

\subsection{Constraints and Discovery Potential}
\label{sec:pheno2} 

Our results in the plane $\lambda'_{1j1}$ versus
$M_{\tilde{q}}$, under the hypothesis that one $\lambda'_{1j1}$ dominates,
are also compared in Fig.~\ref{fig:lim_l1j1}
to the best indirect limits.
The most stringent comes from the non-observation of neutrinoless
double beta decay, but only concerns $\lambda'_{111}$ coupling.
The most severe indirect limits on couplings $\lambda'_{121}$ and
$\lambda'_{131}$, which could allow for the production of
squarks $\tilde{c}$ and $\tilde{t}$ respectively,
come from Atomic Parity Violation. Two constraints are given on
the figure which depend on the experimental input.
H1 direct limits are thus better or comparable to the most 
stringent constraints on $\lambda'_{121}$, $\lambda'_{131}$.
For high masses of $\chi^0_1$ our limit even improves these
indirect contraints by a factor up to $\simeq 4$.

In the case where two couplings $\lambda'_{1j1}$ and
$\lambda'_{3jk}$ are non vanishing,
the only relevant indirect limits~\cite{DAVIDSON} come from the processes
$\tau \rightarrow \pi e$, $\tau \rightarrow K e$ and 
$B \rightarrow \tau e X$, which constrain the products
$\lambda'_{1j1} \times \lambda'_{3jk}$. 
These indirect limits are given in Fig.~\ref{fig:lim_l3jk} 
for $\lambda'_{1j1} = 0.3$.
H1 direct limits improve these contraints by typically one order
of magnitude.
Note that better indirect limits on couplings $\lambda'_{3jk}$
exist, coming e.g. from $\tau \rightarrow \pi \nu$, 
$Z \rightarrow \tau \tau$ or $K^+ \rightarrow \pi^+ \nu \bar{\nu}$.
However these only concern the $\tilde{d}^k_R$ and can thus be
evaded assuming e.g. $\tilde{u}^j_L$ to be much lighter 
than other squarks. \\

Contrary to leptoquarks~\cite{BUCHMULL}, it was seen above that 
the squarks accessible at HERA can naturally have a small branching 
ratio ${\cal{B}}$ in $\Rp$ decay modes and thus can partly avoid the severe
contraints set at the Tevatron for scalar 
leptoquarks~\cite{TEVATRONLQ}.
The difficulty in explaining an excess in $\Rp$-like decay channels
such as that observed in the 1994-96 data set of H1 precisely resides
in the necessity to maintain a sizeable product 
$\lambda'_{1jk} \sqrt{{\cal B}}$ while respecting the Tevatron
scalar leptoquark constraints which implies that ${\cal {B}}<0.5$
for $M \simeq 200 \GeV$ and at the same time the upper limits on 
$\lambda'_{1jk}$ coming from indirect processes.
For example, considering the full amplitude of the NC-like excess observed
in H1, the only viable scenarii are~\cite{HIGHXYSQ}:
\begin{center}
    \begin{tabular}{|c|c|c|}
      \hline
      Production process
       & estimate of $\lambda' \sqrt{{\cal B}}$
        & constraints on ${\cal B}$\\
      \hline
      $e^+_R d_R \rightarrow \tilde{c}_L$ 
       & $\lambda'_{121}\sqrt{{\cal B}}\sim 0.025 - 0.033$
        & $0.1 \lsim {\cal B} < 0.5$ \\
      \hline
      $e^+_R d_R \rightarrow \tilde{t}_L$
       & $\lambda'_{131}\sqrt{{\cal B}}\sim 0.025 - 0.033$
        & $0.1 \lsim {\cal B} < 0.5$ \\
      \hline
      $e^+_R s_R \rightarrow \tilde{t}_L$
       & $\lambda'_{132}\sqrt{{\cal B}} \sim 0.15 - 0.25$
        & $0.07 \lsim {\cal B} < 0.5$ \\
      \hline
    \end{tabular}
\end{center}
Such branching ratios can only be met in small regions of the SUSY
parameter space, regions which are moreover challenged by the search
recently carried at LEP for $\Rp$ decays of charginos which sets a
lower limit on $M_{\chi^{\pm}}$ of $\sim 90 \GeV$.

\newpage

\section{Do we need conserved $R$-parity at LEP?}

\label{chap6.1}
In $e^+e^-$ colliders such as LEP in its first phase i.e. running
at the $Z$ peak, the search for \rpv effects mainly concerned
leptonic topologies \cite{aleph1}.
LEP in its second phase, i.e. LEP 2, is going to higher
center of mass energies than the $Z$ peak and, 
along with a deeper and wider interest to supersymmetry
with \rpv couplings, extends the search
for \rpv effects \cite{aleph2}. 
Assuming that one coupling dominates at one time,
the effects of the \rpv terms
on the phenomenology can
be classified in three parts:
\begin{malist}
\item {\bf effects in the decay} of the supersymmetric particles
produced in pair (in the usual way) in  $e^+e^-$ interactions
in which the $\lambda_{ijk}$, $\lambda_{ijk}^{\prime}$ or 
$\lambda_{ijk}^{\prime \prime}$ couplings can be involved; 
\item {\bf single production} of a neutralino (with a neutrino), a chargino
(with a charged lepton) or a resonant sneutrino,
all involving $\lambda_{ijk}$ couplings and also 
single production of     
a squark in $\gamma e$ interactions that
can occur in $e^+e^-$ collisions via the interaction of a quark
from a resolved $\gamma$ radiated by one of the incoming particle 
($e^+$ or $e^-$) with the other incoming particle, involving 
$\lambda_{ijk}^{\prime}$ couplings;
\item  {\bf t-channel exchange} of a slepton via $\lambda_{ijk}$ couplings,
in the lepton-pair production $e^+e^- \rightarrow l^+l^-$ 
or t-channel exchange of a squark via $\lambda_{ijk}^{\prime}$ couplings,
in the 
quark-pair production 
$e^+e^- \rightarrow q \bar q$  both giving deviation
from the expectation
of the Standard Model.
\end{malist}
These effects have been experimentally searched for at the LEP collider.
In the following, a short description and
a very brief review is given on these experimental searches performed by
the LEP collaborations,
mainly based from  
already published papers and contributions submitted to
conferences \cite{aleph}. 
\subsection{Effects of the $R$-parity violating couplings in the decay}
\begin{figure}[h]
\begin{center}
\epsfig{file=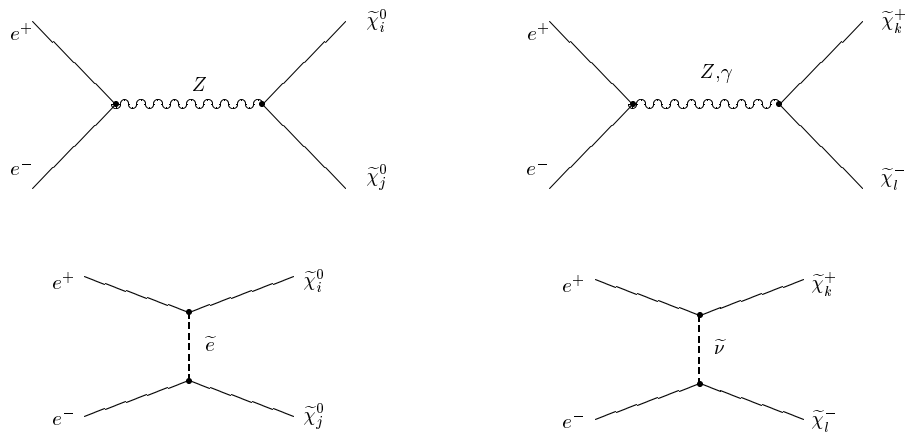,width=12cm}
\end{center}
\caption{Gaugino pair production diagrams ($i,j=1,..4;\ k,l=1,2$)}
\label{cb-prodiag}
\end{figure}

Neutralinos ${\tilde {\chi}}^{o}_{1,2,3,4}$
and charginos ${\tilde {\chi}}^{\pm}_{1,2}$ are gauginos that can be
produced in pair in $e^+e^-$ collisions via the ordinary couplings
from supersymmetry with conserved $R$-parity \cite{gdr1}. 
In the $s$-channel, the gauginos
are produced via the exchange 
of a $\gamma$ or a $\mathrm{Z}$ (figure~\ref{cb-prodiag}); 
in the $t$-channel, they can be produced via a selectron for the neutralinos,
or a sneutrino for the charginos, if the slepton masses
are low enough. When the selectron mass is
sufficiently small (\siminf 100~\GeVcc),
 the neutralino production is enhanced.
On the contrary, if the \snue\ mass is in the same range, 
the chargino cross section can decrease due to
destructive interference between the $s$- and $t$-channel amplitudes. 
If the dominant component of
neutralinos and charginos is the higgsino ($|\mu| \ll M_2$), the production
cross sections are large and insensitive to slepton masses.

Sfermions 
$\tilde f$ i.e. sleptons
and squarks, can also be produced in pair in $e^+e^-$ collisions via
the ordinary couplings
from supersymmetry with conserved $R$-parity provided that their masses are
kinematically accessible which is hoped to be the case for the sfermions of the third
familly, especially stop $\tilde t$ and sbottom $\tilde b$, because of the splitting
in mass of the mass-eigenstates.
\par
In the presence of \rpv terms 
in the superpotential,
the lightest neutralino ${\tilde {\chi}}^{o}_{1}$, usually considered as the LSP
(see \cite{gdr1} et \cite{gdr2}) can decay into a fermion and its virtual supersymmetric
partner which then decays via the \rpv couplings into two
fermions. This decay chain gives rise to 3 fermions in the final state.
For pair produced supersymmetric particles like ${\tilde {\chi}}^{o}_{2}$,
${\tilde {\chi}}^{\pm}_{1,2}$ or $\tilde f$ all heavier than the LSP
${\tilde {\chi}}^{o}_{1}$, the \rpv decays can be classified into 2 categories:  
\begin{malist}
\item {\bf indirect \rpv (or cascade) decays.}
The supersymmetric particle first decays through a $R$-parity conserving
vertex to an on-shell supersymmetric particle till the LSP ${\tilde {\chi}}^{o}_{1}$
which then decays, as described above, via one \rpv coupling.
\item {\bf direct \rpv decays.} 
The supersymmetric particle decays directly to standard particles through 
one \rpv coupling.
\par 
\end{malist}
\begin{figure}[h]
\begin{center}
\epsfig{file=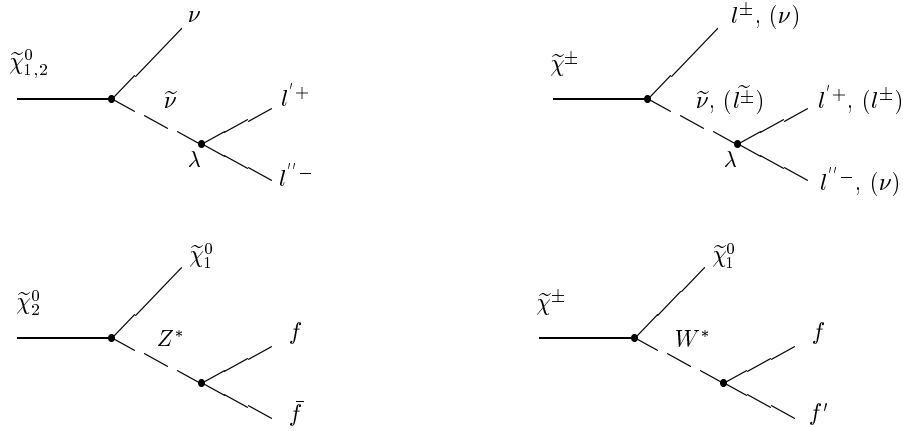,width=12cm}
\end{center}
\caption{Gaugino direct (upper part) and indirect (lower part) decay diagrams}
\label{cb-decadiag}
\end{figure}
Some examples of direct and indirect decays of gauginos,
when $\lambda_{ijk}$ couplings are involved,
are shown in figure~\ref{cb-decadiag}, and the 
corresponding possible signatures are given in 
table~\ref{cb-tab2}.\\
Decay of
supersymmetric particles via $\lambda_{ijk}$ couplings give rise in general to
{\bf leptonic topologies} although
one can see in table~\ref{cb-tab2} that jets may be present in the final states in
case of  
indirect gaugino decays.
%
%
\begin{table}[h]
\begin{center}
\begin{tabular}{|l|l|l|} \hline
final states         &   direct             &  indirect  \\
                     &  decay of            & decay of       \\\hline 
$2 l +$ \Emiss       & \XPI \XMI            &   \\
$4 l +$ \Emiss       & \XOI \XOI, \XPI \XMI &  \XOII \XOI \\
$6 l $               & \XPI \XMI            & \\
$6 l + $  \Emiss     &                      & \XPI \XMI, \XOII \XOI \\
$4 l + 2$ jets + \Emiss &                   & \XOII \XOI \\
$4 l + 4$ jets + \Emiss &                   & \XPI \XMI \\
$5 l + 2$ jets + \Emiss &                   & \XPI \XMI \\ \hline
\end{tabular}
\caption{Final states in gaugino pair production when a $\lambda_{ijk}$ 
coupling is dominant}
\label{cb-tab2}
\end{center}
\end{table}
\par
\vspace*{1.cm}
When $\lambda_{ijk}^{\prime}$ couplings are involved, decays of
supersymmetric particles give rise in general to
{\bf semi-leptonic} topologies.
A typical example of a direct supersymmetric particle decay
into standard particles 
is the semi-leptonic decay of the stop
${\tilde t}_1 \rightarrow l q $ 
giving rise to a 2 leptons + 2 jets signature. Another typical
example is the direct decay of the 
${\tilde {\chi}}^{o}_{1}$ into 3 fermions which are 
one lepton (charged or neutral) and 2 quarks.
The signature of a pair produced
${\tilde {\chi}}^{o}_{1}$ followed by a decay via one violating $R_p$ coupling
$\lambda_{ijk}^{\prime}$
is then 2 leptons + 4 jets, 4 jets and missing energy when the leptons in the
final state are neutrinos or 1 lepton + 4 jets and missing energy.
\par
Still in the $\lambda_{ijk}^{\prime}$ dominance case, 
a typical example of indirect decay is the decay 
${\tilde {\chi}}^{\pm}_{1} \rightarrow {\tilde {\chi}}^{o}_{1} W^{\ast \pm}$
with conserved $R$-parity followed by the \rpv decay of the ${\tilde {\chi}}^{o}_{1}$
as above giving rise to 2 leptons + 8 quarks, 4 leptons + 6 quarks or even
6 leptons + 4 quarks partonic final state in which the leptons 
may be charged or neutral (thus
giving rise to missing energy) and in which the high multiplicity of quarks
leads to a multijet pattern.

\par 
\vspace*{1.cm}
When $\lambda_{ijk}^{\prime \prime}$ couplings are involved, decays of
supersymmetric particles give rise in general to
{\bf multijet hadronic topologies} although, here again, leptons
may be present in the final state in case of indirect decays of gauginos.
\par
\begin{figure}[htb]
\begin{center}
\epsfig{file=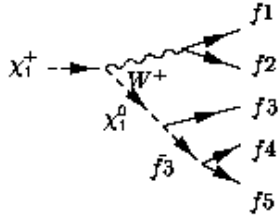,width=4.0cm}
\end{center}
\caption{Example of cascade decay for chargino to LSP neutralino
with $\lambda^{\prime \prime}_{ijk}$ couplings}
\label{fb-deca}
\end{figure}

One can have 4 quarks in the partonic final state when a squark directly decay
into 2 quarks (e.g. ${\tilde t}_1 \rightarrow s b $) or 6 quarks for
the  direct decay 
${\tilde {\chi}}^{o}_{1} \rightarrow q q^{\prime} q^{\prime\prime} $.
One may have 8 or even 10 quarks in the partonic final state 
in case of indirect decay of squarks
or charginos. For example one may have 10 quarks in chargino indirect decay:
${\tilde {\chi}}^{\pm}_{1} \rightarrow {\tilde {\chi}}^{o}_{1} W^{\ast \pm}$
followed by
${\tilde {\chi}}^{o}_{1} \rightarrow q q^{\prime} q^{\prime\prime} $
and $W^{\ast \pm} \rightarrow q_1 q_2$ as shown in figure~\ref{fb-deca} where
$f1$, $f2$, $f3$, $f4$ and $f5$ stand for $q1$, $q2$,
$q$, $q^{\prime}$ and $q^{\prime\prime}$ respectively. 

\begin{figure}[htb]
\begin{center}
\epsfig{figure=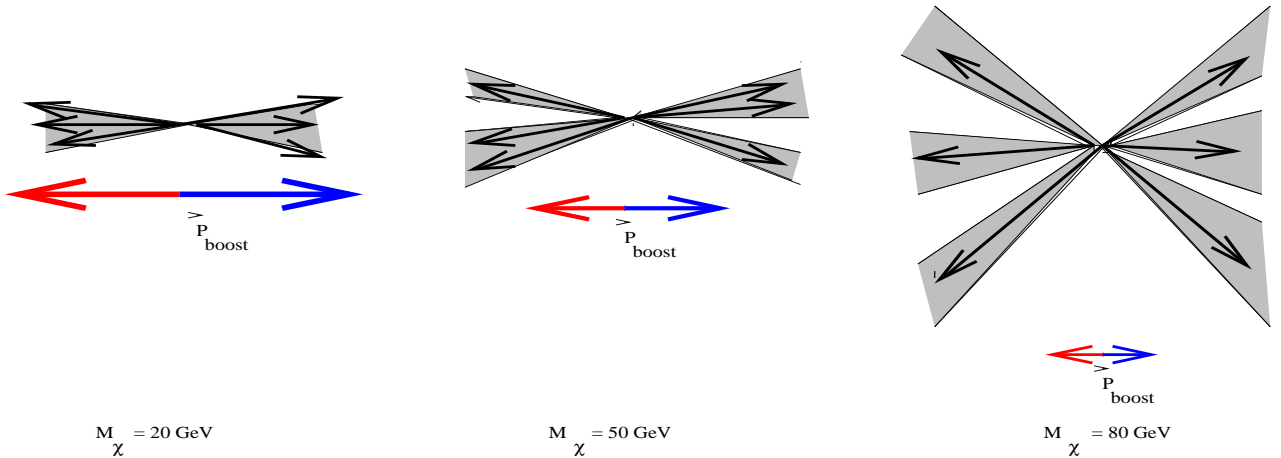,height=17cm,width=6cm,angle=-90}
\end{center}
\caption{Schematic jet patterns for the 6-quarks partonic final states
of neutralino pair decay with $\lambda^{\prime \prime}_{ijk}$ couplings.}
\label{fb-fig2}
\end{figure}

These partonic final state quarks hadronize into jets, giving a multijet 
pattern whose shape depends then on the boost of the initial supersymmetric particle.
The schematic jet patterns for the 6-quarks partonic final states produced
by the \rpv decay 
of the neutralino pair at $\sqrt{s} = 183$ GeV, for three masses of neutralino
are given in figure~\ref{fb-fig2} from which one can see that 
6-quarks partonic final states
can lead to 2-jets, 4-jets and 6-jets topologies, depending here only on the
boost of each primary sparticle \footnote{Actually, the jet topology is also 
widely dependent on the resolution
parameter of the employed jet-reconstructing algorithm}.

Now, still in the $\lambda_{ijk}^{\prime \prime}$ couplings
case for indirect chargino decay,
if $W^{\ast \pm} \rightarrow l \nu$,
one can see that leptons can be present in the topology/partonic
final state which could then
be 6 quarks + 2 charged leptons and missing energy or 8 quarks + 1 charged lepton
+ missing energy. Leptons can also be present in case of a cascade decay
of the type: ${\tilde {\chi}}^{o}_{2} \rightarrow {\tilde {\chi}}^{o}_{1} Z$
followed by $ Z \rightarrow l^+ l^-$.
\vspace*{1.cm}
\par
The requirement that the sparticle
decays within the detector
(typically within 1\,m)  translates, for the energies and masses
of interest at LEP, for a sfermion e.g. the $\tilde {\nu}$ to 
weak lower bounds $\lambda ,\lambda^{\prime} \gg 10^{-8}$ and for
a gaugino e.g. the \achia , 
 to weak lower bound $\lambda\ge 3\times 10^{-6}$.
For values of $\lambda$ between $10^{-5}$ and $10^{-6}$ for the
neutralinos/charginos
and $10^{-7}$-$10^{-8}$ for the sfermions, the \rpv decays appear as
displaced vertices in the detector.
For weaker values of $\lambda$ the \rpv signatures become indistinguishable
from the \rp ones. Very low mass neutralinos decay outside the detector
even for relatively high $\lambda$ values.
A further complication arises when $\lambda^{\prime\prime}$
or $\lambda^{\prime}$ are
involved and when the decay lifetimes to quarks become larger than the
hadronization ones. Then the system hadronizes into a squark hadron before
 decaying
and all the ambiguities in the modelization of the
\rp $\tilde{t}$  decay become relevant for the \rpv decays.
Experimental searches of pair produced gauginos and
pair produced sfermions have been recently performed in the four LEP
experiments 
in the context of 
\rpv couplings. These searches generally assume that there are no
displaced vertices.

\begin{figure}[htb]
\begin{tabular}{||c|c||} \hline
\epsfig{file= 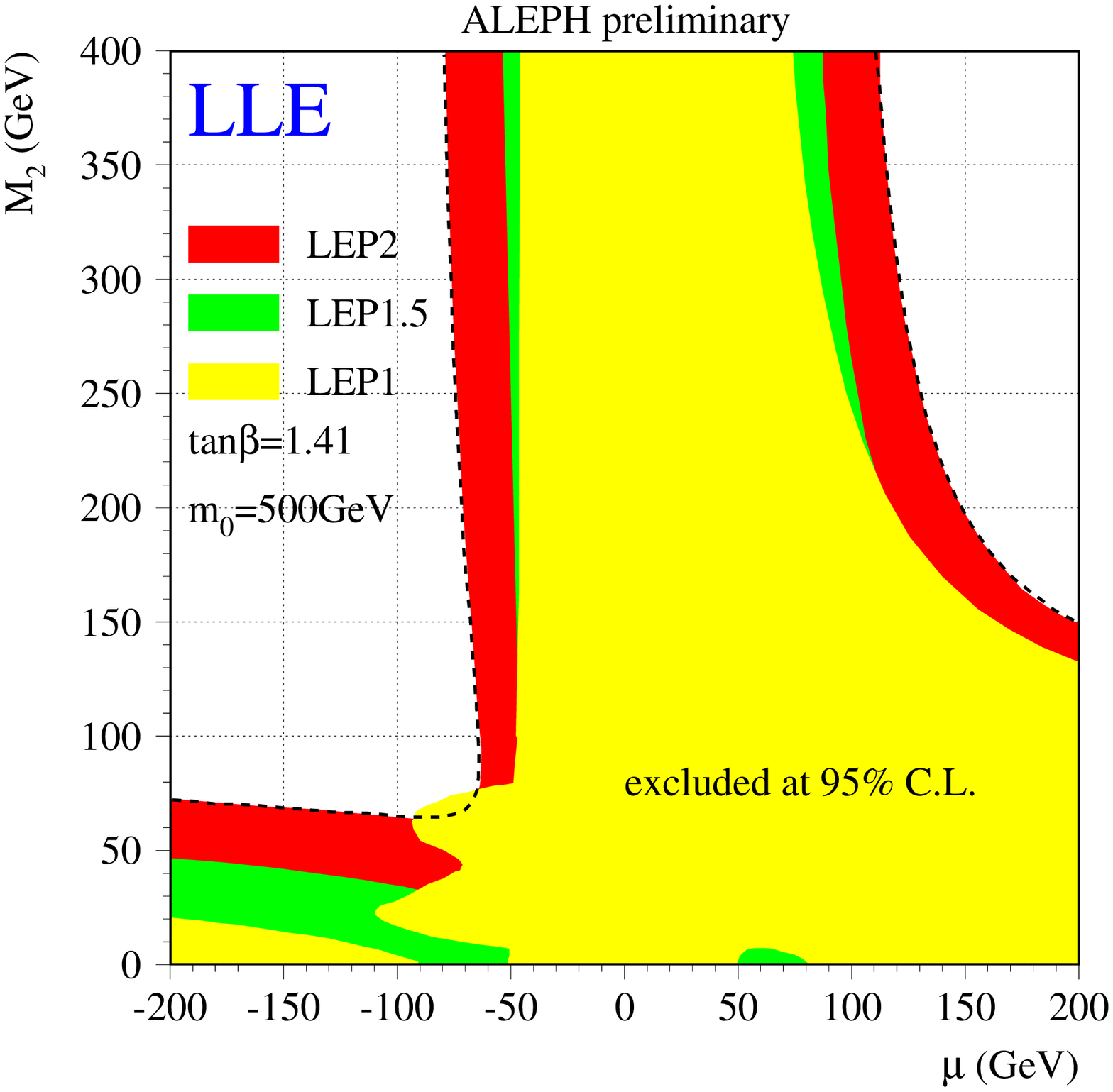,width=8.cm} &
\epsfig{file= 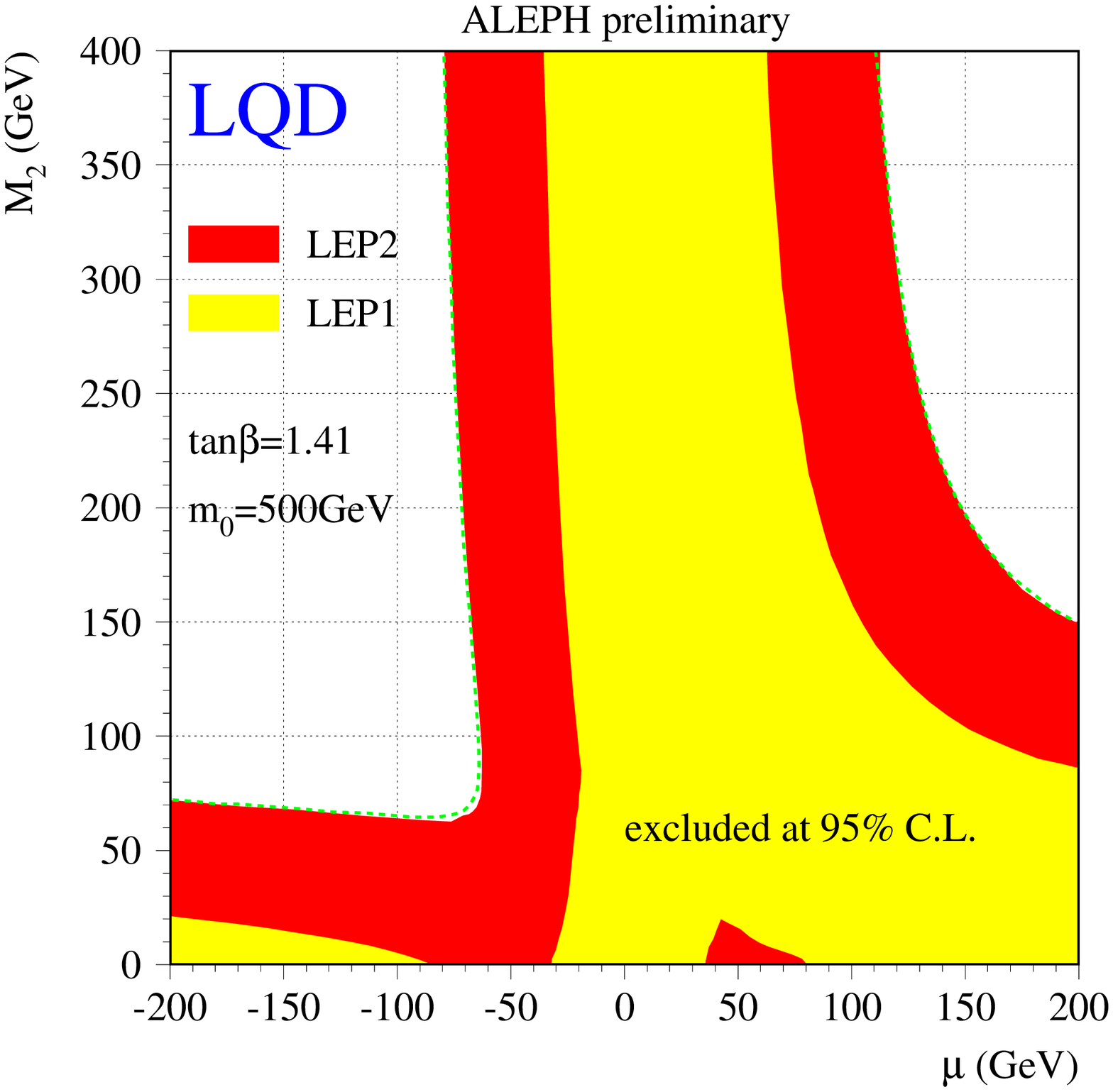,width=8.cm} \\ \hline
    &
\epsfig{file= 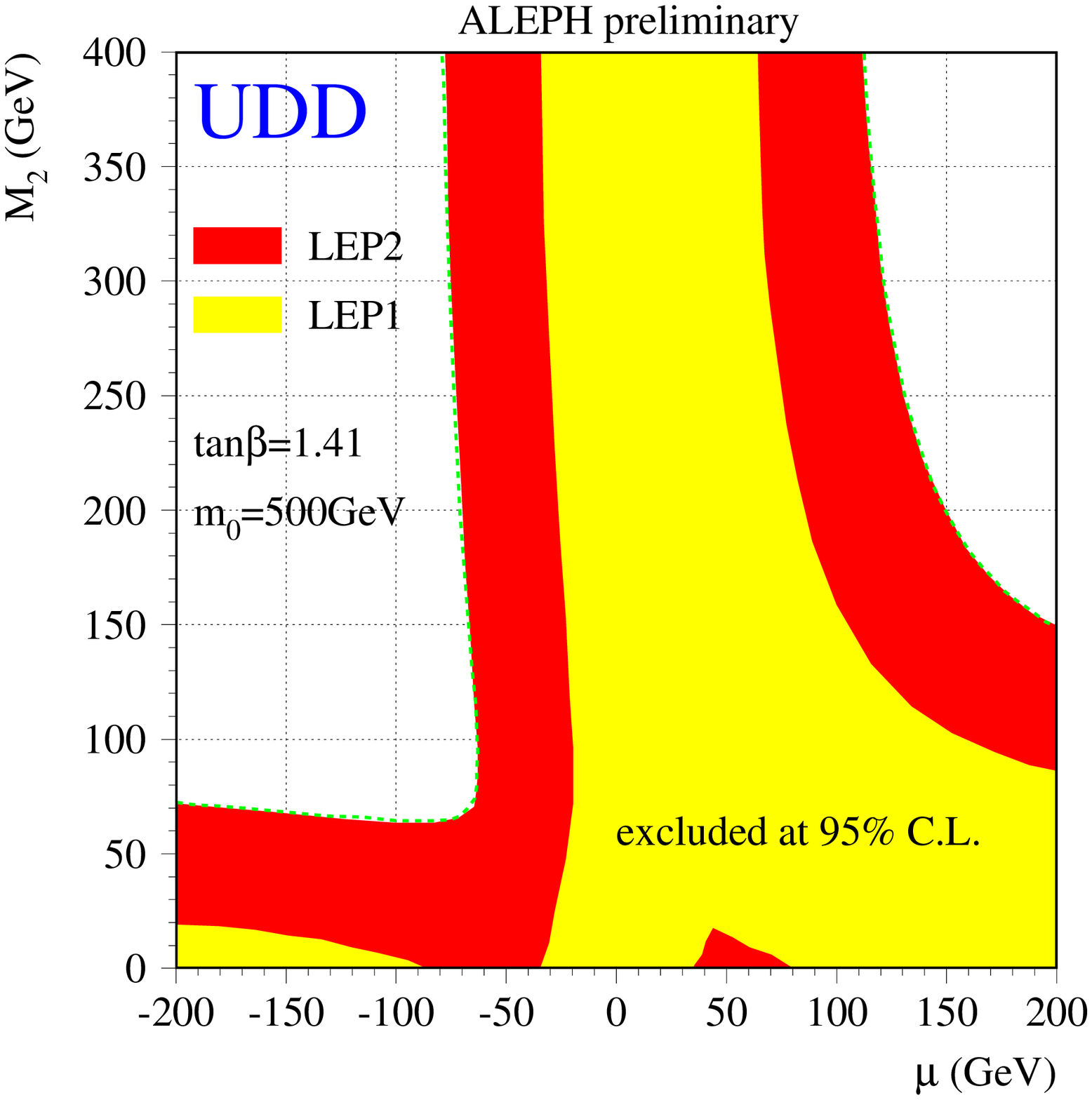,width=8.cm} \\ \hline
\end{tabular}
\caption{Charginos and Neutralinos 95 \% C.L.
exclusion in the $\mu$, $M_2$ plane for $\tan \beta= \sqrt {2} $.
Valid for any choice of generation indices i, j, k of the couplings
$\lambda_{ijk}$,
$\lambda^{\prime}_{ijk}$ and
$\lambda^{\prime \prime}_{ijk}$ and valid for $m_o=500$ GeV$/c^2$ i.e. large
sfermion masses.}
\label{lp1}
\end{figure}

\begin{figure}[htb]
\begin{center}\mbox{\epsfig{file= 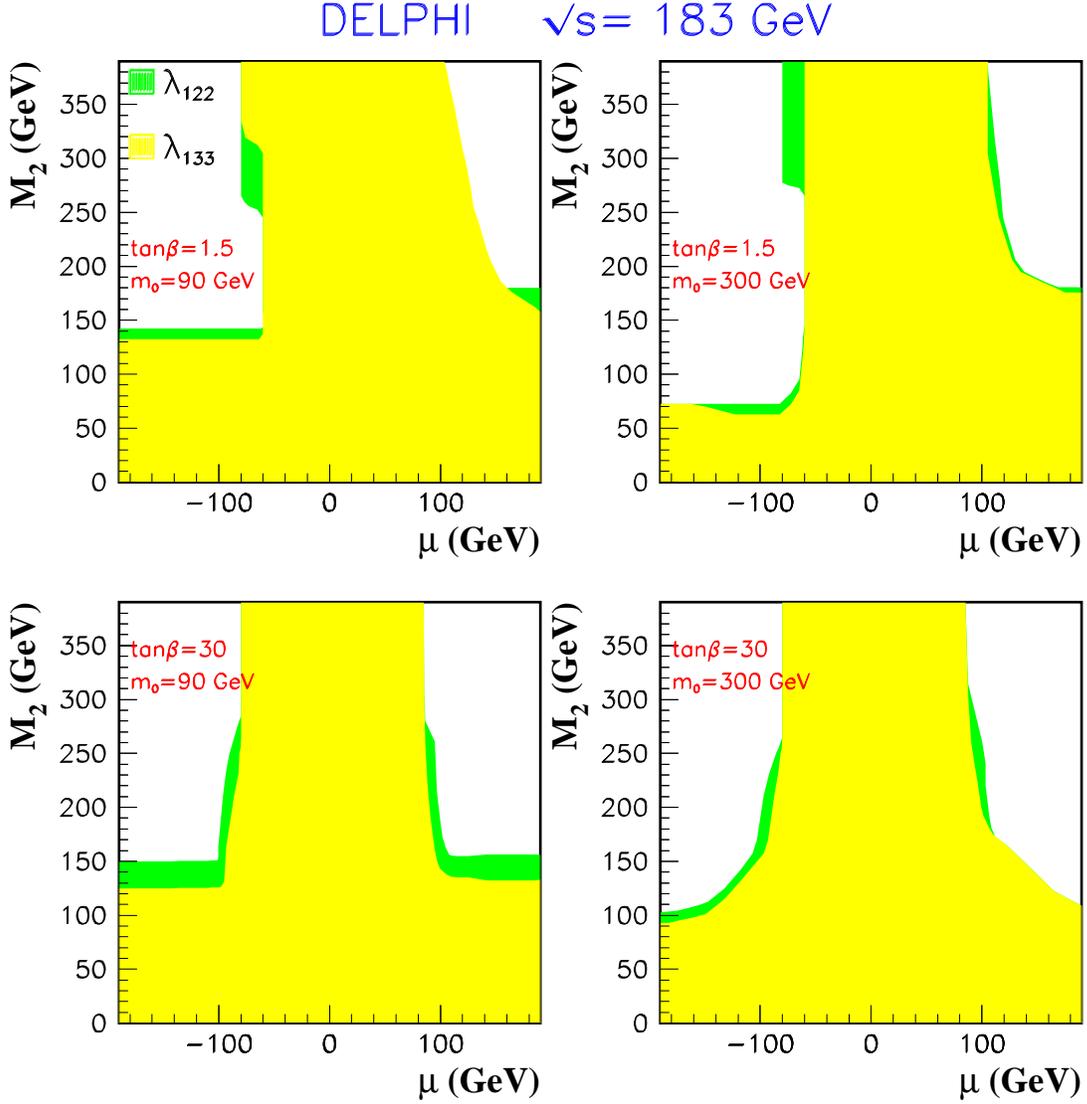,width=16.cm,height=16.cm}}
\end{center}
\caption{Regions in $\mu$, $M_2$ parameter space excluded at 95 \% C.L.
for two values of tan$\beta$ and two values of $m_{0}$. 
The exclusion area obtained from the $\lambda_{133}$ search is shown in 
light grey and the corresponding area for the 
$\lambda_{122}$ search is shown in dark grey. The second exclusion area
includes the first. The data collected in DELPHI at E$_{CM}=$ 183 GeV are used.}
\label{lp3}
\end{figure}

Using the 
1996 and 1997 data of LEP,
no evidence for \rpv signals have been found
and limits on the masses of sfermions have been derived as well as limits on 
MSSM parameters relevant for the gauginos sector. 
Examples of these limits are given in figures~\ref{lp1}
and figures ~\ref{lp3} in cases of both
direct and indirect decay of neutralinos and charginos.

\begin{figure}[htb]
\begin{center}
\leavevmode
\epsfxsize = 18 true cm
\epsfbox{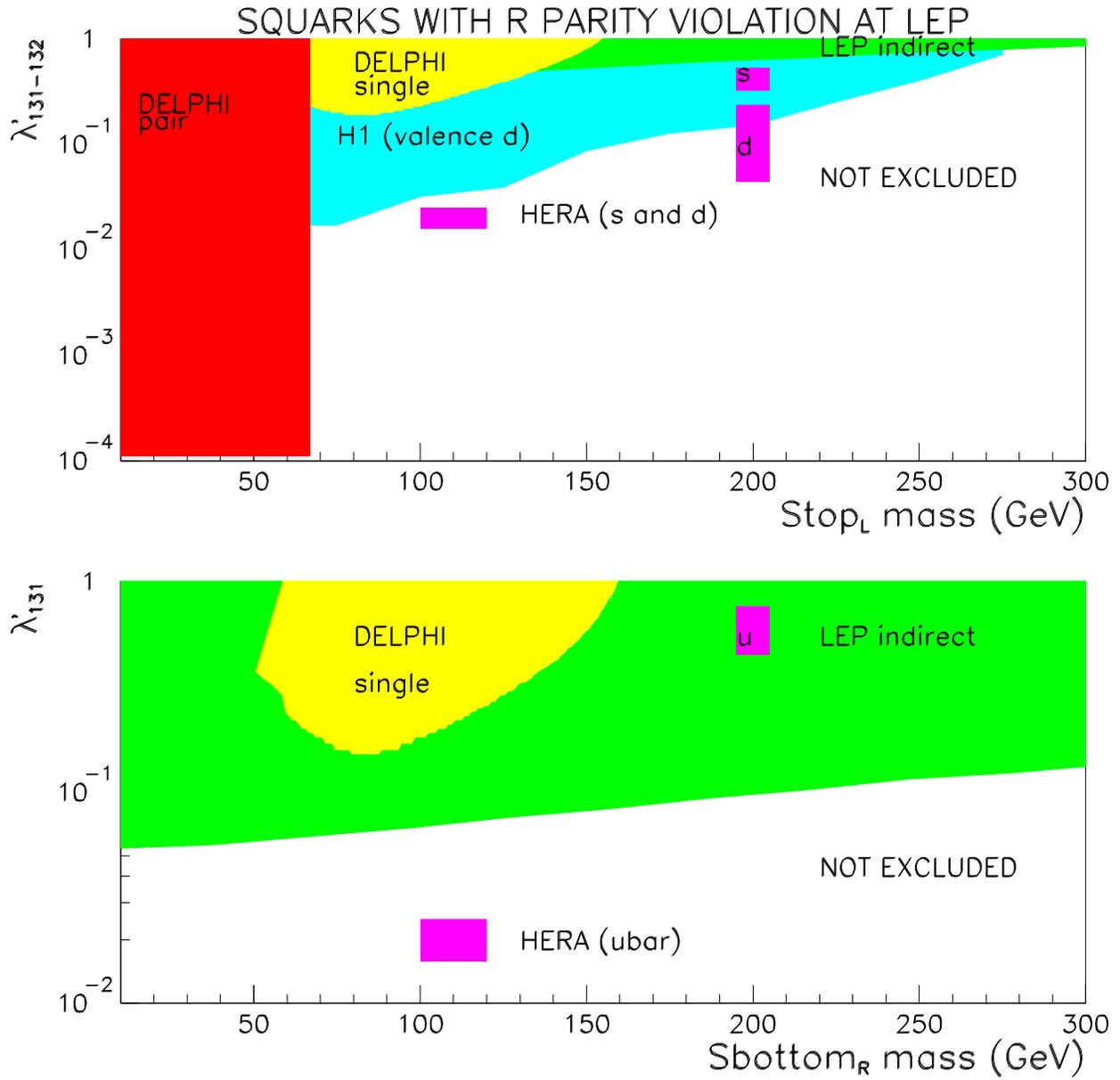}
\caption{Exclusion domain
in the ${\lambda}^{\prime}$ versus $m_{{\tilde q}}$ plane.}
\label{lp2}
\end{center}
\end{figure}

Another example of these limits is given in figure~\ref{lp2} in case of the stop
search (indicated by DELPHI pair) 
with a direct decay together with other searches described in the following.


\clearpage

\subsection{Effects of the $R$-parity violating couplings in the single production}

\begin{figure}[htb]
\begin{center}\mbox{\epsfig{file= 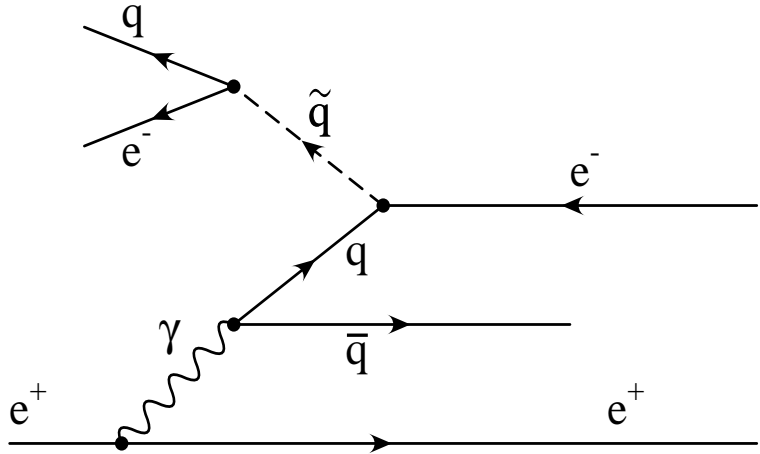,width=10.cm,height=6.cm}}
\end{center}
\caption{Single squark production}
\label{feyn1}
\end{figure}

Estimates for single production 
of a neutralino (with a neutrino), a chargino
(with a charged lepton), a resonant sneutrino,
all involving $\lambda_{ijk}$ couplings, have been given in section 4. 
Single production of     
a squark in $\gamma e$ interactions
can also occur in $e^+e^-$ collisions 
i.e. through the interaction of a quark
from a resolved $\gamma$ radiated by one of the incoming particle 
($e^+$ or $e^-$) with the other incoming particle, involving 
$\lambda_{ijk}^{\prime}$ couplings as shown in
figure~\ref{feyn1}.
The striking differences with the above pair production are 
that in this single production the ${\lambda}^{\prime}$ directly
intervene in the expression of the cross-section and that
higher squark masses are accessible. 
In this case as well, the direct decay gives 
a signature of a single lepton opposite a hadronic jet with a resonant mass or
missing energy and a hadronic jet;  the indirect decay 
(not present in figure~\ref{feyn1}) gives a
signature of a jet opposite to a two other jets and a lepton or 2 jets and
missing energy coming from ${\tilde {\chi}}^{o}_{1}$ decay via
 ${\lambda}^{\prime}_{ijk}$ couplings.
\par
This effect has also been searched for experimentally at LEP and no evidence
have been found for its occurence in the 1997 data from the LEP experiments
and exclusion domains
in a ${\lambda}^{\prime}$ coupling and squark mass plane has been 
derived  
as shown for example in figure~\ref{lp2} (indicated by DELPHI single). 

\subsection{Indirect effects of the $R$-parity ${\lambda}^{\prime}$ violating couplings
in $e^+e^-$ colliders}
The t-channel exchange can in principle give access
to squark masses well beyond the kinematical limits. Deviations
of the SM cross sections for $e^+e^- \rightarrow q \bar q$ processes
depend on the mass and type i.e. u or d,  of the exchanged squark and on the 
${\lambda}^{\prime}$ couplings. 
No deviation has been observed in 1997 LEP data and exclusion domains
in a ${\lambda}^{\prime}$ coupling and squark mass plane have been 
derived as shown in 
figure~\ref{lp2} (indicated by LEP indirect). The relevant exclusion
domain derived from the H1 collaboration is also shown in this figure 
(indicated by H1(valence d) as well as the bands, indicated by u,d,s 
which would have been relevant for the so called HERA anomaly found in 1997
with the pre-1997 HERA data. 

The LEP collider has taken data in 1997 and will continue to take
data until year 2000 thus allowing to pursue the search for
these \rpv effects. In the next section, a study of some of these \rpv effects
is described assuming that the year 2000 scenario will be $\sqrt{s} =$~200~\GeV,
with a high luminosity around 200 pb$^{-1}$ per experiment.
 
\subsection{$R$-parity scenario at LEP2, at $\sqrt{s} =$~200~\GeV,
with a high luminosity}

All the results of $R$-parity violating searches
obtained until now show no evidence for a \rpv\ signal; they
are in agreement with the Standard Model
expectations. In case of pair production studies, they  are used
to constrain domains of the MSSM parameter space and to derive limits
on the mass of supersymmetric particles.
In this section, the possibility of a run at $\sqrt{s} =$~200~\GeV\ with an
integrated luminosity of 200~pb$^{-1}$ per experiment is the basis to
evaluate reachable limits on gaugino masses in the hypothesis of no discovery of
$R$-parity violating sparticle decay. Though it exists 3 terms in
the \rpv\ superpotential, we consider here only the  
$\lambda_{ijk} L_i L_j {\bar E}_k$ term. 

\subsubsection{Pair production of gauginos}
The appropriate MSSM parameters to consider in this study
are tan$\beta$, $m_0$,  $M_2$, $\mu$; they are scanned
over the ranges : 1~$<$~tan$\beta \leq$~30, 
20~\GeVcc~$\leq m_0 \leq$~500~\GeVcc,
0~$\leq M_2 \leq$~400~\GeVcc,
-200~\GeVcc~$\leq \mu \leq$~200~\GeVcc.
Depending on the parameter values, the
cross sections at $\sqrt{s} = $~200~\GeV\ vary typically from 0.1 to 10 pb.
If no \rpv\ signal is observed, it is possible
to rule out some regions in the MSSM parameter space. 
It is usual to present excluded regions in the \Mmuplane\
for different values of tan$\beta$ and $m_0$. 

\begin{figure}[h]
\begin{center}
\epsfig{file=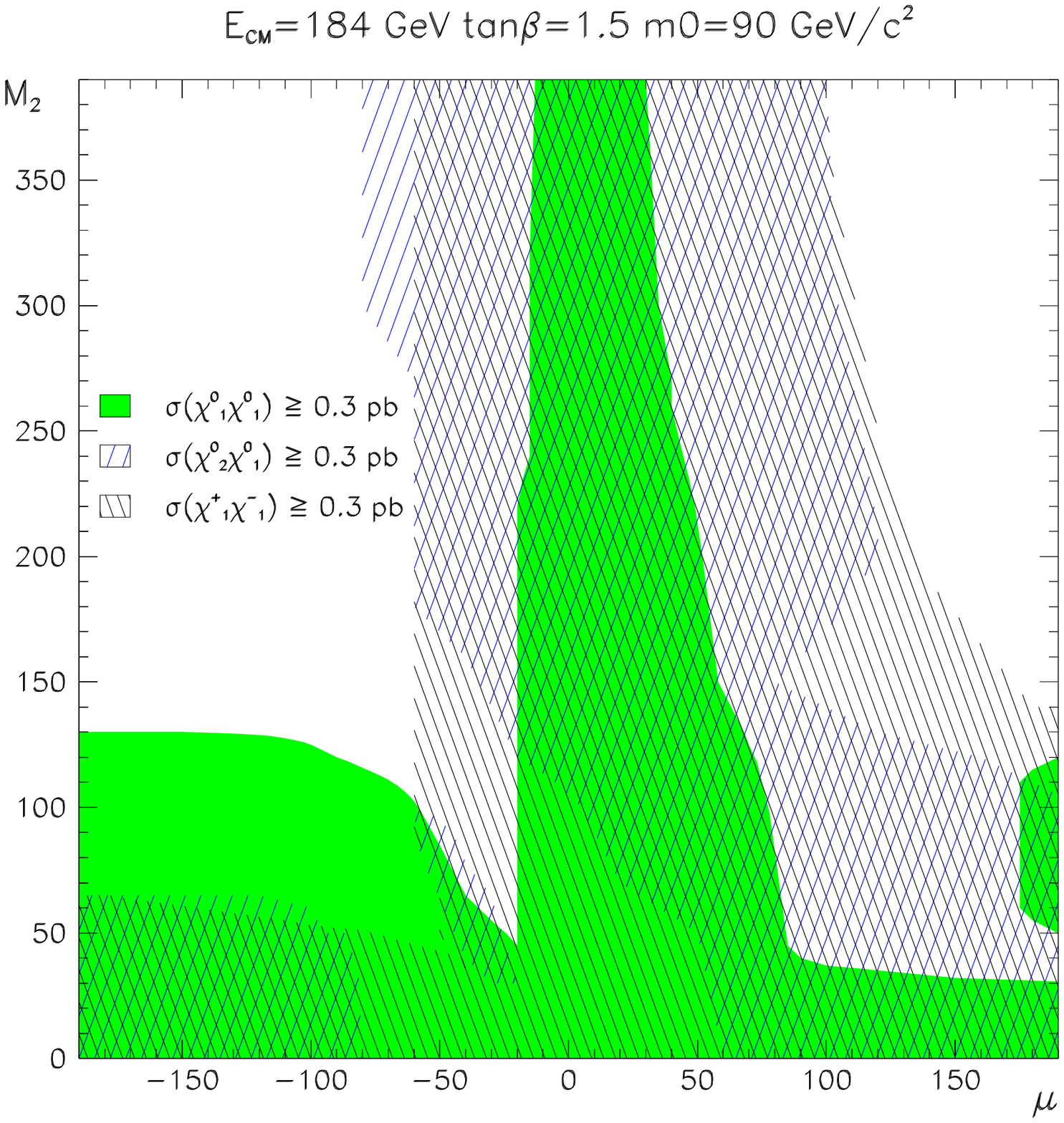,width=8cm}
\label{cb-gaugzone}
\caption{Gaugino cross sections in the ($\mu$,$M_2$) plane for tan$\beta$~=~1.5 ,
m$_0$~=~90~\GeVcc and ${\sqrt s} = 184 $ GeV ; in the dark grey 
area $\sigma$(\epem $\rightarrow$ \XOI\XOI)$> 0.3$~pb, in the hatched area
$\sigma$(\epem $\rightarrow$  \XOII\XOI)$> 0.3$~pb and/or
 $\sigma$(\epem $\rightarrow$  \XPI\XMI)$> 0.3$~pb.}
\end{center}
\end{figure}

In gaugino pair production
several processes leading to the same type of final states have a 
non negligeable cross sections. In figure~\ref{cb-gaugzone}, the 
area where the \XOI\XOI, \XOII\XOI\ and \XPI\XMI\ cross sections
are above 0.3~pb are presented in the  \Mmuplane\ 
for tan$\beta$~=~1.5, $m_0$~=~90, and $\sqrt{s} =$~184~\GeV. 
With $\sigma_{MSSM} = 0.3$~pb, and \lum~$\simeq$~50~pb$^{-1}$, 
15 events are produced, and a 30\% efficiency leads to 4-5 events
to be detected.
As we can see from this plot, regions where the \XOI\XOI\ cross
section is too low can be excluded by looking for 
the \XPI\XMI\ (or \XOII\XOI) processes, provided their production 
cross section is high enough.
Therefore, one has to perform an analysis sensitive to most of the
possible final states occuring in the  \XOI\XOI, \XOII\XOI\ and
\XPI\XMI\ production.

\subsubsection{Hypothesis on the $R$-parity violating couplings}
In case of pair production of supersymmetric particles, $R_p$ is
conserved at the production vertex; the cross section does not 
depend on the \rpv\ couplings. On the contrary, the \rpv\ decay
width depends on the $\lambda$ coupling strength, which then 
determines the mean decay length of the LSP~\cite{dawson,
dreiner1}. Searching for the sparticles with the hypothesis that their 
lifetimes is negligible leads to $\lambda \geq 10^{-5} - 10^{-6}$ in 
case of gauginos, which is below the upper limits derived from Standard
Model processes~\cite{bargerg, reviews2, ledroit}. 
The most stringent upper bound on the $\lambda_{ijk}$ couplings is the one
applied to \Lacc~\cite{reviews}:
$$
\lambda_{133} < 0.003 {{M_{\widetilde{l}}}\over{100\ \mathrm{GeV}/c^2}}
$$
which is over the sensitivity limit. 
A LSP with a low mass, 
and then with a high boost, can escape the detection before decaying. 
So the assumption of a negligible LSP lifetime restricts the 
sensitivity of this study to $M_{LSP}$ \simsup 10~GeV, 
when considering the lowest upper bound on the couplings.

To perform the search, it is also assumed that only one among the 45
possible
couplings is dominant. According to the considered dominant coupling,
the decay of the pair-produced gauginos and sfermions will lead to 
different topologies, which could be purely leptonic (from
2 to 8 leptons, with or without missing energy), purely
hadronic (from 2 to 10 jets), or mixed (leptons + jets).
The analyses performed by the LEP experiments are aimed to select
specific topologies coming from the large number of possibilities of
final states.

\subsubsection{Direct and indirect decays of gauginos}

In case of a dominant $\lambda_{ijk}$ coupling, the sleptons couple to
the leptons, and the gauginos decay into charged leptons and neutrinos.
The decay of the lightest neutralino leads to one neutrino and two
charged leptons (figure~\ref{cb-decadiag}, upper part).
The heavier neutralinos and the charginos, depending on their mass
difference with \XOI, can either decay directly into 3 standard
fermions,
or decay to \XOI, via for example virtual
$Z $ or $W$, as illustrated on figure~\ref{cb-decadiag}, lower part.
Note that, even if the $\lambda$ couplings lead to purely leptonic
decay modes of the lightest neutralino, in case of chargino and heavier 
neutralino, the final state may contain some hadronic activity.

Decay types and branching ratios depend on the set of MSSM parameters
and on the value of the considered $\lambda_{ijk}$ coupling.
In case of  pair production of gaugino, 
the final states listed in table~\ref{cb-tab2} have to be considered.

\subsubsection{Expected limits at $\sqrt{s} =$~200~\GeV}
{\bf We start from an example: search at $\sqrt{s} =$~183~\GeV}\\
As already mentionned,  it is generally assumed
that only one coupling is dominant, and to be conservative when
establishing limits, the results are derived from the study done with the
coupling giving the lowest detection efficiency. For example, in case
of $\lambda_{ijk}$ coupling, if the $\lambda_{133}$ is dominant, 
the leptons from \rpv decay are mainly taus, and electrons. 
This case should have the worst efficiency due to the presence of several
$\tau$ in the final state, and will give the most conservative limits.
To determine efficiencies, events with \rpv\ decay of gauginos
are produced using the SUSYGEN generator~\cite{susygen} coupled to the detector
simulation program. For example, considering the DELPHI experiment~\cite{delphidet},
selection efficiencies in all the considered \Mmuplanes\ 
 are in the range 22-34\% for \XOI\ pair produced,
20-37\% for \XPI \XMI, and 20-25\% for \XOII \XOI.
The background is mainly du to  \epem~$\rightarrow$~\WW\ and
 \epem~$\rightarrow$ ~\ZZ\ events.
Since the results of the analyses are in agreement with the Standard
Model expectation, regions in the studied MSSM parameter space are
excluded, which allows to derive a lower limit on neutralino mass.\\
{\bf The high energy, high luminosity case.} \\
\begin{figure}[h]
\begin{center}
\epsfig{file=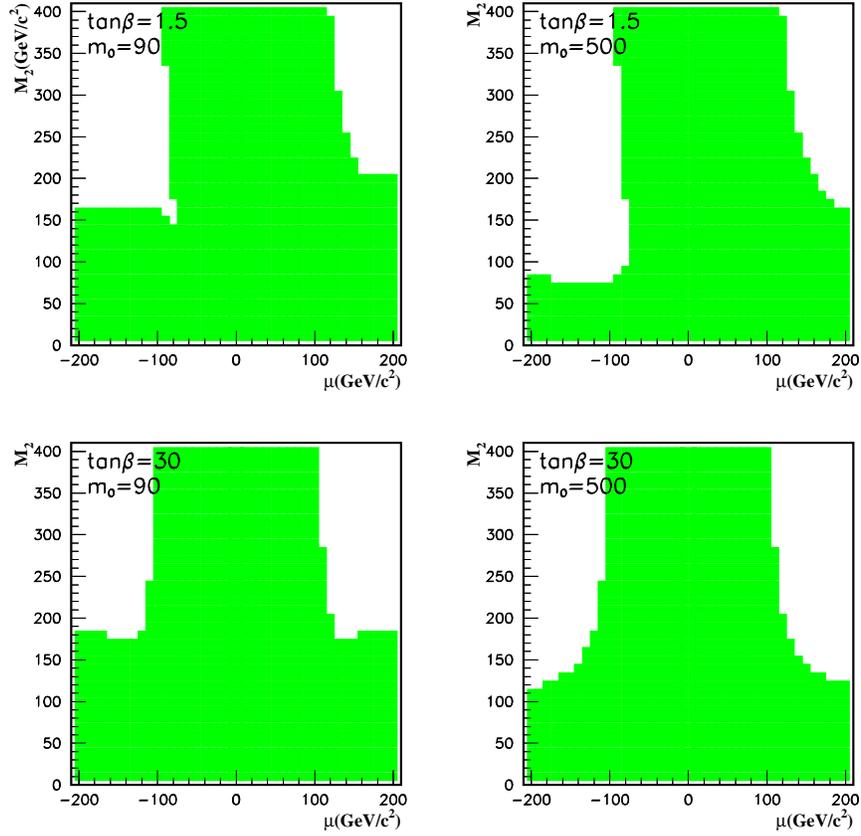,width=13cm}
\end{center}
\caption{Regions in \Mmuplane\ excluded at 95 \% C.L.
for two values of tan$\beta$ and two values of $m_{0}$. 
The exclusion area are obtained with the hypothesis that the selection
efficiencies are 30\%, 20\% and 25\% for \XOI\XOI, \XOII\XOI, and
\XPI\XMI\ respectively, and that the 95\%C.L. upper limit on the number
of signal events is 5.}
\label{cb-exclu200}
\end{figure}
In order to evaluate the limits reachable if 200~pb$^{-1}$ of \epem\
events are collected at 200~\GeV, several parameters have to be
considered.
\begin{malist}
\item {\it The production cross sections at 200~GeV}:
they are determined with the SUSYGEN program. There is no
serious variation of the cross sections in the range of
MSSM parameters considered between 184 and 200~\GeV.
\item {\it The luminosity}:
compared to the luminosity accumulated in 1997, there is
a factor of $\sim$~4 increase, which  is favorable to extend the 
excluded areas.
\item {\it The Standard Model processes}: 
in most of the \rpv\ analyses, the main background contributions
come from the four-fermion processes and the \Zg\ events. From 184 to
200~\GeV, the cross sections 
of the four-fermion processes increase~\cite{yellow},
 especially the \epem~$\rightarrow$ ~\ZZ\
cross section; the \epem~$\rightarrow$ ~\Zg\ cross section decreases.
\item {\it The selection efficiencies}:
it is assumed that the selection efficiencies will be in the same
range than those obtained at $\sqrt{s}=$~184~\GeV. 
\end{malist}

The processes contributing to the selected final states are combined             
to give the exclusion contours at $95 \% $~C.L. in the \Mmuplane.
The maximum number of signal events in presence of background
is given by the standard formula \cite{pdg}.
 All the points in the  \Mmuplane\ which satisfy the condition:
 \begin{center}
 $N \leq$ (  $\sum_{i=1}^{3}\epsilon_{i} \sigma_{i}$) \lum  \hspace{0.3cm}
 where $i$ runs for the contributing processes
  ( \XOI \XOI, \ \XOII \XOI,
   \ \XPI\XMI)
   \end{center}
are excluded at $95 \%$~C.L.

Instead of considering separately the number of observed events
and the number of expected background events one could obtained
at a high energy run, we consider several values for the 
95\%~C.L. upper limit on the number of signal event ($N_{95} = 3, 5, 10, 15$).
We consider also several sets of efficiencies for the 3~considered
processes.  The exclusion area obtained with the hypothesis that the selection
efficiencies are 30\%, 20\% and 25\% for \XOI\XOI, \XOII\XOI, and
\XPI\XMI\ respectively, and that the 95\%C.L. upper limit on the number
of signal events is equal to 5, are presented in figure~\ref{cb-exclu200}, for 
two values of $m_0$ (90 and 500~\GeVcc) and two values of tan$\beta$
(1.5, 30). 
\begin{figure}[h]
\begin{center}
\epsfig{file=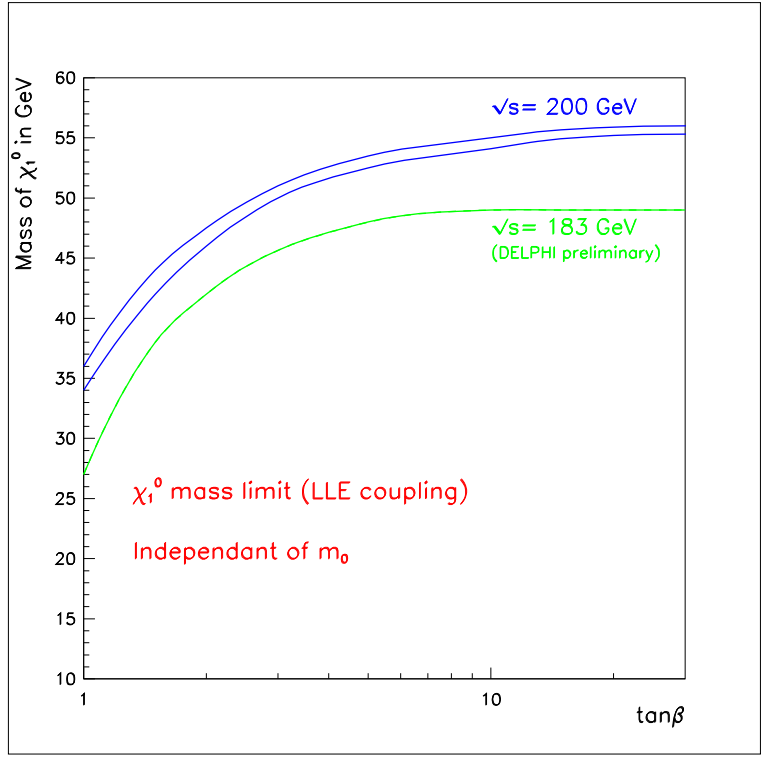,width=10cm}
\end{center}
\caption{\XOI\ mass limit as a function of tan$\beta$. The light grey curve
presents the preliminary current limit obtained by DELPHI with 
the 1997 data. The 2 dark grey curves show the range for the 
expected \XOI\ mass limit in case of 200~pb$^{-1}$ collected 
at $\sqrt{s} =$~200~\GeV: 
the upper curve is obtained assuming
$N_{95}=$~5, $\epsilon_{\tilde{\chi}_1^0 \tilde{\chi}_1^0}=$~30\%,
 $\epsilon_{\tilde{\chi}_2^0 \tilde{\chi}_1^0}=$~20\%,
 $\epsilon_{\tilde{\chi}_1^+ \tilde{\chi}_1^-}=$~25\%;
the lower curve is obtained assuming
$N_{95}=$~15, $\epsilon_{\tilde{\chi}_1^0 \tilde{\chi}_1^0}=$~30\%,
 $\epsilon_{\tilde{\chi}_2^0 \tilde{\chi}_1^0}=$~25\%,
 $\epsilon_{\tilde{\chi}_1^+ \tilde{\chi}_1^-}=$~15\%}
\label{cb-chi0lim}
\end{figure}

The results can be translated into a lower limit on \XOI\ mass, as a 
function of  tan$\beta$.
In order to obtain this limit, for each value of 
tan$\beta$ we scanned over $m_{0}$ values from 20 up to 500~\GeVcc\
and we determined the limit on the \XOI\ mass. This limit decreases
as $m_0$ increases, then it becomes constant for $m_0
>$~200-300~\GeVcc.
So, at a given tan$\beta$, the lowest mass limit is given by the
highest value of $m_{0}$. Considering this,
we can set a mass limit independantly of the choice of $m_{0}$. This
procedure has been repeated for different sets of efficiency and
$N_{95}$ values. 
At $m_0$=500~\GeVcc, the lowest \XOI\ mass not excluded
is in an area where the dominant contribution comes from the chargino
pair production. Moreover, since the cross section of this process 
rapidly increases, the neutralino mass limit is not very sensitive
to the efficiency and $N_{95}$ variations (figure~\ref{cb-chi0lim}).

\subsection{Conclusion}
The LEP is now running at $\sqrt{s}=$~189~\GeV, and an integrated
luminosity of around 150~pb$^{-1}$ is expected. In this case,
combining 1997 and 1998 data, and in the hypothesis that no evidence for \rpv
is found in the 1998 data, the limits on the \XOI\ mass will be
almost the same as those expected at $\sqrt{s}=$~200~\GeV. 
Since the luminosity plays a major role in the \XOI\ mass limit
determination, we will obtain a higher limit by combining the data
taken in 1997 to 2000, at the different center of mass energies
from 183 to 200~\GeV. 
The high energy run will allow explore the chargino mass up to
a value close to the kinematical limit  
and to explore sfermions masses beyond the present
limits. 
All these \rpv effects may also be search for within the future projects
of higher energies (center of mass energies of 500 GeV
or more) linear leptonic colliders. 
 
\newpage

\section{$R$-parity violation at LHC}

The search for SUSY at hadron colliders has been mainly investigated 
within the minimal supergravity model \cite{r1}. 
This leads to the canonical missing $E_T$ signature
which  follows from the $R$-parity conservation assumption. 
However, $R$-parity need not to be conserved in supersymmetric extensions 
of the Standart Model (see introduction). Therefore, it is important to 
considere the new phenomenology associated with $R$-parity broken models, 
in order to design LHC experiments the most efficient way to search 
for new physics.
Studies of $R$-parity violation can be approached by considering 
two extreme cases :
\begin{itemize}
\item \rpv couplings are dominant with respect to the gauge couplings.
Then, the running of these new Yukawa couplings has to be taken into account
into RGE evolution. The mass spectrum and branching ratio of the 
supersymmetric particle could be affected depending on the magnitude of the 
couplings. In that case, one generally searches for specific processes 
involving the couplings at the production. 
\item \rpv couplings are small enough, so that the usual decay pattern 
of sparticles remains unchanged. Then, mSUGRA predictions are still 
valid, the only difference coming from the decay of the LSP.  
\end{itemize}
\par
In the results described below, we have followed the second approach. 
Other simplifying assumptions have been considered : 
i) The mSUGRA framework is still used in order to 
predict mass spectrum and branching ratio. In a large region of parameters,
the LSP is the $\chio_1$. We restrict our conclusions to these regions.
ii) Among the \rpv couplings, only one dominates. The theoritical motivation 
being that in analogy with the Standard Model, a hierarchical structure 
is also expected between the Yukawa couplings violating $R$-parity.
In the following, we have treated more specifically the purely leptonic 
decays of equ.(\ref{equ3}).
Assuming slepton mass degeneracy, the branching
ratio in each of the four final states of equ.(\ref{equ3}) is 25\%
independent of the $\chio_1$ composition.  

\begin{equation} 
\chio_1 \stackrel{\lb_{ijk}}{\longrightarrow} \left\{
\begin{array}{c}
\bar{\nu_i} e^+_j e^-_k \\
e^+_i \bar{\nu_j} e^-_k \\
\nu_i e^-_j e^+_k \\
e^-_i \nu_j e^+_k 
\end{array}
\right.
\label{equ3}
\end{equation} 
In table \ref{tab1}, we display the $\chio_1$ branching ration in 0, 1
or 2 leptons according to the $\lb_{ijk}$ selected. The mean number of
additional isolated leptons varies between 1 and 2 per $\chio_1$
decay. The two extreme cases correspond to $\lb_{133} \ne 0$ (or      
equivalently $\lb_{233} \ne 0$) and $\lb_{121} \ne 0$ (or equivalently
$\lb_{122} \ne 0$). 
\begin{table}[h]
\centering
\begin{tabular}
{|c|c|c|c|c|c|c|}\hline
\multicolumn{1}{|c|}{$\lb_{ijk}$} & 
\multicolumn{1}{|c|}{Decay channel} &
\multicolumn{3}{|c|}{Fraction of leptons} &
\multicolumn{2}{|c|}{Mean number of} \\
$ijk$ & & $0 l$ & $1 l$ & $2 l$ & $<l>$ & $<\nu>$ \\
\hline
\hline
$121$ & $e^- \nu_\mu e^+ + \nu_e \mu^- e^+ + c.c$ & 0\% & 0\% & 100\% 
& 2.0 & 1.0 \\ 
$122$ & $e^- \nu_\mu \mu^+ + \nu_e \mu^- \mu^+ + c.c$ & 0\% & 0\% & 100\%
& 2.0 & 1.0 \\ 
$123$ & $e^- \nu_\mu \tau^+ + \nu_e \mu^- \tau^+ + c.c$ & 0\% & 65\% & 35\%
& 1.3 & 2.3 \\ 
$131$ & $e^- \nu_\tau e^+ + \nu_e \tau^- e^+ + c.c$ & 0\% & 32.5\% & 67.5\%
& 1.7 & 1.7 \\ 
$132$ & $e^- \nu_\tau \mu^+ + \nu_e \tau^- \mu^+ + c.c$ & 0\% & 32.5\% &
67.5\% & 1.7 & 1.7 \\ 
$133$ & $e^- \nu_\tau \tau^+ + \nu_e \tau^- \tau^+ + c.c$ & 21.1\% & 55.3\%
& 23.6\% & 1.0 & 3.0\\ 
$231$ & $\mu^- \nu_\tau e^+ + \nu_\mu \tau^- e^+ + c.c$ & 0\% & 32.5\% &
67.5\% & 1.7 & 1.7 \\ 
$232$ & $\mu^- \nu_\tau \mu^+ + \nu_\mu \tau^- \mu^+ + c.c$ & 0\% & 32.5\%
& 67.5\% & 1.7 & 1.7 \\ 
$233$ & $\mu^- \nu_\tau \tau^+ + \nu_\mu \tau^- \tau^+ + c.c$ & 21.1\% &
55.3\% & 23.6\% & 1.0 & 3.0 \\ \hline
\end{tabular}
\caption{$\chio_1$ decay channels and branching ratio in 0, 1 and 2
  leptons ($e$ or $\mu$).
  $BR(\tau^- \to e^- \bar{\nu_e} \nu_\tau) =
  BR(\tau^- \to \mu^- \bar{\nu_\mu} \nu_\tau) = 17.5 \%$ has been used.}
\label{tab1}
\end{table}
For $\lb_{121} \ne 0$, the gain is obvious : there are always 2
additional isolated leptons per $\chio_1$ decay. For $\lb_{133} \ne
0$, however, due to the branching ratio of the $\tau$ in hadronic
modes, in 21.1\% of the cases, there are no additional leptons. 
In order to cover most of event's topologies, we have studied these two 
extreme cases : $\lb_{121} = 0.05$ and $\lb_{233} = 0.06$. The magnitude of
the couplings is taken from the present limit~\ref{chap2}.  
In order to study the $\chio_1$ decay,
one has to provide the decay width of the 3 body process (via virtual
$\tilde{f}$) in each final states. The details of the calculation is 
based on the theoritical paper \cite{rdecay} and is given in \cite{mypaper}. 
It takes into account effects due to sfermion mixing without neglecting
final states masses. However, these effects are small for the 
pure leptonic decay of the neutralino and a branching ratio of 25\% 
in each of the 4 possible final states remains an excellent 
approximation (this is not true for $\lbp$ or $\lbpp$ coupling).
\begin{figure}[h]
  \begin{center}
    \includegraphics[width=8cm]{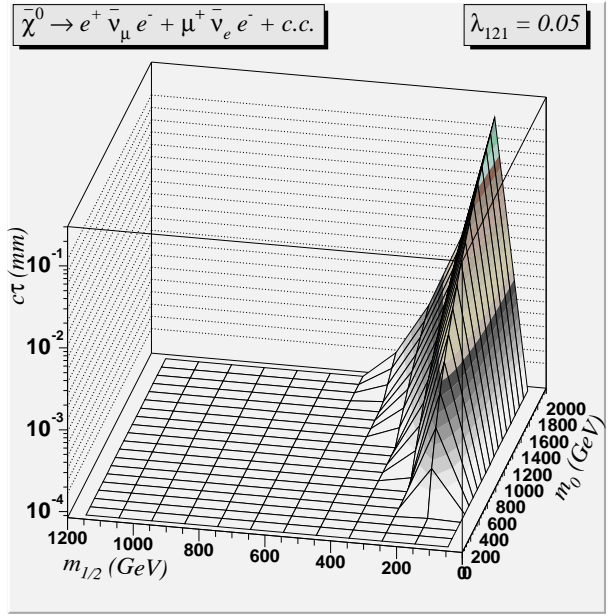}
    \caption{Neutralino $c\tau$ as function of $m_0$ and $m_{1/2}$. 
	Parameters used are $\lb_{121}=0.05$, $A_0=0$, $\mu<0$ and 
	$\tan{\beta}=2.0$.}
    \label{lifetime}
  \end{center}
\end{figure}
As an illustration, in figure~\ref{lifetime}, the neutralino lifetime 
is plotted in the $m_0-m_{1/2}$ plane. The parameters used are 
$\lb_{121}=0.05$, $A_0=0$, $\mu<0$ and $\tan{\beta}=2.0$. 
At first approximation, the lifetime varies as $m_{\tilde{f}}^4/m_{\chio_1}^5$ 
(usual 3-body decay). Therefore, the lifetime reaches its maximum for large
values of $m_0$ and small $m_{1/2}$ since $m_{\tilde{f}}$ and $m_{\chio_1}$
scale respectively as $m_0$ and $m_{1/2}$.

\subsection{ATLAS discovery potential}

     The ATLAS Collaboration has demonstrated earlier in the
framework of the Supergravity (SUGRA) Model 
and assuming $R$-parity conservation that Supersymmetry
(SUSY) can be ruled out experimentally at the LHC, 
for masses less than about 1 TeV, if it is 
not realized in Nature. On the other hand it was found that
if SUSY exists, one can not
only discover it experimentally, but one can also constrain the
model and determine its parameters with high precision
(see \cite{LHC1}-~\cite{LHC5}). 

     If $R$-parity is not conserved, some SUSY signals disappear
(or at least become weaker), e.g. the missing energy will be
reduced since in this case the lightest supersymmetric particle
(LSP) is allowed to decay. On the other hand, the same decay
would enable the experimental determination of the mass of the
LSP and thus would provide more information on the parameters
of the SUGRA model. It is therefore important to repeat the
previous study on SUSY to see how the above mentioned results are
modified in case of the violation of the $R$-parity. This study
has been carried out assuming that one of the
$\lambda_{ijk}\ (i,j,k=1,2,3)$ coupling is nonzero. 
Inclusive and exclusive reactions in the LHC 
points 1, 3 and 5 of the earlier
study (see ~\cite{LHC1}) have been considered.

      Since there was no existing generator which produced
events in hadron colliders {\it and} violated $R$-parity,
ISAJET~\cite{isajet} and SUSYGEN~\cite{susygen} have been merged.
After detailed testing this new program~\cite{isarpv} has been used
to produce over 1 million of events which were subsequently analysed. 
The response of the ATLAS detector has been simulated 
by ATLFAST~\cite{ATLFAST}. The event selection will be described
in a forthcoming report~\cite{rparLHC}. The Standard Model
background, mainly $t\bar t$ and $W/Z$ pair production 
has been generated by ISAJET~\cite{isajet}.

\begin{figure}
\begin{center}
  \mbox{\epsfig{file=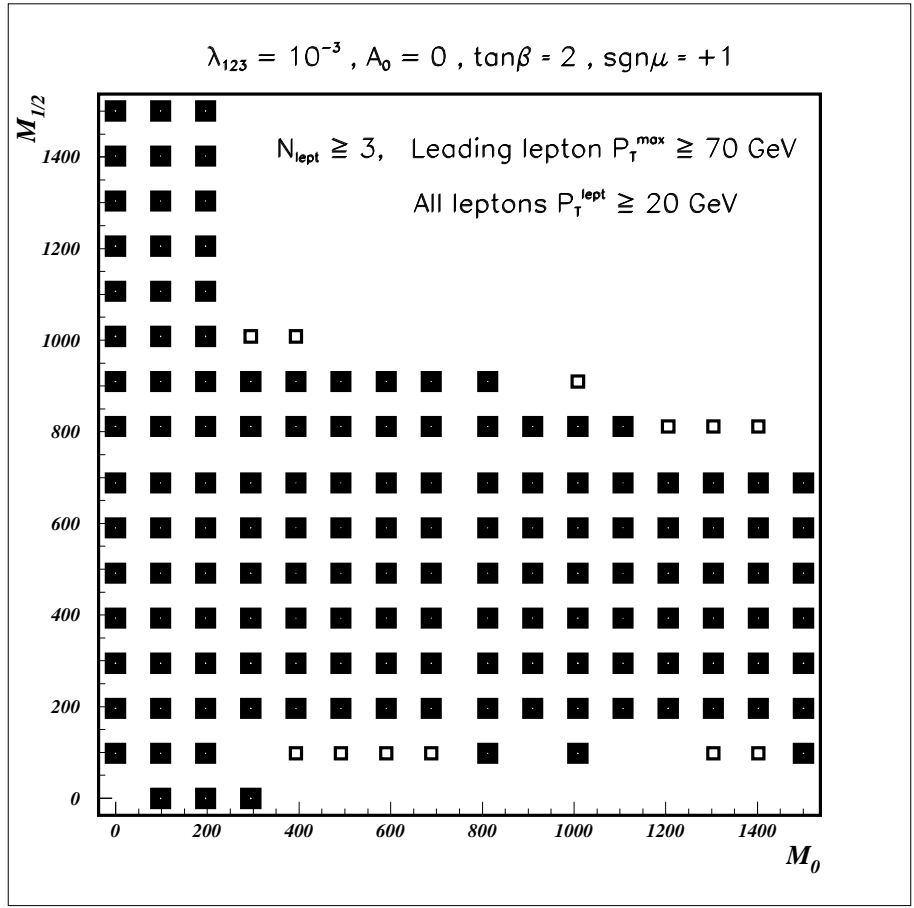,width=12cm  }}
\end{center}
\caption{\small ATLAS expected excluded region in the $M_{1/2}\ vs.\ M_0$ plane
for 1 year of LHC low luminosity run. Boxes correspond to
SUGRA signals above the Standard Model with more than 5 
(full boxes) or less than 5 (empty boxes) st. deviations}
\label{fig:sugraspace}
\end{figure}

      The study of the inclusive reactions confirmed the 
above mentioned result in case of
$R$-parity conservation: if SUSY is realized in Nature it can
be safely discovered by ATLAS in the complete parameter space
of interest. This is demonstrated in case of the non-vanishing
$\lambda_{123}$ coupling in figure~\ref{fig:sugraspace}.

\vskip 1cm

\begin{figure}[htb]
\begin{center}
  \mbox{\epsfig{file=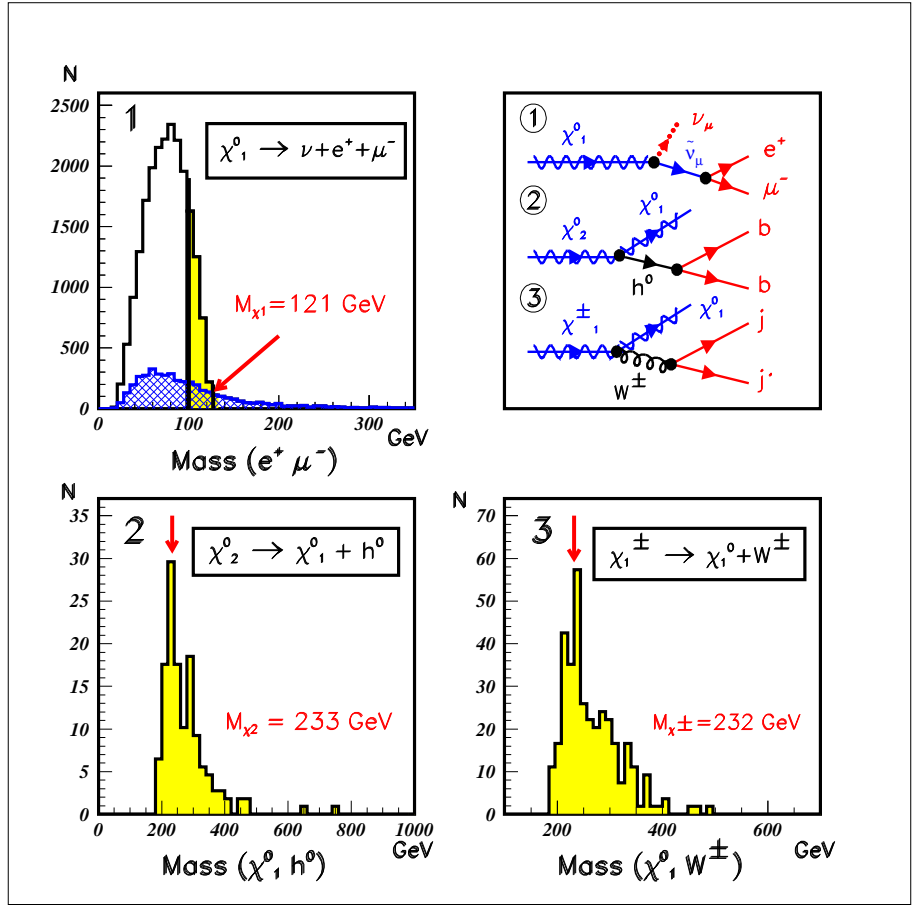,width=12cm  }}
\end{center}
\caption{\small ATLAS reconstruction of gauginos: the endpoint of the
distribution in inlet 1 provides the mass of the $\chi^0_1$.
The combination of the $\chi^0_1$ with the $h^0$ shows the mass
peak of the  $\chi^0_2$ (inlet 2). Finally the combination of the 
$\chi^0_1$ with the $W^{\pm}$ shows the mass
peak of the  $\chi^{\pm}_1$ (inlet 3).
The events around the endpoint of inlet 1 (indicated by the
dotted area) are used to reconstruct the
gauginos shown in inlets 2 and 3. The hatched area in
inlet 1 corresponds to the combinatorial background
from SUSY particles other than the $\chi_1^0$.
 }
\label{fig:reconstr}
\end{figure}

     It was also found that for most of the $\lambda_{ijk}$ couplings
(whose values have been chosen to be $10^{-3}$ taking into account
existing limits and considering event topologies without displaced vertices),
the parameters of the SUGRA model can be determined at least
with the same precision than in case when $R$-parity is 
conserved~\cite{rparLHC},~\cite{these}. In figure~\ref{fig:reconstr}
one can see an example of the reconstruction of the gauginos
in the LHC point 1 for $\lambda_{122}=10^{-3}$.

\vskip 1cm

    These results have been presented in several meetings of
the ATLAS collaboration and of the GdR SUSY.

\subsection{CMS discovery potential}

\subsubsection{SUSY signal simulation}

The analysis is performed within the framework of the minimal
supergravity model (mSUGRA) where only five parameters need to be
specified : $m_0$, $m_{1/2}$, $A_0$, $\tan{\beta}$ and $sign(\mu)$. 
PYTHIA 6.1 \cite{rpyth} including supersymmetric processes 
\cite{rspyth} has been used with its default structure function
(CTEQ2L) to generate both supersymmetric signals and Standard Model
background. PYTHIA allows to generate either MSSM
supersymmetric models (the user providing the mass spectrum) or mSUGRA
models. For mSUGRA, the mass spectrum at the electroweak symmetry   
breaking is computed form the parameters specified at the GUT scale     
, using approximate analytic
formulae. The difference with the exact numerical resolution has been checked
to lie 
within 10 \%.
An obvious interesting feature of PYTHIA is the possibility to use at
hadron colliders, the initial and final state radiation and
fragmentation models of PYTHIA/JETSET.
In the present version of PYTHIA $R$-parity violation is not
implemented. Therefore an interface between PYTHIA and the theoritical calculation
of the $\chio_1$ decay width described in the previous section has been
incorporated \cite{mypaper}.

SUSY signals are generated in the $m_0-m_{1/2}$ plane 
with the other parameters set to $\mu<0$ $A_0=0$ and $\tan{\beta}=2.0$.
We consider only squarks and gluinos production ($pp \to
\tilde{g}\tilde{g},\ \tilde{g}\tilde{q},\ \tilde{q}\tilde{q}$) which is
the dominant source of SUSY events at the LHC. 
The production of isolated leptons arises form cascade decays of squarks 
and gluinos and from the $\chio_1$ decay.

\subsubsection{SM background simulation}

The Standard Model processes generated are those which produced
isolated leptons. We have considered :
i) $t\bar{t}$ production where leptons may arise from $t \to b W$ 
followed by $W \to l \bar{\nu}$.
ii) single boson production ($Z/\gamma^*,W$).  
iii) double boson production ($ZZ,\ ZW,\ WW$).
iiii) QCD jet production (where leptons arise from heavy flavors).   
Additionnal jets come from initial and final state QCD radiations
using parton showers approach.

\subsubsection{CMS detector simulation}

We use the fast non-geant simulation package CMSJET 4.3 \cite{rcmsjet}.
It is well adapted in view of the huge statistic needed. 
The main features of CMSJET relevant to this analysis are :
\begin{itemize}  
\item Charged particles are tracked in a 4 T magnetic field with 100\% 
reconstruction efficiency per track.
\item ECAL calorimeter up to $|\eta|=2.6$ with a granularity of 
$\Delta \phi \times \Delta \eta = 0.0145 \times 0.0145$. The energy 
resolution is parametrized as $\Delta E/E=5\%/\sqrt{E} \oplus 0.5\%$.
\item HCAL calorimeter up to $|\eta|=3$ and $|\eta|=5$ in the very forward.
The granularity is $\Delta \phi \times \Delta \eta = 0.087 \times 0.087$ 
(for $|\eta|<2.3$) and the energy resolution depends on $\eta$ (equals to 
$82\% \sqrt{E} \oplus 6.5 \%$ at $\eta =0$).
\item  Lepton's momenta are smeared (both $\mu$ and $e$) according to 
parametrization obtained from detailed GEANT simulations with an angular 
coverage going up to $|\eta| < 2.4$.   
\end{itemize}

\subsubsection{Events selection}

In the presence of $R$-parity violation via leptonic couplings 
($\lb$ and $\lbp$), an excess of isolated leptons is expected coming from
the $\chio_1$ decay. Even if a few amount of $\et$ is still expected (due 
mainly to neutrinos from $\chio_1$ decay), one cannot use any longer this 
criteria without cutting too much signal. Therefore, severe constraints 
are required on isolated leptons. 

The lepton isolation is defined by :
i) no charged particle with $p_T > 2~GeV/c$ in a cone radius 
$R=\sqrt{\Delta \eta^2 + \Delta \phi^2} = 0.3$ about the lepton direction. 
This criteria mainly suppresses the background from $t\bar{t}$.
ii) The transverse energy deposited in the calorimeter in a cone 
radius $R=0.3$ about the lepton direction should not exceed $10\%$ of 
the lepton transverse energy.

Then, events satisfying the following cuts are selected :
i) At least 3 isolated leptons with $p_T > 20~GeV/c$ for $e$ and 
$10~GeV/c$ for $\mu$, and $|\eta_l|<2.4$.
ii) At least 2 jets with $p_T > 50~GeV/c$ and $|\eta_{jet}|<4.5$.

\begin{figure}[h]
  \begin{center}
    \includegraphics[width=6.5cm]{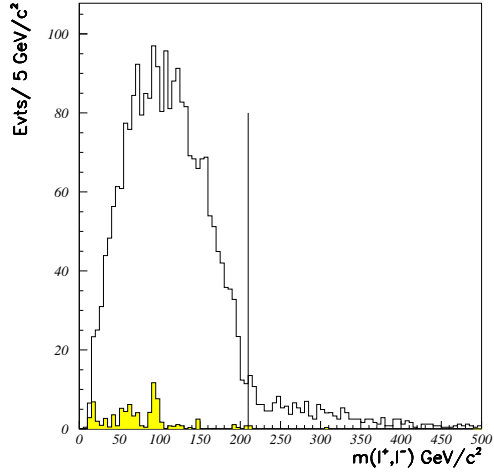}
    \caption{CMS dilepton invariant mass distribution for $\lb_{121} = 0.05$,  
	$m_0=1000$, $m_{12}=500$, 
	$\tan{\beta}=2$, $A_0=0$ and $\mu<0$. The shaded area corresponds 
	to the SM background. The straight line indicates the $\chio_1$ mass
        equal to 210 GeV/$c^2$.
	Signal and background are normalized to an integrated luminosity
	of $10^4~pb^{-1}$.}
    \label{spectrum}
  \end{center}
\end{figure}
The invariant mass $m_{l^+l^-}$ is then reconstruted for opposite sign
leptons. When several combinations per event are allowed, the one with the 
minimal angular separation is chosen. A significant deviation from the Standard
Model spectrum provides evidence for SUSY. Moreover, the specific shape of
the mass distribution with its sharp edge, allows to measure directly 
$m_{\chio_1}$. This measurement could then be used to determine the mSUGRA 
parameters. In figure~\ref{spectrum}, an example is shown for 
$\lb_{121} = 0.05$, $m_0=1000$, $m_{12}=500$, $\tan{\beta}=2$, $A_0=0$ 
and $\mu<0$.

\subsubsection{Results and conclusion}

\begin{figure}[h]
  \begin{center}
    \includegraphics[width=7cm]{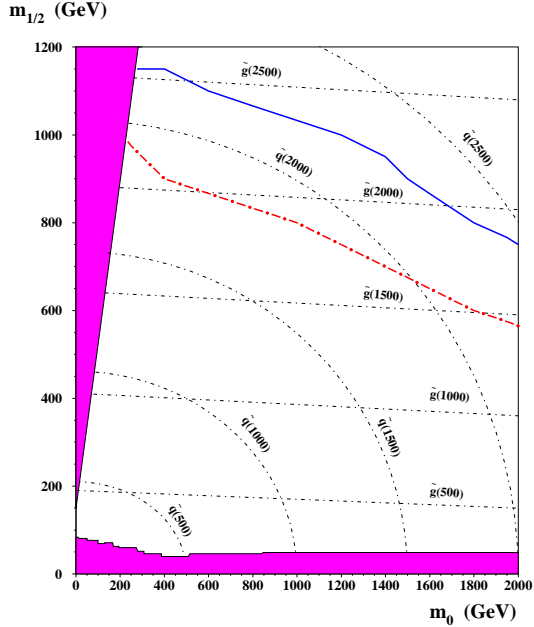}
    \caption{$5 \sigma$ discovery potential of CMS in mSUGRA for $\mu<0$, 
	$\tan{\beta}=2$ and $A_0=0$. In the region below the black curve
  	a signal of \rpv-supersymmetry via $\lb_{121} = 0.05$ would be 
	discovered ($5 \sigma$ for an integrated luminosity  of 
	$10^4~pb^{-1}$). 
	The dotted curve corresponds to the discovery potential for a 
	signal of \rpv-supersymmetry via $\lb_{233} = 0.06$.	
	In the shaded regions mSUGRA model is 
	not valid or $\chio_1$ is no longer the LSP.}
    \label{exclu}
  \end{center}
\end{figure}

We define the signal significance as $N_S/\sqrt{N_S+N_B}$ where $N_S$ and $N_B$
are respectively the number of signal and background events. The mSUGRA plan 
is scanned and for each points the signal significance is computed. The 
$5 \sigma$ contour is then determined and the result is shown in figure~\ref{exclu} 
for $\lb_{121} = 0.05$ (continuous curve) and $\lb_{233} = 0.06$
(dotted curve) with an integrated luminosity of $10^4~pb^{-1}$. 
Since these couplings correspond to the two extreme scenarii for a 
pure leptonic decay of $\chio_1$, the discovery potential for any 
$\lb_{ijk}$ should lie between the two curves. 
In presence of \rpv-supersymmetry via any $\lb_{ijk}$ terms, 
with one year at low luminosity ($10^4~pb^{-1}$) the maximum gluino mass reach 
varies from 1.5 to 2.5 TeV depending on $m_0$ and $\lb_{ijk}$ while the squark
mass varies from 1.8 to 2.5 TeV. The discovery potential with \rpv-SUSY is 
found to be better, or of the same order, than with $R$-parity 
conserved scenarii (in case of pure leptonic decays). 
More details of this analysis could be found in \cite{mypaper}.
Analyses for more pessimistic couplings ($\lbp$ and $\lbpp$) are underway.

\newpage

\section{Neutrino masses and $R$-parity violation}

Neutrinos are massive in numerous extensions of the Standard Model. In
supersymmetric models, three types of contributions to neutrino masses
and mixings can be present\footnote{Of course, neutrino masses can
receive contributions from other (non supersymmetric) mechanisms, such
as the well-known seesaw mechanism \cite{seesaw}, which involves heavy
right-handed neutrinos.}: (i) neutrino-zino mixing via sneutrino vacuum
expectation values \cite{aulak,rossvalle,nu-zino}; (ii) neutrino-higgsino mixing via
bilinear $R$-parity violating terms $\mu_i\, L_i H_u$
\cite{hallsuz,dawson}; (iii) fermion-sfermion loops induced by the
trilinear $R$-parity violating terms $\lambda_{ijk}\, L_i L_j \bar e_k$
and $\lambda'_{ijk}\, L_i Q_j \bar d_k$ 
\cite{hallsuz,dimohall,nu-lambda,godbole,decarlos}. These
contributions are generally expected to be large, if present, and
have to match the experimental limits on neutrino masses 
\begin{equation}
  m_{\nu_e} < 4.5\ eV \qquad m_{\nu_\mu} < 160\ keV \qquad m_{\nu_\tau}
< 23\ MeV 
\end{equation}
and the cosmological bound on stable, doublet neutrinos, $\sum_i
m_{\nu_i}\ \leq\ {\cal O} (10\, eV)$. Note that any of these mechanisms
require $R$-parity violation, since a sneutrino vev induces $R$-parity
breaking. Note also that the introduction of a right-handed neutrino is
not required; it follows that only Majorana masses are generated: 
\begin{equation}
  -\, \frac{1}{2}\, M^{\nu}_{ij}\: \bar \nu_{Li}\, \nu^c_{Rj}\ +\ h.c.
\end{equation}
where $\nu^c_{Rj}$ is the $CP$ conjugate of $\nu_{Lj}$, and $M^{\nu}$ is
a symmetric matrix. The relative rotation between charged lepton and
neutrino mass eigenstates defines a lepton mixing matrix, which is the
analog of the CKM matrix in the quark sector, and is responsible for
neutrino oscillations.

Let us first consider mechanism (i). Since the squared masses of the
sneutrinos receive negative contributions from the $D$-terms, it is not
unlikely that they assume a vev - actually radiative corrections ensure
that only the tau sneutrino has a vev. As a result, the tau neutrino
mixes with the zino, and we end up with a $5 \times 5$ neutralino mass
matrix. Its diagonalization yields a mass eigenstate which can be
identified with a massive tau neutrino \cite{barbieri}: 
\begin{equation}
  m_{\nu_3}\ =\ \frac{M^2_Z\, (M_1 c_w^2 + M_2 s_w^2)\, \mu}{M_1\, M_2\,
\mu\, -\, M^2_Z\, \sin 2 \beta\, (M_1 c_w^2 + M_2 s_w^2)}\ \left(
\frac{\vev{\widetilde{\nu_{\tau}}}}{v} \right)^2 
\end{equation}
where $c_W \equiv \cos \theta_W$, $s_W \equiv \sin \theta_W$, $v = 174\,
GeV$ and $M_1, M_2$ are the $U(1)_Y \times SU(2)_L$ gaugino soft
masses. Thus $m_{\nu_3}$ is typically of the order of the weak scale,
unless $\vev{\widetilde{\nu_{\tau}}} \ll v$. The LEP limit on
$m_{\nu_{\tau}}$ actually requires $\vev{\widetilde{\nu_{\tau}}}
\lesssim 1\, GeV$, while if the tau neutrino does not decay quickly
enough, the cosmological bound turns into $\vev{\widetilde{\nu_{\tau}}}
\lesssim 1\, MeV$. Note that if the theory initially preserves
$R$-parity, it is spontaneously broken by the sneutrino vev, and
trilinear couplings $\lambda_{3jk}\, L_3 L_j \bar e_k$ and
$\lambda'_{3jk}\, L_3 Q_j \bar d_k$ are generated. As will be shown
below, this gives rise to nonzero masses and mixings for the electron
and muon neutrinos. In addition, if lepton number is a global symmetry
of the theory, its spontaneous breaking yields a massless Goldstone
boson called Majoron. Energy loss from red giant stars via Majoron
emission then leads to the constraint \cite{Majoron}
$\vev{\widetilde{\nu_{\tau}}} \lesssim 100\, keV$, which would require a
large amount of fine-tuning in the scalar potential.

Let us now consider mechanism (ii) (we refer to Ref. \cite{BDLS,note_GDR} for
more details). The bilinear terms $\mu_i\, L_i H_u$ in the
superpotential induce a mixing between the neutral leptons and up
higgsino that cannot be completely rotated away in the presence of
generic soft terms. Indeed, the sneutrinos acquire vevs together with
the Higgs bosons. By redefining the lepton and down Higgs superfields in
such a way that only $H_d$ assumes a vev, one finds that one lepton
superfield (say $L_3$) still couples to $H_u$. As a consequence, the
diagonalization of the neutralino mass matrix yields one massive
neutrino \cite{Hempfling}: 
\begin{equation}
 m_{\nu_3}\ =\ \frac{M_Z^2\, \cos^2\! \beta\, (M_1 c_w^2 + M_2 s_w^2)\,
\mu \cos \xi}{M_1\, M_2\, \mu \cos \xi - M_Z^2\, \sin {2 \beta}\, (M_1
c_w^2 + M_2 s_w^2)}\ \tan^2 \xi 
\label{eq:m_nu}
\end{equation}
where $\xi$ is the angle between the vectors ${\bf v} = \{ v_0 \equiv
\vev{H^0_d};\, v_i \equiv \vev{L^0_i} \}$ and $\mbox{\boldmath{$\mu$}} =
\{ \mu_0 \equiv \mu;\, \mu_i \}$; thus $m_{\nu_3}$ vanishes in the limit
where the $v_{\alpha}$ are proportional to the $\mu_{\alpha}$. The
factor in front of $\tan^2 \xi$ is typically of the order of the weak
scale; therefore, the LEP limit on $m_{\nu_\tau}$ requires a strong
alignment ($\sin \xi \ll 1$) of the $v_{\alpha}$ along the
$\mu_{\alpha}$, typically $\sin \xi \lesssim 10^{-2}$, while the
cosmological bound is satisfied for $\sin \xi \lesssim 10^{-5}$. Such an
alignment could follow from some GUT scale-universality in the soft
terms \cite{nilles} or from horizontal symmetries \cite{banksnir,Borzumati}. 

Consider finally mechanism (iii). Trilinear $R$-parity violating
couplings contribute to each entry of the neutrino mass matrix through
lepton-slepton or quark-squark loops: 
\begin{equation}
 M^{\nu}_{ij} |_{\lambda}\ \simeq\ \sum_{k,l,m,n} \frac{\lambda_{ikl}
\lambda_{jmn}}{8 \pi^2}\, \frac{M^e_{kn} (\widetilde{M}^{e\,
2}_{\scriptscriptstyle{LR}})_{ml}}{\widetilde{m}^2_e} 
\label{eq:m_lambda}
\end{equation}
\begin{equation}
 M^{\nu}_{ij} |_{\lambda'}\ \simeq\ \sum_{k,l,m,n} \frac{3\,
\lambda'_{ikl} \lambda'_{jmn}}{8 \pi^2}\, \frac{M^d_{kn}
(\widetilde{M}^{d\, 2}_{\scriptscriptstyle{LR}})_{ml}}{\widetilde{m}^2_d} 
\label{eq:m_lambda'}
\end{equation}
where $M^e$ ($M^d$) is the charged lepton (down quark) mass matrix,
$\widetilde{M}^{e\, 2}_{\scriptscriptstyle{LR}} = M^e_{ij}\, (A^e_{ij} +
\mu \tan \beta)$ ($\widetilde{M}^{d\, 2}_{\scriptscriptstyle{LR}} =
M^d_{ij}\, (A^d_{ij} + \mu \tan \beta)$) is the left-right slepton (down
squark) mass-squared matrix, $\widetilde{m}_e$ ($\widetilde{m}_d$) is an
averaged scalar mass, and $3$ is a colour factor. Assuming no strong
hierarchy among the $A^e_{ij}$ and $\lambda_{ijk}$ (the $A^d_{ij}$ and
$\lambda'_{ijk}$) - or a flavour structure that is linked to the fermion
mass hierarchy \cite{Banks,note_GDR} -, the contributions with $k,l,m,n
= 2 \mbox{ or } 3$ dominate. To get an order of magnitude estimate, let
us assume that the dominant diagrams involve tau-stau and bottom-sbottom
loops, respectively, i.e. 
\begin{equation}
M^{\nu}_{ij} |_{\lambda}\ \simeq\ \frac{\lambda_{i33} \lambda_{j33}}{8
\pi^2}\, \frac{m^2_{\tau}\, (A_{\tau} + \mu \tan
\beta)}{m^2_{\widetilde{\tau}}}\ \sim\ \lambda_{i33}\, \lambda_{j33}\,
(4 \times 10^5\, eV) \left( \frac{100\, GeV}{\widetilde{M}_{\tau}}
\right) 
\label{eq:tau_stau}
\end{equation}
\begin{equation}
M^{\nu}_{ij} |_{\lambda'}\ \simeq\ \frac{3\, \lambda'_{i33}
\lambda'_{j33}}{8 \pi^2}\, \frac{m^2_b\, (A_{b} + \mu \tan
\beta)}{m^2_{\widetilde{b}}}\ \sim\ \lambda'_{i33}\, \lambda'_{j33}\,
(8 \times 10^6\, eV) \left( \frac{100\, GeV}{\widetilde{M}_b}  \right) 
\label{eq:bottom_sbottom}
\end{equation}
where $\widetilde{M}_{\tau}$ ($\widetilde{M}_b$) is a combination of
$\mu$, $\tan \beta$ and soft parameters. Generic $R$-parity violating
couplings would lead to a large electron neutrino mass; the present
experimental bound on $m_{\nu_e}$ therefore provides indirect limits on
$\lambda_{133}$ and $\lambda'_{133}$: 
\begin{equation}
\lambda_{133} \; \lesssim \; 3.10^{-3}\
 \left(\frac{\widetilde{M}_{\tau}}{100\ GeV}\right)^{\frac{1}{2}}
\hskip 1cm  \mbox{and}  \hskip 1cm
\lambda'_{133} \; \lesssim \; 7.10^{-4}\
 \left(\frac{\widetilde{M}_b}{100\ GeV}\right)^{\frac{1}{2}}
\end{equation}
Let us mention, however, that if the bottom-sbottom contribution
(\ref{eq:bottom_sbottom}) were dominant for each entry, the matrix
$M^{\nu}_{ij}$ would be singular at leading order since $M^{\nu}_{ii}
M^{\nu}_{jj} \simeq M^{\nu}_{ij} M^{\nu}_{ji}$, resulting in a
suppression of the light neutrino masses (as usual, the above limits are
obtained by assuming that only one coupling is nonzero). In any case, a
small $R$-parity violation, with $\lambda$ and $\lambda'$ couplings
comparable in strength with Yukawa couplings, could induce neutrino
masses in the phenomenologically interesting range, namely $10^{-3}\, eV
\lesssim m_{\nu} \lesssim 10\, eV$. This, of course, strongly depends on
the model. 

Let us stress, finally, that in any of the cases considered above,
contributions (\ref{eq:m_lambda}) and (\ref{eq:m_lambda'}) are
present. Indeed, in both cases (i) and (ii), the lepton and down Higgs
superfields have to be redefined in such a way that only $H_d$ assumes a
vev; it follows that trilinear $R$-parity violating couplings are
generated even if they were absent from the initial theory.  

We conclude that supersymmetry without $R$-parity implies neutrino
masses and oscillations, and that a small $R$-parity violation could be
of great relevance for neutrino phenomenology.

\newpage

\section{Conclusions and perspectives}

The group of \rpv of the GDR for supersymmetry,
after about two years of regular and successful running,
has covered a wide range of activities in the fields of \rpv effects.
\par
These activities have started with reviews of the state of the art and updates
concerning indirect effects and bounds on $R$-parity odd interactions as well
as reviews on the rich phenomenology and handful results obtained at the HERA and LEP
colliders for which we have benefited of the work done in the 
collaborations i.e. H1, ALEPH, DELPHI, and L3,
where members of the group of \rpv are active. 
The exploration of the phenomenology and discovery potentiel
of \rpv effects has started in the LHC 
collaborations  i.e. ATLAS and CMS,
where members of the group of \rpv are also active.
The study of the discovery potential  mainly concerned at the moment
the $\lambda_{ijk}$ coupling.
At the same time, simulation tools
are developped in the group of \rpv in order to include \rpv effects
in hadronic machines such as TEVATRON and LHC and these tools extend
existing code like ISASUSY, SUSYGEN, SPYTHIA and EUROJET.
\par
Theoretical contributions to the group of \rpv of the GDR for supersymmetry
have covered fundamental aspects such as fermion mass models based 
on abelian family symmetries, leading to a hierarchy among \rpv
couplings that mimics, in order of magnitude, the existing hierarchy
among Yukawa couplings or models in which neutrino masses can be understood
in terms of \rpv couplings effects. Theoretical activities also
concerned  important and more phenomenological aspects such as the production
of single supersymmetric particles with
the help of \rpv couplings at colliders like LEP and TEVATRON.
In addition, we have benefited of the contributions from guests 
of the group of \rpv on various theoretical and phenomenological
aspects of supersymmetry with \rpv (see \cite{guest}). 
\par
In the near future, we will continue to benefit from the work
done at the HERA and LEP colliders for the search of \rpv effects
and we will remain tuned for the update of
new indirect effects on $R$-parity odd interactions. We hope to
extend further the exploration of the phenomenology of \rpv effects at the LHC
by considering
$\lambda^{\prime}_{ijk}$ and $\lambda^{\prime\prime}_{ijk}$ couplings.
Future work can also include the study of 
\rpv effects at the next leptonic linear collider (${\sqrt s} \sim$ 500 GeV or more).
We will certainly have to face important questions concerning neutrino masses
in terms of \rpv interactions
and this will generates activities in the group of \rpv.
Further theoretical developpments on the fundamental side 
can include a better understanding of the possible hierarchy among 
\rpv couplings, this, along the lines described above or, why not,
along new lines unfully explored yet, with the hope for explicit models
to be derived and testable experimentally.
On a more phenomenological side,
the difficult task of developping tools for
RGE includings the \rpv couplings is still desirable but may wait for strong
physical motivations.
However, one cannot exclude that one of these motivations may come from
the possibility that neutrino have mass that can be understood
in terms of \rpv effects.

\newpage

\section{Appendix A}

We  adopt the summation convention  over dummy indices.
 The  conventions in the review of Haber and Kane are followed throughout \cite{haber}.
 We work in a metric of signature $(+---)$ and use: $P_{L,R}=(1\mp \g_5), \ 
 (g,g')= (g^2+{g'}^2)^\ud (\cos \t_W, \sin \t_W),\ \ 
e/g=\sin \t_W, \ e/g '=\cos \t_W, m_W^2={g^2v^2\over 4}, \  m_Z^2={(g^2+g^{'2} ) 
v^2\over 4}, \
(Z \ W )=\pmatrix {\cos \t_W &  -\sin \t_W \cr \sin \t_W   &  \cos  \t_W }\  ( W\ B) ,\
 L= g' Y^\mu B_\mu + g  J^{\mu  a} 
W_\mu  ^a+ g_3 J^{\mu \a } G_\mu ^\a , \ \tan \b =v_u/v_d$.
Often, one uses  the alternate  notations: $g'=g_1$, $g =g_2 \, g_3=g_s$.
Frequent notations  used in   GUT or SGUT discussions are:
$ g^{'2}_a=g_a^2  k_a, \ k_a=[1,1, {5\over 3}] , \ {1\over k_a g_a ^2} = {1-\b _a t 
\over g_X^2 }, \
M_a (t)= m_\ud  (1-\b_a t), \ \b_a = b_a g_X^2 /(4\pi )^2,  \ t=\log m_X^2/Q^2,\  b_a=[
 3, -1,-11],  \ \ [a=3,2,1] $.
Numerical values for some familiar parameters are: $ m_P= \sqrt {8\pi } / \kappa  =   
 1.22 \ 10^{19} GeV, \
 m_X= 2\ 10^{16 } GeV, \, k_a g_a^2 =4\kappa ^2/\a ' = 32\pi /(\a ' M_P^2) , M_{string}=
g_X 5.27 \ 10^{17} GeV .$  
The following notations for  the Standard Model classical (tree level)  parameters are 
also used: 
$a(f_H)\equiv a_H(f)=a(\tilde f_H)=2T_3^H (f)-2Qx_W, 
[H=L,R], \ x_W= \sin^2 \t_W$;
$a(\tilde f_H^\star ) =-a(\tilde f_H), \  a_L(f^c)=-a_R(f), 
\  a_R(f^c)=-a_L(f), \   {G_\mu \over \sqrt 2} = {g^2\over 8M_W^2}. $    
Recall that the input parameters employed in high precision tests  of the standard 
model are chosen  as the subset of  
best experimentally  determined  parameters among the  basic  set,  
$[\a ^{-1} = 137.036, \
\a _s= 0.122 \pm 0.003 ,\ m_Z= 91.186 (2),\  G_\mu = 1.16639(1) \ 10^{5} GeV^{-2} , \ 
m_t (pole) = 175.6 \pm 5.0 , m_H]$.  The remaining parameters  are then 
  deduced by means of  fits to  the familiar basic 
data (Z-boson lineshape and decay widths, $\tau $ polarization, forward-backward (FB)  
or polarization asymmetries, APV, beta decays, masses,  ...).  
Experimental data can be consulted from the PDG compilation \cite{pdg}.  

Useful auxiliary parameters for the \rpv coupling constants are, 
 $ r_{ijk} (\tilde e_{kR})= {M_W^2\over g_2^2\tilde 
m^2_{e_{kR}}  } \vert \l_{ijk}\vert^2$. 
In presenting numerical results for coupling constants,
we  distinguish between the fisrt two families and the third by using
middle  and begining alphabet indices,
respectively, such that,   $ l, m, n    \in [1,2] $ and
$   i, j , k  \in  [1,2,3]$. The following list of abbreviations 
is used: \rpv for $R$-parity violation, NC for neutral current, 
CC for charged current, BF for branching fraction, SM  for Standard Model, 
and EDM for electric dipole moment.
A factor $d_{kR}^p$  in a numerical equation, such as, $ \l '_{11k}= n\times
{d^p_{kR}  } $,  
stands  for the notation,  $n\times ({ m_{d_{kR}} \over 100 GeV})^p $.
The following notations for quadratic products of coupling constants are used: 
$F_{abcd}=\sum_i \l _{iab} \l _{icd} ({\tilde m_i\over 100 GeV })^{-2} ,\  
F'_{abcd}=\sum_i \l ' _{iab} \l' _{icd} ({\tilde m_i\over 100 GeV })^{-2} $. 
\section{Appendix B}
The lagrangian in four-component Dirac notation describing the \rpv Yukawa
interaction terms (i.e. couplings scalar-spinor-spinor) is
 \cite{bargerg,rdecay} :
\begin{equation}
\begin{array}{lll}
{\cal L}_{{\not R}_p} & = & \lb_{ijk} \left[ 
                                 \nuL^i \bar{e}_R^k e_L^j +
                                 \eL^j \bar{e}_R^k \nu_L^i + 
                                 (\eR^k)^* (\bar{\nu}_L^i)^c e_L^j - 
                                 (i \leftrightarrow j) 
                               \right] +\\
               &   & \lbp_{ijk} \left[ 
                                  \nuL^i \bar{d}_R^k d_L^j + 
                                  \dL^j \bar{d}_R^k \nu_L^i +
                                  (\dR^k)^* (\bar{\nu}_L^i)^c d_L^j -
                                  \eL^i \bar{d}_R^k u_L^j -
                                  \uL^j \bar{d}_R^k e_L^j - 
                                  (\dR^k)^* (\bar{e}_L^i)^c u_L^j
                                \right] +\\
               &   & \lbpp_{ijk} \left[
                                   (\bar{u}_R^i)^c d_R^j \tilde{d}_R^k +
                                   (\bar{u}_R^i)^c \tilde{d}_R^j d_R^k +
                                   \tilde{u}_R^i (\bar{d}_R^j)^c d_R^k
                                \right]\\
               &   & + h.c.
\end{array}
\label{equ2}
\end{equation} 
Here, the superscripts $^c$ stand for the charge conjugate spinors and
the $^*$, for the complex conjugate of scalar fields.
The coupling constant $\lb$ is antisymmetric under
the interchange of the first two indices while $\lbpp$ is
antisymmetric under the interchange of the last two. Therefore, there
are 9+9 such couplings and 27 $\lbp$ leading to 45 new coupling
constants. The first two terms of equ.(\ref{equ2}) induce a lepton number
violation while the last one violates baryon number conservation. 

 
\end{document}